\documentclass[a4paper,11pt]{article}
\usepackage{jheppub} 
\usepackage{lineno}
\usepackage{arydshln}
\usepackage{textcomp}
\usepackage{xcolor}
\usepackage[frozencache,cachedir=minted-cache-new]{minted} 
\usepackage{comment}
\usepackage{graphicx}
\usepackage{hyperref}
\usepackage{lipsum}
\usepackage{mathtools}
\usepackage[normalem]{ulem}
\usepackage{tikz}
\usepackage{booktabs}
\usetikzlibrary{arrows.meta, positioning, calc, decorations.pathreplacing}
\usepackage{placeins}

\definecolor{ForestGreen}{rgb}{0.13, 0.55, 0.13}

\title{Transforming Calabi-Yau Constructions: Generating New Calabi-Yau Manifolds with Transformers}

\author[a]{Jacky H. T. Yip,$^{\spadesuit,}$}
\author[b]{Charles Arnal,$^{\spadesuit,}$}
\author[c,d]{François Charton,}
\author[a]{Gary Shiu}
\affiliation[a]{Department of Physics, University of Wisconsin-Madison}
\affiliation[b]{FAIR, Meta}
\affiliation[c]{CERMICS, École Nationale des Ponts et Chaussées}
\affiliation[d]{Axiom}

\emailAdd{hyip2@wisc.edu}
\emailAdd{charlesarnal@meta.com}
\emailAdd{francois.charton@enpc.fr}
\emailAdd{shiu@physics.wisc.edu}

\abstract{Fine, regular, and star triangulations (FRSTs) of four-dimensional reflexive polytopes give rise to toric varieties, within which generic anticanonical hypersurfaces yield smooth Calabi-Yau threefolds. We introduce CYTransformer, a deep learning model based on the transformer architecture, to automate the generation of FRSTs. We demonstrate that CYTransformer efficiently and unbiasedly samples FRSTs for polytopes across a range of sizes, and can self-improve through retraining on its own output. These results lay the foundation for \href{https://aicy.physics.wisc.edu}{\texttt{AICY}}: a community-driven platform designed to combine self-improving machine learning models with a continuously expanding database to explore and catalog the Calabi-Yau landscape.}

\footnotetext{$^\spadesuit$ Equal contribution, randomized order.}

\begin{document}
\maketitle
\flushbottom

\section{Introduction}

The vastness of the string landscape presents a serious computational challenge. This immensity stems from the multitude of choices for the internal manifolds on which string theory is compactified (or for non-geometric constructions, choices of conformal field theory). Even with a fixed compactification manifold, additional discrete choices---such as bundle or brane configurations, and the quantized fluxes threaded through internal cycles---further enlarge the space of solutions.

Despite its vastness, the string landscape is conjectured to be finite, in the sense that there are only finitely many low energy effective field theories with a fixed, finite energy cutoff that are consistent with quantum gravity~\cite{Douglas:2003um,Vafa:2005ui,Acharya:2006zw}. The finiteness of the landscape is both an important premise in the program of landscape statistics~\cite{Douglas:2003um} and argued to be a universal property of quantum gravity~\cite{Vafa:2005ui}. It is however only when we restrict to very small regions of the landscape, e.g., intersecting D-brane models in a specific Calabi-Yau orientifold, that an exact number of solutions is known~\cite{Loges:2022mao} (though it was shown earlier that the number is finite~\cite{Douglas:2006xy}).

Compactifications of string theory on Calabi-Yau manifolds stand out as an especially well-motivated class of solutions for data mining the landscape. In particular, Calabi-Yau threefolds  yield four-dimensional vacuum configurations of superstring theory that can potentially accommodate realistic particle physics coupled to gravity~\cite{Candelas:1985en}.\footnote{For recent review of the state-of-the-art in constructing the Standard Model from string compactifications, see~\cite{Marchesano:2024gul}.} While Calabi-Yau threefolds are neither fully classified nor known to be finite in number, those that can be realized as hypersurfaces in toric varieties are, at least in principle, amenable to combinatorial enumeration. As we review below, Batyrev's construction~\cite{Batyrev:1993oya} of toric Calabi-Yau threefolds begins with fine, regular, star triangulations (FRSTs) of four-dimensional relexive polytopes. There are in total $473{,}800{,}776$ such polytopes which have been fully enumerated by Kreuzer and Skarke~\cite{Kreuzer:2000xy}. The finiteness of this class of Calabi-Yau manifolds stems from the fact that the number of reflexive polytopes is finite. However, the formidable combinatorics lie in the FRSTs whose number grows exponentially with the number of vertices of a polytope. On the other hand, different FRSTs can give rise to topologically equivalent Calabi-Yau threefolds. As a result, a complete enumeration of inequivalent Calabi-Yau threefold hypersurfaces in toric varieties is currently unfeasible, though their number was shown to be bounded by $N_{\rm CY} < 1.65 \times 10^{428}$~\cite{Demirtas:2020dbm}. A recently developed software package \texttt{CYTools}~\cite{Demirtas:2022hqf} has strong capabilities in triangulating polytopes and in computing topological data for toric Calabi-Yau manifolds. However, even with the impressive speedup, it would take  longer than the age of the universe to uncover a sizable fraction of toric Calabi-Yau threefolds, not to mention a complete enumeration.

The current state of affairs calls for a scalable, self-improving learning algorithm to automate the generation of toric Calabi-Yau manifolds. Non-learning algorithms such as those involving bistellar flips~\cite{10.5555/1952022}, random walks, or both, do not scale well with the number of vertices of the polytope in exploring the full space of FRSTs~\cite{Demirtas:2020dbm}. Genetic algorithms~\cite{MacFadden:2024him} and reinforcement learning~\cite{Berglund:2024reu} have been used to perform targeted searches for toric Calabi-Yau threefolds that yield favorable phenomenological features in the resulting four-dimensional theory. However, these algorithms have not been used to significantly expand the search space of FRSTs. For instance, it has not been demonstrated that the lessons learned from triangulating a given polytope are transferrable to other polytopes. It is also not clear that the FRSTs generated by these algorithms are representative of the full ensemble.

In light of these considerations, we develop \textbf{CYTransformer}---an encoder-decoder transformer model~\cite{Vaswani:2017lxt}---to automate the generation of toric Calabi-Yau spaces. Transformers are machine learning models designed to process and generate sequential data in an autoregressive manner. Their attention mechanisms allow them to capture complex dependencies within the data. They have proved in recent years to be widely successful at sequence modeling, and in particular at natural language processing. The transformer architecture underpins all state-of-the-art large language models~\cite{Touvron2023LLaMAOA,Jiang2023Mistral7,Achiam2023GPT4TR,Llama3} and has made the recent generative AI revolution possible. While transformers' ability to model natural language and programming code is now well-established, they have also been applied to more exotic data distributions coming from mathematics~\cite{Funsearch, charton2024patternboost, Alfarano, hashemi2025transformersenumerativegeometry}, physics~\cite{Geneva2020TransformersFM, Janny2023EagleLL,  Cai:2024znx, Smatrix}, or chemistry~\cite{jumper2021highly}. Unlike natural and programming languages, which have shared grammatical and logical structures~\cite{Inductionheads, Iterationhead, FunctionVI} that transformers are known to be well-suited for, in contrast, data distributions from new scientific problems vary widely, creating unique challenges. Finding an efficient and transformer-compatible representation for such problems is a non-trivial task.

In our approach, the coordinates of the vertices of a given polytope are used as inputs to the transformer's encoder, and the transformer's decoder outputs a token sequence which represents a candidate triangulation of the polytope. Within the sequence, each token corresponds to a simplex. With this setup, the probabilistic generation of the next simplex in the sequence is conditioned on all previous ones, as well as on the input polytope. CYTransformer successfully learns the intricate structure of FRSTs, which allows it to \emph{efficiently} generate new FRSTs for polytopes unseen during training for constructing new Calabi-Yau manifolds. Moreover, we observe that the FRSTs generated are \emph{representative} of the entire ensemble. Finally, we show that CYTransformer can continuously improve beyond its initial training data by repeatedly generating new FRSTs and retraining on them. This method, which we term \textit{self-improvement}, is a form of reinforcement learning with rejection sampling, a technique that has recently become a staple in large language model training~\cite{R1}. This approach helps mitigate the difficulty of obtaining diverse training data for large polytopes, offering a promising path toward scaling CYTransformer to larger polytopes.

These promising results motivate us to launch \texttt{AICY}: \href{https://aicy.physics.wisc.edu}{https://aicy.physics.wisc.edu}---\emph{living} software and data repositories for Calabi-Yau manifolds. Existing databases for data mining the toric Calabi-Yau landscape such as~\cite{KS-Database,Altman-Database}, while documenting the state-of-the-art at a given point in time, are static. We envision \texttt{AICY} as a community-driven platform integrating a software repository of machine learning models, like the CYTransformer, which self-improves by training on a data repository continuously enriched by user contributions. As importantly, \texttt{AICY} will enable targeted Calabi-Yau searches by leveraging its repository models with techniques such as reinforcement learning, which act on user-supplied reward functions to guide an adaptive and physics-informed exploration of the string landscape.

This paper is organized as follows. Section~\ref{sec:CY} reviews how Calabi-Yau manifolds arise from FRSTs of reflexive polytopes. Section~\ref{sec:databases} surveys existing triangulation algorithms, focusing on \texttt{CYTools} which we use for data preparation and verification. Section~\ref{sec:machine_learning} details CYTransformer's architecture, data encodings, and experimental setup. Section~\ref{sec:results} evaluates CYTransformer's performance on generation efficiency and representativeness, proposes a hybrid method, and explores the model's self-improvement capability. Finally, section~\ref{sec:AICY} presents our proposal for \texttt{AICY}, and section~\ref{sec:dis} discusses future research directions.

\section{Calabi-Yau manifolds from reflexive polytopes via FRSTs}\label{sec:CY}
The task of generating FRSTs of reflexive polytopes arises from their role in constructing Calabi-Yau manifolds. As shown by Batyrev~\cite{Batyrev:1993oya}, an FRST of a reflexive polytope $\Delta^\circ$ gives rise to a sufficiently smooth, projective, and compact toric variety. This toric variety can then serve as the ambient space for a smooth Calabi-Yau hypersurface, defined by monomials associated to lattice points in the polytope $\Delta$ dual to $\Delta^\circ$. We provide in this section the necessary mathematical background for this construction.

\subsection{Toric varieties}
Standard references on toric varieties are~\cite{Tadao2012-bj,Fulton1993-hj,Cox2024-vp}. Consider a full-dimensional fan $\Sigma$ consisting of cones $\sigma_i$ in an $n$-dimensional lattice $N\cong \mathbb{Z}^n$. Associate to each ray ($1$-dimensional cone) $\sigma_i(1)\in\Sigma$ a homogeneous coordinate $z_i\in\mathbb{C}$. These coordinates are subject to the continuous scaling symmetries
\begin{equation}
\label{eq:hcoor}
    (z_1,z_2,\dotsc,z_r)\sim(\lambda^{q_1} z_1, \lambda^{q_2} z_2,\dotsc,\lambda^{q_r} z_r)    
\end{equation}
for $\lambda\in\mathbb{C}^*\coloneq\mathbb{C}\backslash\{0\}$ and integers $q_i$ satisfying
\begin{equation}
\label{eq:Geq}
    \sum_{i=1}^{r}q_iv_i=0,
\end{equation}
where $v_i\in N$ is the primitive generator of $\sigma_i(1)$. \eqref{eq:Geq} has $r-n$ independent solutions, giving the rank of the continuous scaling group. The associated \emph{toric variety} $\mathcal{A}(\Sigma)$ is obtained via the Cox construction:
\begin{equation}
    \mathcal{A}(\Sigma)\coloneq\frac{\mathbb{C}^r\backslash Z(\Sigma)}{(\mathbb{C}^*)^{r-n}\times G},
\end{equation}
where $(\mathbb{C^*})^{r-n}$ mods out the continuous scalings in~\eqref{eq:hcoor}, and $G=N/\text{span}_\mathbb{Z}\{v_i\}$ mods out the discrete symmetries from finite root-of-unity rescalings of the $z_i$ that arise when the $v_i$ span only a sublattice of $N$. The exceptional set
\begin{equation}
    Z(\Sigma)\coloneq\bigcup_{\text{primitive collections }I}\{z_i=0\;\forall i\in I\},
\end{equation}
with a primitive collection defined as a minimal set of rays not contained in any single cone of $\Sigma$, removes points whose vanishing coordinate pattern does not correspond to any cone of the fan, ensuring every remaining point lies in some affine chart $U_{\sigma_i}$ of $\mathcal{A}(\Sigma)$.

\subsection{Calabi-Yau manifolds in toric varieties}
The anticanonical divisor in a toric variety is $-K_{\mathcal{A}(\Sigma)}=\sum_i D_{\sigma_i(1)}$, where $D_{\sigma_i(1)}$ is the toric divisor corresponding to the ray $\sigma_i(1)$, i.e., the hypersurface $\{z_i=0\}$. If $X\subset\mathcal{A}(\Sigma)$ is a hypersurface whose divisor class satisfies $[X]=-K_{\mathcal{A}(\Sigma)}$, then the adjunction formula gives
\begin{equation}
    K_X=(K_{\mathcal{A}(\Sigma)}+[X])|_X=(K_{\mathcal{A}(\Sigma)}-K_{\mathcal{A}(\Sigma)})|_X=0,
\end{equation}
i.e., the canonical divisor $K_X$ of $X$ vanishes. This implies that the canonical bundle of $X$ is trivial, and $X$ has vanshing first Chern class $c_1(X)=0$. If $X$ is also compact and projective (hence Kähler), then $X$ is a \emph{Calabi-Yau manifold}.

\subsection{From reflexive polytopes to toric Calabi-Yau manifolds}
An $n$-dimensional lattice polytope $\Delta\subset M_\mathbb{R}\coloneq M\otimes_\mathbb{Z}\mathbb{R}$ is the convex hull of a set of vertices in a lattice $M\cong \mathbb{Z}^n$. Its dual $\Delta^\circ\subset N_\mathbb{R}$ associated to the dual lattice $N\cong \mathbb{Z}^n$ is defined as
\begin{equation}
    \Delta^\circ\coloneq{\rm{Conv}}(\{y\in N\mid\langle m,y\rangle\geq -1\;\forall m\in\Delta\}).
\end{equation}
If $\Delta^\circ$ is also a lattice polytope, then both $\Delta$ and $\Delta^\circ$ are \emph{reflexive polytopes}. A reflexive polytope has exactly one interior lattice point, which is taken to be the origin.

The collection of cones generated by each face of $\Delta^\circ$ together with the origin forms the face fan $\Sigma$ of $\Delta^\circ$. If $\Delta^\circ$ is reflexive, then all of its facets lie at lattice distance 1 from the origin. This condition ensures that the anticanonical divisor $-K_{\mathcal{A}(\Sigma)}$ of the toric variety $\mathcal{A}(\Sigma)$ constructed from $\Sigma$ (as described in the previous subsections) is Cartier, and hence corresponds to the line bundle $\mathcal{O}(-K_{\mathcal{A}(\Sigma)})$. The lattice points of $\Delta$ index the global sections of $\mathcal{O}(-K_{\mathcal{A}(\Sigma)})$; choosing a generic section $s$ in the space of global sections of $\mathcal{O}(-K_{\mathcal{A}(\Sigma)})$ over $\mathcal{A}(\Sigma)$ and setting $s=0$ defines an anticanonical hypersurface $X\subset\mathcal{A}(\Sigma)$, which is a Calabi-Yau manifold. Concretely,
\begin{equation}
    s\coloneq\sum_{m\in\Delta\cap M}c_mp_m=0,
\end{equation}
where the coefficients $c_m$ specify the complex structure moduli of $X$, and
\begin{equation}
    p_m=\prod_{i=1}^{r}z_i^{\langle m,v_i\rangle+1}
\end{equation}
is the monomial from the homogeneous coordinates $z_i$

Up to lattice automorphisms in $GL(n,\mathbb{Z})$, the number of reflexive polytopes in any given dimension $n$ is finite; e.g., there are only $16$ distinct reflexive polytopes in $2$ dimensions. Kreuzer and Skarke developed an algorithm to enumerate \emph{all} reflexive polytopes for any $n$~\cite{Kreuzer:1998vb}. Of particular relevance to this work, they completed the classification for $n=4$~\cite{Kreuzer:2000xy}, finding $473{,}800{,}776$ polytopes from which as many as $10^{428}$ topologically inequivalent Calabi-Yau threefolds may be constructed~\cite{Demirtas:2020dbm}. This finite yet vast set of geometries enables large-scale combinatorial and statistical studies of the string landscape.

\subsection{FRSTs for desingularized, projective, and compact toric varieties}
\label{subsec:cyFRST}
A triangulation $\mathcal{T}$ of an $n$-dimensional polytope is a partition of the polytope into a finite collection of $n$-simplices, such that the intersection of any two simplices is either empty or a shared proper face. A \emph{star} triangulation of a reflexive polytope $\Delta^\circ$ is a triangulation whose every simplex contains the origin as a vertex. Coning over these simplices yields a complete simplicial fan $\Sigma$ covering $N_\mathbb{R}$, thereby ensuring the compactness of the associated toric variety $\mathcal{A}(\Sigma)$ and the Calabi-Yau hypersurface $X$ defined within it.

Singularities in $\mathcal{A}(\Sigma)$ arise when the cones in $\Sigma$ are not smooth. A cone is not smooth if its set of minimal generators cannot be extended to a full $\mathbb{Z}$-basis of the lattice. These singularities may be inherited by the hypersurface $X$ to be defined within $\mathcal{A}(\Sigma)$, rendering $X$ non-smooth. To resolve these singularities, the original $\Sigma$ must be refined by additionally requiring $\mathcal{T}$ to be \emph{fine}. Fineness requires that all maximal simplices in $\mathcal{T}$ are as subdivided as possible with respect to the lattice, so that the resulting $\Sigma$ has cones that are generated by elementary rays and are closer to being smooth. In other words, $\mathcal{T}$ should include all lattice points on the polytope, not just the vertices, as potential simplex vertices.

However, lattice points lying strictly in the interior of codimension-1 faces need not be included as vertices in the triangulation. While omitting them may induce mild singularities in $\mathcal{A}(\Sigma)$, $X$ remains smooth, as the toric divisors corresponding to those points do not intersect $X$, a generic anticanonical hypersurface. This is known as the maximal projective crepant partial desingularization. The set of points consisting of all lattice points in $\Delta^\circ$, excluding those strictly interior to codimension-1 faces, together with the origin, is referred to as the set of \emph{resolved vertices}.

The final condition we impose on $\mathcal{T}$ is \emph{regularity}, requiring that it arises from the projection of the lower convex hull of lifted resolved vertices via a height function. The lifting procedure induces a piecewise linear convex function over the resolved vertices, which guarantees that the associated $\Sigma$ is polytopal (the normal fan of a convex polytope $\subset M_\mathbb{R}$), hence the resulting toric variety is projective. Consequently, the embedded $X$ is also projective, and thus Kähler, and has finite-dimensional cohomology groups.

To summarize, we seek fine, regular, and star triangulations (FRSTs) of the resolved vertices of a reflexive polytope in order to construct sufficiently desingularized, projective, and compact toric varieties, so that the generic anticanonical hypersurface is a smooth Calabi-Yau manifold:
\begin{itemize}
    \item \textbf{Fine (F):} Every resolved vertex appears as a vertex of some simplex in $\mathcal{T}$; i.e., all relevant lattice points are used in the triangulation.
    \item \textbf{Regular (R):} $\mathcal{T}$ arises as the projection of the lower faces of a convex polytope constructed from lifting each resolved vertex $p_i$ to $(p_i,h_i)\in N\times\mathbb{R}$ with some height $h_i\in\mathbb{R}$.
    \item \textbf{Star (S):} Every simplex in $\mathcal{T}$ contains the origin as a vertex.
\end{itemize}

\subsection{Hodge numbers, lattice points, and resolved vertices}
Two important invariants for classifying a Calabi-Yau threefold $X$ are its \emph{Hodge numbers}, $h^{1,1}$ and $h^{2,1}$. They count ways in which the Kähler metric and the complex structure can be deformed, respectively. Mirror symmetry states that
\begin{equation}
\label{eq:ms}
    h^{1,1}(X)=h^{2,1}(X^\circ)\;{\rm{and}}\;h^{1,1}(X^\circ)=h^{2,1}(X),
\end{equation}
where $X$ is constructed from $\Delta$, and $X^\circ$ from $\Delta^\circ$. Baytrev proved~\eqref{eq:ms} by providing the formulas~\cite{Batyrev:1993oya}
\begin{equation}
\label{eq:h11h21eqs}
\begin{split}
    h^{1,1}(X)&=\ell(\Delta^\circ)-4-1-\sum_{\Gamma^\circ}\ell^*(\Gamma^\circ)+\sum_{\Theta^\circ}\ell^*(\Theta^\circ)\ell^*(\hat{\Theta}^\circ)\\
    h^{2,1}(X)&=\ell(\Delta)-4-1-\sum_\Gamma\ell^*(\Gamma)+\sum_\Theta\ell^*(\Theta)\ell^*(\hat{\Theta}),
\end{split}
\end{equation}
where $\ell(\alpha)$ is the number of lattice points in $\alpha$, $\ell^*(\alpha)$ is the number of lattice points in the interior of $\alpha$, $\Gamma$'s ($\Gamma^\circ$'s) are codimension-$1$ faces of $\Delta$ ($\Delta^\circ$), $\Theta$'s ($\Theta^\circ$'s) are codimension-$2$ faces of $\Delta$ ($\Delta^\circ$), and $\hat{\Theta}$ ($\hat{\Theta}^\circ$) is the face of $\Delta^\circ$ ($\Delta$) dual to $\Theta$ ($\Theta^\circ$). \eqref{eq:ms} follows directly from the fact that $(\Delta^\circ)^\circ=\Delta$ for reflexive polytopes. 

The reflexive polytopes are primarily organized by $h^{1,1}$. As $h^{1,1}$ increases, enumerating all FRSTs of a polytope using non-learning algorithms becomes computationally prohibitive, since the number of resolved vertices typically grows, reaching up to 496 for a polytope with $h^{1,1}=491$. In this work, we restrict our attention to triangulations of \emph{favorable} reflexive polytopes, defined as those $\Delta^\circ$ for which the final term in the first equation of~\eqref{eq:h11h21eqs} vanishes. That is, the number $N_{\rm vert}$ of resolved vertices satisfies the relation $N_{\rm vert}=\ell(\Delta^\circ)-\sum_{\Gamma^\circ}\ell^*(\Gamma^\circ)=h^{1,1}+4+1$. For clarity to all audiences, we refer to each polytope configuration throughout the paper by its $(h^{1,1},N_{\rm vert})$ tuple.

\section{Triangulation software and the non-learning fast sampler}
\label{sec:databases}

\texttt{CYTools}~\cite{Demirtas:2022hqf} is a software package with polytope triangulation as one of its primary design focuses. We describe below how it is utilized in our setup.

\subsection{Fetching reflexive polytopes}
The Kreuzer-Skarke (KS) database~\cite{KS-Database} contains the complete list of $473{,}800{,}776$ four-dimensional reflexive polytopes. Following Batyrev's prescription as detailed in section~\ref{sec:CY}, this database enables the construction of companion databases of polytope FRSTs and the associated Calabi-Yau threefolds, such as~\cite{Altman-Database}. In contrast, \texttt{CYTools} functions not as a static database but as a tool that fetches polytopes from the KS database on which it performs computations in real time.

\subsection{FRST enumeration}
\label{sec:all}
\texttt{CYTools} uses \texttt{TOPCOM}~\cite{Topcom} as a backend to construct FRSTs of a given point set in Euclidean space. If the point set is not full-dimensional, an affine transformation is first applied to embed it into full-dimensional space. \texttt{TOPCOM} then enumerates all FRSTs by exploring the connected subgraph corresponding to regular triangulations within the flip graph of all possible triangulations. In this flip graph, each node represents a triangulation, and each edge corresponds to a bistellar flip: a local move that replaces a subcomplex with a complementary one sharing the same boundary. In other words, the set of FRSTs can be exhaustively traversed through successive bistellar flips.

In building the training datasets, we use \texttt{CYTools} to obtain the complete sets of FRSTs of polytopes with configurations $(h^{1,1},N_{\rm vert})=(5,9+1)$, $(6,10+1)$, $(7,11+1)$, and $(8,12+1)$. For polytopes with $N_{\rm vert}>12+1$, exhaustive enumeration becomes prohibitively slow and memory-intensive, as some of these polytopes admit an extremely large number of triangulations. Complete FRST sets for selected test polytopes are also required for evaluating the representativeness of transformer-generated FRST samples (see section~\ref{subsec:perfmet}).

Notably, we have verified that for polytopes with $h^{1,1}=5$ and $6$, \texttt{CYTools} generates the same FRSTs as those stored in~\cite{Altman-Database}. However, the polytopes, or more precisely, their resolved vertex sets, are expressed in different bases in the two sources. We find that the triangulations match only after transforming the resolved vertices into a normal form, as defined in~\cite{Kreuzer:1998vb}. This basis discrepancy has implications for learning; our transformer model exhibits distinct loss curves when trained on data from~\cite{Altman-Database} versus \texttt{CYTools}, suggesting that certain bases may be more favorable for learning. We leave a detailed investigation of basis choice and its impact on learning performance to future work.

\subsection{The fast sampler}
\label{sec:fast}
We describe a non-machine-learning algorithm available in \texttt{CYTools}, known as the \emph{fast sampler}, which offers a low-cost method for obtaining FRSTs. It serves as a benchmark for evaluating the performance of our transformer model.

As described in section~\ref{subsec:cyFRST}, a regular triangulation $\mathcal{T}$ is one for which there exists a \emph{height vector} $\boldsymbol{h}$ assigned to the resolved vertices, such that lifting each vertex $i$ by $h_i$ and projecting the lower codimension-$1$ faces of the resulting convex hull yields $\mathcal{T}$. A natural choice is the \emph{Delaunay triangulation}, obtained by setting $h_i=\left|\boldsymbol{p}_i\right|^2$, where $\boldsymbol{p}_i$ denotes the coordinates of vertex $i$. 

The fast sampler samples FRSTs near the Delaunay triangulation through the following procedure:
\begin{enumerate}
    \item Initialize the height vector as $h_i=\left|\boldsymbol{p}_i\right|^2$ for each resolved vertex $i$.
    \item Sample $\epsilon_i$ from a Gaussian distribution with standard deviation $c$ for each vertex. Update the height vector according to $h_i\rightarrow h_i+\epsilon_i$.
    \item If the resulting triangulation is not fine, repeat the sampling step. Once a fine triangulation is found, lower the origin to obtain an FRST.
\end{enumerate}

The parameter $c$ controls the deviation from the Delaunay seed and thus governs the diversity of sampled triangulations. If $c$ is too large, some vertices may be lifted excessively, causing them to lie above the lower convex hull and produce non-fine triangulations. For fair comparisons, we tune $c$ carefully to optimize the fast sampler's performance.

We also use the fast sampler to generate training triangulations for the $(h^{1,1},N_{\rm vert})=(9,13+1)$ and $(10,14+1)$ configurations. While computationally efficient, the fast sampler is not expected to yield fair samples representative of the full population. Prior work~\cite{Demirtas:2020dbm} has shown that it introduces sampling bias in geometric and topological quantities.\footnote{In contrast, \texttt{CYTools} also includes a fair sampler that is designed to generate more representative samples by thoroughly exploring the height vector space and applying bistellar flips. However, for the small $(h^{1,1}, N_{\rm vert})$ polytopes considered here, the fair sampler's wall-finding step frequently fails. Improving its success rate through parameter fine-tuning on a per-polytope basis lies beyond the scope of this study.}

In summary, we make extensive use of \texttt{CYTools} to fetch polytopes from the KS database, obtain FRSTs on which our models are trained, and analyze the transformer-generated candidate triangulations.

\section{Machine learning FRSTs with CYTransformer}
\label{sec:machine_learning}


Our goal is to train a model that takes as input a reflexive polytope, represented by some encoding of its set of resolved vertices, and outputs FRSTs of that polytope. To this end, we propose the \textbf{CYTransformer}, an \emph{encoder-decoder transformer}, that encodes the input polytope as a sequence of high-dimensional vectors, one per resolved vertex, and uses an autoregressive decoder to produce a \emph{candidate triangulation}: a sequence of tokens representing the simplices of an FRST, ending with a special end-of-sequence token. The model is trained on examples, pairs of polytopes and FRSTs, generated using \texttt{CYTools}. The trained model is then used to generate FRSTs of new polytopes. Since the model only learns from examples, it has no a priori knowledge of what an FRST is. Therefore, there is no guarantee that the output of a trained CYTransformer describes a valid triangulation, let alone an FRST. We use an external tool to verify that the model's output are FRSTs.

Our experiments consist of two phases: \emph{training} and \emph{inference}. During the training phase, CYTransformer is provided with polytopes with a fixed number of resolved vertices, and associated FRSTs. The model is trained to minimize a cross-entropy loss that rewards it for predicting the same simplices as the FRSTs in the training data. CYTransformer learns a probability distribution for the next output tokens, conditional on the input polytope and the previous output tokens. This allows a trained model to output, for any input polytope with $N_{\rm vert}$ resolved vertices, a sequence of tokens that is likely to represent a valid FRST.

During the inference phase, the trained CYTransformer is used to generate candidate triangulations for new polytopes with the same number of vertices. Candidate solutions are verified with \texttt{CYTools}. If training has succeeded, we expect the model to generate new FRSTs, for polytopes that were not seen during training. When evaluating the model, we ensure that the test polytopes have not been used during training.

This two-phase process can be iterated, in a manner of \emph{self-improvement}. The FRSTs generated during the inference phase and verified by \texttt{CYTools} are collected into a new training set that is used to further train the model. The improved model is then used to generate new FRSTs, starting a new cycle. We expect this iterated fine-tuning procedure to improve model performance beyond what is possible with the initial training data, because model-generated data can drastically increase the size and diversity of the training set. We detail this process in subsection~\ref{subsec:selfim_method}.


\subsection{Model architecture}


CYTransformer uses the original encoder-decoder architecture introduced in the landmark paper~\cite{Vaswani:2017lxt}. In this model, two transformer stacks coexist: an encoder, which processes the input polytope, and an auto-regressive decoder which outputs the FRST, conditional on the encoded polytope. Modern state-of-the-art large language models use a decoder-only architecture, where input and output sequences are concatenated and are processed by a unique auto-regressive transformer stack. We believe the encoder-decoder architecture is well-suited to our problem, because a separate bidirectional encoder can learn a richer representation of input polytopes, and can be cached at inference for faster generation.

The high-level architecture, which we illustrate in figure~\ref{fig:full_architecture}, is as follows. The model takes as input a sequence representing the input polytope, a list of four-dimensional vectors representing its resolved vertices, and a sequence representing the simplices of the candidate triangulation produced so far. Both of these sequences are used to predict the next simplex of the triangulation, or a special end-of-sequence token that indicates that the prediction is complete. The input polytope sequence is transformed by the encoder into a sequence of high-dimensional vectors, the encoder output. The decoder processes the sequence of tokens predicted so far (i.e., the previous simplices in the triangulation), and the encoder output (via a cross-attention mechanism) to calculate a probability distribution $\boldsymbol{P}$ for the next output token. 

To generate an FRST, the model begins with an input polytope and an effectively empty output, samples a token from the predicted probability distribution, and adds it to the output sequence. Then, it repeats the process with the same input polytope and the extended output sequence, predicting the next token. This is repeated until the end-of-sequence token is sampled, or the sequence reaches a predefined maximal length. Note that because the output tokens are sampled from the many predicted probability distributions, the generative procedure is non-deterministic: running the same model twice on the same polytope will generally produce different candidate triangulations. This stochasticity is a key component of our design: it allows for generating many candidate triangulations of one polytope.

\begin{figure*}[t]
    \centering
    \resizebox{0.75\textwidth}{!}{\begin{tikzpicture}[
    >=Latex, thick,
    encoder/.style={
        draw=blue!50, fill=blue!15, rounded corners=8pt,
        minimum width=3.5cm, minimum height=2.5cm, align=center
    },
    decoder/.style={
        draw=orange!70!red, fill=orange!20, rounded corners=8pt,
        minimum width=3.5cm, minimum height=2.5cm, align=center
    },
    io/.style={font=\small, align=center}
  ]

  \node[encoder] (enc) at (0,0) {Encoder};
  \node[decoder] (dec) at (7,0) {Decoder};

  \node[io] (enc_in) at (0,-3.5) {%
    $\displaystyle
    \begin{bmatrix} -1 \\ 0 \\ 0 \\ 0 \end{bmatrix}
    \quad
    \cdots
    \quad
    \begin{bmatrix} -1 \\ 1 \\ -1 \\ 0 \end{bmatrix}
    $};

  \node[io, below=0cm of enc_in] (legend) {Input polytope};

  \coordinate (in_left) at ($(enc_in.west) + (0,1.0)$);
  \coordinate (in_right) at ($(enc_in.east) + (0,1.0)$);

  \draw[decorate,decoration={brace,amplitude=6pt},thick]
    (in_left) -- (in_right);

\node[io, above=0.75cm of enc] (enc_out) {%
  $\displaystyle
\scalebox{1.2}{$\begin{bmatrix} \phantom{0} \\ \vdots \\ \phantom{0} \end{bmatrix}$}
    \quad
    \cdots
    \quad
    \scalebox{1.2}{$\begin{bmatrix} \phantom{0} \\ \vdots \\ \phantom{0} \end{bmatrix}$}
  $
};

  \node[io, above=0.cm of enc_out] {Encoder output};

  \node[io, below=1.5cm of dec] (dec_in) {%
  $\langle \text{sos} \rangle \;\; \langle 20 \rangle \;\; \langle 32 \rangle \;\; \langle 90 \rangle \;\; \text{ \large\textcolor{red}{ ?}}$
};

 \node[io, below=0.0cm of dec_in] {Partially constructed sequence of tokens};

  \coordinate (in_dec_left) at ($(dec_in.west) + (0,0.5)$);
  \coordinate (in_dec_right) at ($(dec_in.east) + (0,0.5)$);

  \draw[decorate,decoration={brace,amplitude=6pt},thick]
    (in_dec_left) -- (in_dec_right);

  \node[io, above=0.75cm of dec] (dec_out) {%
  $P = \left[ p_1, \dots, p_M \right]$
};

  \node[io, above=0.0cm of dec_out] {%
  Probability distribution\\
  over tokens
};

\node[io, right=2.2cm of dec_out] (sampled_token) {%
  $\langle 34 \rangle$
};

\node[io, above=0.0cm of sampled_token] {%
  New token
};

\draw[->, thick] (dec_out) -- (sampled_token);

\node[io, below=0.0cm of $(dec_out)!0.5!(sampled_token)$] {sampling};

  \draw[->, thick] ($(enc_in.north) + (0,0.45)$) -- ($(enc.south) + (0,-0.1)$);

  \coordinate (p1) at ($(enc_out) + (2.5,0)$);
  \coordinate (p2) at ($(p1) + (0,-5)$);
  \coordinate (p3) at ($(p2) + (4,0)$);
  \coordinate (dec_enc_point) at ($(dec.south) + (-0.5,0)$);

  \draw[->, thick] (enc) -- (enc_out);
  \draw[->, thick] ($(dec_in.north) + (0,0.55)$) -- (dec.south);
  \draw[->, thick] (dec) -- (dec_out);

  \draw[->, thick, rounded corners=8pt]
    (enc_out) -- (p1) -- (p2) -- (p3) -- (dec_enc_point);

\coordinate (s1) at ($(sampled_token) + (0,-1.5)$);  

\path let \p1 = (s1), \p2 = (dec_in.east) in
  coordinate (s2) at (\x1, \y2);

\coordinate (dec_in_right) at (dec_in.east);

\draw[->, thick, rounded corners=8pt]
  (sampled_token) -- (s1) -- (s2) -- (dec_in_right);

\end{tikzpicture}}
    \caption{\textbf{CYTransformer architecture.} The high-level pipeline for our model in inference mode. The encoder processes the input polytope, as a sequence of four-dimensional vertex vectors, into a latent representation. The decoder autoregressively generates tokens, representing simplices, conditioned on both the encoder output and previously generated tokens, sampling from the predicted token distribution $\boldsymbol{P}$ until the end-of-sequence token \texttt{<eos>} is drawn.
    }
    \label{fig:full_architecture}
\end{figure*}

\begin{figure*}[t]
    \centering
    \includegraphics[width=0.9\textwidth]{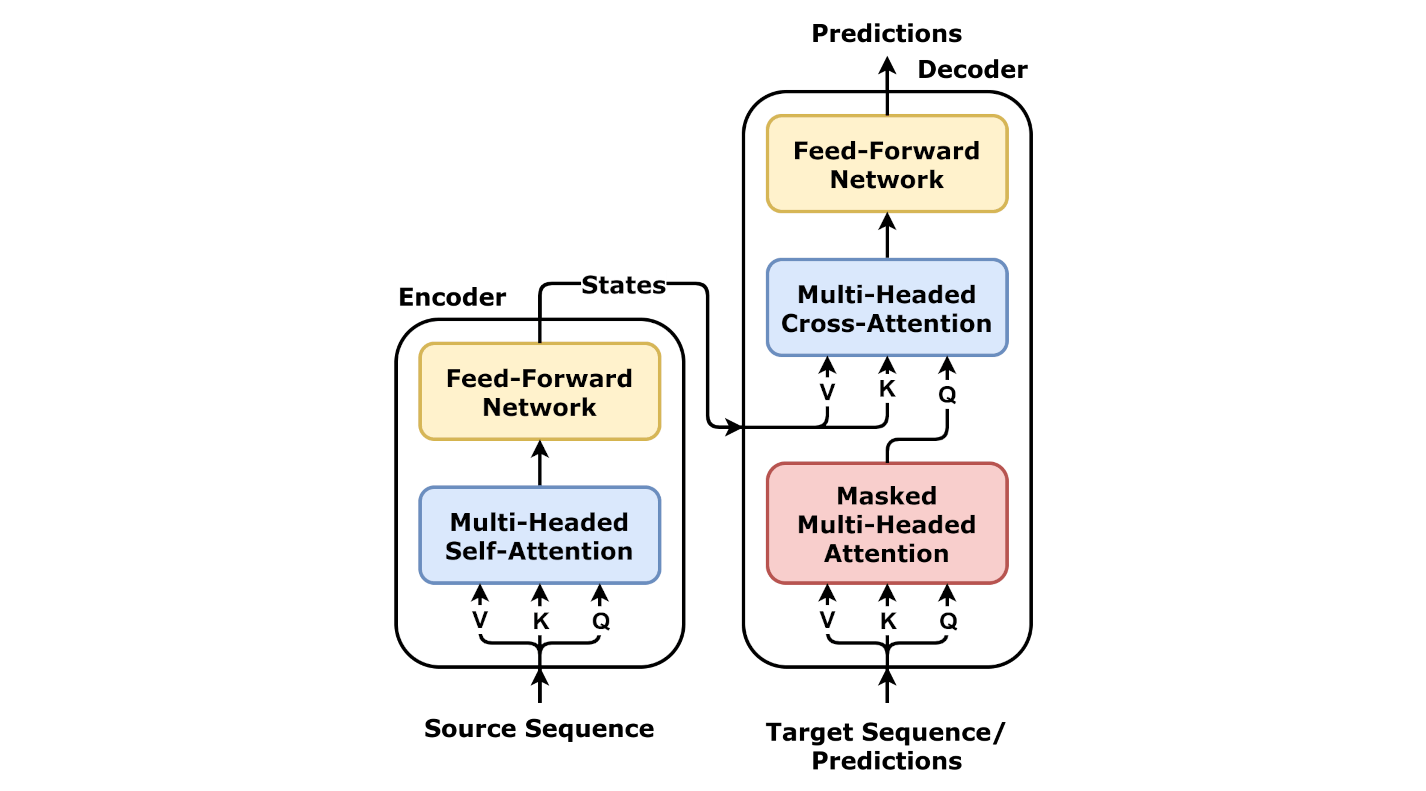}
    \caption{\textbf{Encoder and decoder layers} (illustration from~\cite{encoderdecodersource}) \textbf{.} The encoder layer applies multi-headed self-attention over the input sequence, followed by a feed-forward network. The decoder layer applies masked self-attention (to preserve autoregressive generation), followed by cross-attention to the encoder outputs, and then a feed-forward network. This structure enables conditioning the output sequence on the full input polytope while generating tokens autoregressively.
    }
    \label{fig:encoder_decoder}
\end{figure*}

The encoder input, a sequence of four-dimensional coordinate vectors (see section~\ref{sec:tokenization}) is transformed into a sequence of high-dimensional vectors by a position-aware multilayer perceptron and, before being fed into the encoder, a stack of transformer layers with bidirectional attention. Each transformer layer uses a self-attention layer, which processes relations between different elements in the input sequence (the vertices of the polytope), and a feed-forward network (a multilayer perceptron, with one hidden layer four times\footnote{This ratio is a well-accepted convention in the transformer architecture.} the size of the transformer dimension), built over residual connections that merely add a copy of the input vectors to their transformed versions (figure~\ref{fig:encoder_decoder}). 
Decoder layers have the same architecture, but use causal self-attention (i.e., attending only to previous vectors in the sequence), and a cross-attention mechanism that links to the encoder output.

The self-attention mechanism captures dependencies between elements in the sequence by computing normalized dot products of the transforms of sequence vectors by three learnable attention matrices, known as key $K$, query $Q$, and value $V$. Specifically, the vector currently processed is transformed by $Q$, and other vectors in the sequence by $K$ and $V$. In the encoder, the attention is bidirectional: keys and values are computed for all elements in the sequence. The auto-regressive decoder uses causal self-attention, which only computes the keys and values over ``past'' tokens (those before the sequence element being processed). Decoder layers also use a cross-attention mechanism, where the keys and values are computed from the encoder output, which allows for decoder output to be conditional on the model input (the polytope). In a decoder-only setting, this is handled by the self-attention.


\subsection{Encodings, tokenization, and vocabulary}\label{sec:tokenization}


We propose standard encodings for polytopes and triangulations that can be applied to different tasks. We encode four-dimensional reflexive polytopes as arrays of size $N_{\rm vert}-1$, where $N_{\rm vert}$ is the number of resolved vertices of the polytope. Each entry in the array is a non-zero vertex of the polytope, represented by its $4$-dimensional integer coordinates. The origin, which must be a vertex for the star requirement to be met, is omitted. For instance, a polytope with $N_{\rm vert}=9+1$ vertices (nine non-zero vertices plus the origin) would be represented as follows. The order of the vertices matters, as explained in the next paragraph.
\begin{center}
\begin{minipage}{0.2\linewidth}
\begin{verbatim}
[[-1  0  0  0]
 [-1  2  4 -1]
 [-1  1 -1  1]
 [-1  1  2  0]
 [ 1 -1 -1  0]
 [-1  0 -1  1]
 [ 0  0 -1  0]
 [-1  0 -1  0]
 [-1  1 -1  0]]
\end{verbatim}
\end{minipage}
\end{center}

Triangulations (FRST or not) are represented as sequences of 4-simplices containing the origin. For polytopes of fixed size $N_{\rm vert}$, there are ${N_{\rm vert}-1\choose 4}$ possible such simplices, which we encode as as many discrete tokens (with \texttt{<i>} standing for the $i$-th element in the enumeration of possible 4-simplices). This enumeration, and the mapping between tokens and simplices, is contingent on a pre-existing ordering of the vertices in the polytope, and an agreed method for enumerating possible choices of $4$ vertices (e.g., lexicographic).

For example, for a polytope with $N_{\rm vert}=10+1$, with vertices encoded as $V_0$ to $V_9$, using a lexicographic ordering for simplex selection, the token \texttt{<0>} represents the $4$-simplex $\{\text{origin},V_0,V_1,V_2,V_3\}$, \texttt{<1>} represents $\{\text{origin},V_0,V_1,V_2,V_4\}$, and \texttt{<99>} represents $\{\text{origin},V_2,V_3,V_6,V_8\}$. More generally, the simplex $\{\text{origin},V_{i_1},V_{i_2},V_{i_3},V_{i_4}\}$ (where $0\leq i_1<i_2<i_3<i_4\leq N_{\rm vert}-2$) is mapped to the token \texttt{<j>}, where $j$ is the position of the simplex among all such simplices with respect to the lexicographic order on $(i_1,i_2,i_3,i_4)$. The output vocabulary also includes three special control tokens: a start-of-sequence token, an end-of-sequence token, and a padding token, denoted henceforth as \texttt{<sos>}, \texttt{<eos>}, and \texttt{<pad>}. Padding tokens are added after the end-of-sequence token to guarantee that all sequences have the same predefined length.

Here is the tokenization of an FRST of the polytope presented above:
\begin{verbatim}
[<sos> <20> <32> <90> <108> <36> <121> <47> <62> <2> <54> <125> <69> <84>
<101> <91> <56> <6> <5> <57> <13> <7> <eos> <pad> <pad> <pad> <pad> <pad>
<pad> <pad>].
\end{verbatim}

For a given polytope, the tokenization scheme (i.e., the meaning of the tokens) is contingent on the ordering of the ${N_{\rm vert}-1}$ vertices. Therefore, the same triangulation of the same polytope can be represented by different sequences, depending on the ordering of the vertices. This has two practical consequences: there are many synonymous representations of the same polytope-triangulation pair, and the polytope encoding must keep track of the order of the vertices. Although we do not address these questions in this paper, we note that the ambiguity could be lifted by agreeing on a fixed ordering of the polytope vertices (e.g., sorting them using the natural total order over $\mathbb Z^4$). Then, each polytope-triangulation pair would have a unique representation. Alternatively, a permutation-invariant architecture, such as the set transformer~\cite{lee2019set}, could be used for the encoder.

The size of the output vocabulary grows as the fourth power of $N_{\rm vert}$. Transformers can routinely handle vocabularies of up to 50,000 words, which suits the values of $N_{\rm vert}$ considered here. For larger $N_{\rm vert}$, a two-token encoding of simplices should be considered.


\subsection{Training}

We follow standard practices for the supervised training of encoder-decoder transformers. We create a training set by fetching reflexive polytopes with a fixed number $N_{\rm vert}$ of resolved vertices. For each polytope we compute (using \texttt{CYTools}) a set of FRSTs, and generate as many pairs of polytopes and associated triangulations. These form our initial training set.

For a given pair of polytope and FRST, CYTransformer is trained to minimize the cross-entropy loss between the predicted sequence (the triangulation it predicts for this polytope) and the FRST from the training set. Specifically, if for some position $i$ in the output sequence the model predicts the distribution $\boldsymbol{P}_i$, and the correct token (in the training set) is \texttt{<j>}, then the contribution of this token to the cross-entropy loss is
\begin{equation}\label{eq:per_token_loss}
    \mathcal{L}_{\text{model}}\text{(polytope, FRST, $i$)}=-\log\left(P_{ij}\right),
\end{equation}
where $P_{ij}$ is the $j$-th entry in $\boldsymbol{P}_i$, i.e., the predicted probability of the token \texttt{<j>}. The loss is accumulated over all tokens in the output sequence (except the padding tokens), and over all examples in a randomly selected mini-batch of $800$ to $1024$ examples. Its gradient is then used to update the model parameters (also known as weights), using the Adam optimizer~\cite{Adam2014}.




As discussed in the previous section, all polytopes and associated FRSTs are defined up to a permutation of the vertices. Furthermore, an FRST is invariant under permutations of the order of its simplices. We want our model to be invariant to these symmetries. To this end, every time a training example is selected, we apply the following procedures:

\begin{enumerate}
    \item The polytope vertices are randomly permuted, and the tokens in the output sequence are replaced accordingly.
    \item The simplices in the output sequence are randomly permuted.
\end{enumerate}

This guarantees that the trained CYTransformer is invariant to these transformations, which are mere artifacts of our tokenization scheme. Other known symmetries of reflexive polytopes, involving combinations of reflections, rotations, and shears, are not considered.

\subsection{Inference}


In this phase, models trained to predict FRSTs of polytopes with $N_{\rm vert}$ vertices are used to generate candidate triangulations for new polytopes with the same number of vertices. These triangulations are generated one token at a time, for a random polytope not present in the training set. Starting with the input polytope and the initial ``empty'' triangulation \texttt{[<sos>]}, the model predicts a probability distribution, which is used to sample a token, for instance \texttt{<21>}. This token is appended to the output sequence, and the process is repeated for the same input and the partial output \texttt{[<sos> <21>]}. The process is repeated until the \texttt{<eos>} token is sampled, or we reach a fixed maximal length. The candidate solution is not guaranteed to be a triangulation or an FRST. We use \texttt{CYTools} to verify it. If it is an FRST, we add it to a list of CYTransformer-generated FRSTs.

More efficient sampling procedures for transformers have been developed, notably \emph{beam search}. Instead of generating solutions one at a time, it maintains a list of good partial predictions and their associated probabilities. For instance, instead of generating a single one-token sentence, it will generate three, then nine two-token sequences, and keep the three most likely (by multiplying the probabilities of successive tokens), and so on. Our experiments with beam search suggest that it tends to generate solutions that only differ by a reordering of simplices or vertices, i.e., it fails to generate \emph{distinct} FRSTs.\footnote{This behavior is not unexpected and is analogous to a language model generating semantically equivalent answers using different phrasings.}

A hybrid strategy to further improve the generative method is discussed in section~\ref{sec:hybrid}.

\subsection{Datasets}\label{subsec:datasets}
Each four-dimensional reflexive polytope $\Delta$ is associated with an $h^{1,1}$ value of the Calabi-Yau manifolds it generates (see~\eqref{eq:h11h21eqs}), a quantity of particular importance to physicists. On the other hand, from a machine learning perspective, it is more reasonable to classify polytopes $\Delta$ according to the numbers $N_{\rm{vert}}$ of resolved vertices of their dual polytope $\Delta^\circ$ (which is the one being triangulated), as this number conditions the length of the encoder input sequence. In particular, each model that we train is specialized for a specific value of $N_{\rm{vert}}$. Consider the set $\mathcal{P}(h^{1,1}, N_{\rm vert})$ of all reflexive polytopes $\Delta$ with a given $h^{1,1}$ value such that their dual $\Delta^\circ$ has $N_{\rm vert}$ resolved vertices: we let $\mathcal{D}(h^{1,1}, N_{\rm vert})$ be the set of all $(\Delta^\circ, T)$ pairs such that $\Delta \in \mathcal{P}(h^{1,1}, N_{\rm vert})$, and $T$ is an FRST of $\Delta^\circ$. We train and evaluate our models on (subsets of) $\mathcal{D}(h^{1,1}, N_{\rm vert})$ for various $(h^{1,1},N_{\rm vert})$.

The two quantities $h^{1,1}$ and $N_{\rm{vert}}$ are loosely related through~\eqref{eq:h11h21eqs}: for polytopes with $N_{\rm{vert}}$ within the range we consider, we find that $N_{\rm vert}=(h^{1,1}+4)+1$ holds predominantly. These are the favorable polytopes, and we restrict this work to these polytopes, primarily for simplicity in classification. Precisely, we consider in our experiments subsets of $\mathcal{D}(h^{1,1}=5, N_{\rm vert} = 9+1)$, $\mathcal{D}(6,10+1)$, $\mathcal{D}(7,11+1)$, $\mathcal{D}(8,12+1)$, $\mathcal{D}(9,13+1)$, and $\mathcal{D}(10,14+1)$. Note that we write $N_{\rm vert}=a+1$ (where $a$ is a positive integer) to draw attention to the fact that the reflexive polytopes considered have $a$ non-zero resolved vertices, which appear as inputs to the model, in addition to the origin, which does not.

We manufacture our datasets using \texttt{CYTools}. When creating a training or test set for a given $(h^{1,1}, N_{\rm vert})$ configuration, we first pick $X$ polytopes in $\mathcal{D}(h^{1,1}, N_{\rm vert})$ uniformly at random, then $Y$ FRSTs for each polytope. For $N_{\rm vert}\in\{9+1,10+1,11+1,12+1\}$, the polytopes are small enough such that \texttt{CYTools} can enumerate all their FRSTs inexpensively (see section~\ref{sec:all}), and we can draw FRSTs uniformly at random among all the FRSTs. For $N_{\rm vert}\in\{13+1,14+1\}$, the polytopes are too large for \texttt{CYTools} to efficiently enumerate all their FRSTs. Instead, we use \texttt{CYTools}' fast algorithm (see section~\ref{sec:fast}) to generate the desired number of FRSTs per polytope.

\subsection{Self-improvement}\label{subsec:selfim_method}
The quality of a machine learning model is constrained by the size and diversity of its training data, and this applies to our models (see section~\ref{subsec:training_dynamics}). For small values of $N_{\rm vert}$, \texttt{CYTools} can generate all FRSTs of a polytope, or a large enough sample to provide a representative subset. However, for larger numbers of vertices, this process becomes prohibitively costly. To reduce dependence on this expensive and uncertain data preparation step, we explore a form of self-improvement. The core idea is simple: use the model's own generated FRSTs to augment its training data for retraining, and thus alleviate the data scarcity. While our current experiments are conducted on moderate values of $N_{\rm vert}$, this serves as a proof of concept for a strategy that becomes especially valuable when scaling to larger configurations.

Concretely, we begin by training CYTransformer on a small initial dataset. After a fixed number of training steps, we pause and use the current model to generate candidate triangulations for polytopes drawn from the training set.\footnote{Note that there is no test set contamination here, as this remains part of the training procedure.} Valid FRSTs among the candidates are identified and added to the training set, expanding its size. The model is then retrained for another iteration on this augmented data, and this cycle is repeated until its training-time FRST generation performance starts stagnating. Since the model incrementally generates more training data for itself, this approach is expected to require only a modest amount of initial data to achieve appreciable performance.

As the model is trained on samples of its own generation, this can be seen as a form of off-policy reinforcement learning, an approach that has become popular for training transformers (and in particular large language models) in recent years \cite{R1, arnal2025asymmetricreinforceoffpolicyreinforcement,  faircodegenteam2025cwmopenweightsllmresearch}.

\subsection{Implementation details}\label{subsec:experimental_setup}

\paragraph{Model hyperparameters.}
All models comprise $16$ encoder and $16$ decoder layers. It uses an embedding dimension of $512$ and $16$ attention heads in each attention layer. Both the feed-forward network and the encoder input embedding layer are multilayer perceptrons, each with a single hidden layer scaled to four times the embedding dimension. Sinusoidal positional encoding is employed, and all activations are ReLUs. The resulting model contains $\sim120$ million learnable parameters, a configuration chosen to balance performance and complexity, as suggested by preliminary experiments.

\paragraph{Training.}
Training is performed by minimizing a cross-entropy loss using the Adam optimizer. The initial learning rate is set to $5\times10^{-5}$. Decoder sequences are padded to various fixed lengths according to the polytope configuration $(h^{1,1},N_{\rm vert})$. Specifically, the sequence length is set to $30$ for $(5,9+1)$, $35$ for $(6,10+1)$, $45$ for $(7,11+1)$, $55$ for $(8,12+1)$, and $65$ for $(9,13+1)$ and $(10,14+1)$. We utilize batch sizes of either $800$ or $1{,}024$, and apply a learning rate exponential decay factor of either $1$ (no decay) or $0.8$ every $100{,}000$ optimization steps depending on $N_{\rm vert}$. Training continues until performance on the validation set stagnates (see section~\ref{subsec:training_dynamics}). A training instance typically spans $200{,}000$ to $1{,}500{,}000$ optimization steps, equating to a few days of parallel training on $8$ Nvidia V100 GPUs.

\paragraph{Datasets.}
For each training instance, we create three disjoint datasets of polytopes: a training set, a validation set used to measure model performance during training and to decide the stopping point, and a test set reserved for evaluating the trained model. The specific details regarding the sizes and compositions of these datasets for each polytope configuration are listed in table~\ref{tab:datasets}.
\begin{table}[h!]
    \small
    \centering
    \begin{tabular}{ccccc}
        \toprule
        &\# of polytopes in&\# of polytopes in&Max. \# of FRSTs\\
        $(h^{1,1},N_{\rm vert})$&validation and test sets&training set&per polytope\\
        \midrule
        $(5,9+1)$ & $500$ & $1000$ - $3000$ & $11$\\
        $(6,10+1)$ & $1000$ & $2000$ - $16000$ & $35$\\
        $(7,11+1)$ & $3000$ & $2000$ - $50000$ & $125$\\
        $(8,12+1)$ & $1000$ & $2500$ - $8000$ & $505$\\
        $(9,13+1)$ & $1000$ & $3000$ - $16000$ & $300$ (fast sampler)\\
        $(10,14+1)$ & $1000$ & $4000$ - $16000$ & $300$ (fast sampler)\\
       \bottomrule
    \end{tabular}
    \caption{\textbf{Datasets.} For each $(h^{1,1},N_{\rm vert})$ configuration, we report the number of polytopes in the validation and test sets, the number of polytopes used for training, and the maximum number of FRSTs per polytope used during training. For the first four configurations, the listed maximum number of FRSTs corresponds to the largest number of FRSTs admitted by any single polytope in the dataset. For $(9,13+1)$ and $(10,14+1)$, FRSTs are generated using the fast sampler, capped at 300 per polytope.}
    \label{tab:datasets}
    \end{table}

\FloatBarrier
\section{Results}
\label{sec:results}

In this section, we present a comprehensive evaluation of CYTransformer and its derived methods. We begin by introducing the performance metrics used throughout, which assess both the \emph{efficiency} of generating new FRSTs for unseen polytopes and the \emph{representativeness} of the generated samples. We then evaluate CYTransformer in isolation, followed by a comparison with \texttt{CYTools}' fast sampling algorithm. Motivated by the observation that CYTransformer generates unbiased sets of FRSTs, we also propose the \emph{CYTransformer-seeded fast sampler}, a hybrid strategy that achieves both improved performance and high inference speed. Finally, we explore CYTransformer’s \emph{self-improvement} capability, showing that it can begin training on a dataset of modest size, then iteratively retrain on its own validated outputs, ultimately achieving representative performance approaching that of models trained on much larger datasets.

\subsection{Training metrics}
We monitor the training of our CYTransformer models as follows: every few thousand steps, we randomly pick $X$ polytopes from the validation set and have the model generate $Y$ candidate triangulations for each. Among the resulting $X\times Y$ candidates, we count the number of distinct FRSTs. We refer to this count as a function of training step as the \hypertarget{met:ttfrstgencurve}{\textbf{training-time FRST generation curve}}. While this metric does not capture every aspect of model performance, unlike the more refined evaluations conducted post-training, it provides a useful proxy for training progress and is therefore our primary diagnostic during training.

As part of tracking CYTransformer's self-improvement capability, we monitor the growth of the training set across self-improvement iterations, where at each step, all self-generated FRSTs are incorporated into the training set.

\subsection{Performance metrics}
\label{subsec:perfmet}
We use the following metrics to evaluate the performance of our trained CYTransformer models and, for comparison, the fast sampler. Unless otherwise noted, all statistics are averaged over candidate triangulations from $N_{\rm guess}=20{,}000$ inference calls per test polytope. Each test set consists of $200$ randomly selected polytopes, fixed for each $(h^{1,1},N_{\rm vert})$ configuration, none of which are used during training or seen before inference. The metrics we consider are:
\begin{itemize}
    
    \item \hypertarget{met:frstgencurve}{\textbf{FRST generation curve.}} This primary performance metric plots the average number of valid FRSTs generated by the model as a function of $N_{\rm guess}$; it can be seen as a refined version of the training-time FRST generation count. For each test polytope, the $N_{\rm guess}$ candidate triangulations are checked for validity as FRSTs, and a cumulative count is maintained. The curves are then simply averaged across all test polytopes. This metric reflects the model's raw ability to produce valid FRSTs, while the steepness of the curve indicates how efficiently the model lands in the FRST space with each inference call. We plot three variants of this curve:
    \begin{enumerate}
        \item \textbf{All FRSTs.} Counting all valid FRSTs, including duplicates if the same FRST is generated more than once.
        \item \textbf{Distinct FRSTs.} Counting only distinct FRSTs, which reflects the model's ability to discover new FRSTs.
        \item \textbf{NTFE FRSTs.} Counting only non-two-face-equivalent (NTFE) FRSTs, where candidate triangulations are compared against previously generated FRSTs under two-face equivalence. The NTFE curve captures the model’s ability to generate structurally distinct FRSTs, as measured by differences in two-face triangulations. Since TFE FRSTs yield topologically equivalent Calabi-Yau manifolds (by Wall's theorem~\cite{WALL1966}), this metric serves as a proxy for the model’s capacity to explore beyond topologically redundant desingularizations.
    \end{enumerate}

    \item \hypertarget{met:frstgenrate}{\textbf{FRST generation rate.}} Defined as the \hyperlink{met:frstgencurve}{FRST generation curve} divided by $N_{\rm guess}$, this metric reflects how quickly valid FRSTs accumulate as sampling progresses. When only distinct FRSTs are counted, a high production rate indicates consistent discovery of new FRSTs, while a declining rate suggests increasing redundancy in the model's outputs.

    \item \hypertarget{met:frstreccurve}{\textbf{FRST recovery curve.}} This metric normalizes the \hyperlink{met:frstgencurve}{FRST generation curve} to reflect coverage of the full FRST space. For each test polytope $i$, the cumulative number of distinct FRSTs found as a function of $N_{\rm guess}$ is divided by $N_{\text{max},i}$, the total number of distinct FRSTs admitted by that polytope.\footnote{The total number is obtained by enumerating all FRSTs, as described in section~\ref{sec:all}.} Denote this recovery fraction by $f_i$. The resulting recovery curve $f_i(N_{\rm guess})$ captures how efficiently the model recovers the full space of FRSTs. As $N_{\rm guess}$ increases, $f_i$ may approach a limiting value. Saturation well below 100\% indicates that the model fails to cover parts of the FRST space, regardless of sampling budget. Such poor coverage suggests mode collapse or an incomplete learned distribution of FRSTs. Comparing recovery curves across individual polytopes can reveal consistent failure modes tied to geometric features of the input polytopes.
    
    For polytopes with large $N_{\text{max},i}$, even a strong sampler will naturally yield a small $f_i$, not due to poor performance but because the sampling budget is limited. Consequently, when computing the average recovery curve over the test polytopes, a simple average of $f_i$ would unfairly penalize such polytopes and distort the overall assessment of model quality. This issue is especially important for configurations with $(h^{1,1},N_{\rm vert})\geq(8,12+1)$, where the test polytopes vary widely in $N_{\text{max},i}$, including many with very large counts. To correct for this, we assign each polytope a weight inversely proportional to its FRST count, $w_i=1/N_{\text{max},i}$, compensating for the fact that high-$N_{\text{max},i}$ polytopes are inherently harder to recover fully. This yields a weighted average recovery curve that more faithfully reflects relative coverage:
    \begin{equation}
        \text{(weighted) avg. \% of all FRSTs recovered}=\frac{\sum_if_i/N_{\text{max},i}}{\sum_i1/N_{\text{max},i}}\cdot100\%.
    \end{equation}
    
    \item \hypertarget{met:hsfrstdist}{\textbf{Height-space FRST distribution.}} This metric explores how unbiasedly the model traverses the space of FRSTs. For each test polytope, we consider the set of distinct FRSTs recovered by the model and compute a height vector $\boldsymbol{h} \in \mathbb{R}^{N_{\rm vert}}$ for each using \texttt{CYTools}. Height vectors can be seen as functions of the resolved vertices (see section~\ref{subsec:cyFRST}). Two height vectors $\boldsymbol{h}_1$ and $\boldsymbol{h}_2$ whose difference $\boldsymbol{h}_1 - \boldsymbol{h}_2 = \boldsymbol{h}_{\rm aff}\in \mathbb{R}^{N_{\rm vert}}$ is an affine function (affine with respect to the coordinates of the vertices) define the same triangulation, which makes the choice of height vector non-canonical. We remove this ambiguity by projecting each height vector onto the orthogonal complement of the subspace spanned by affine functions. The result is the affine-normalized height vector
    \begin{equation}
        \boldsymbol{h}_{\rm proj}=\boldsymbol{h}-\boldsymbol{h}_{\rm aff}=\boldsymbol{h}-\boldsymbol{Pc},
    \end{equation}
    where
    \begin{equation}
        \boldsymbol{P}=
        \begin{bmatrix}
        1 & \boldsymbol{p}^T_{1} \\
        1 & \boldsymbol{p}^T_{2} \\
        1 & \boldsymbol{p}^T_{3} \\
        \vdots & \vdots \\
        1 & \boldsymbol{p}^T_{N_{\rm vert}}
        \end{bmatrix}
    \end{equation} 
    encodes the vertex coordinates $\boldsymbol{p}_i$ and the vector $\boldsymbol{c}$ is an array of $5$ coefficients that solves the least square problem
    \begin{equation}
        \arg\min_{\boldsymbol{c}\in\mathbb{R}^{5}}|\boldsymbol{Pc}-\boldsymbol{h}|^2.
    \end{equation}
    This yields the solution $c=(\boldsymbol{P}^T\boldsymbol{P})^{-1}\boldsymbol{P}^T\boldsymbol{h}$. In addition, since scaling the height vector does not affect the resulting FRST, we must use a scale-invariant similarity metric. For example, the $\ell^2$-norm would undesirably penalize differences in overall magnitude. We therefore use cosine similarity between each projected height vector and that of the unique Delaunay triangulation, providing a scale-invariant comparison relative to a consistent reference. This yields a distribution of similarity scores for each test polytope. Comparing these distributions across models reveals how differently they sample the FRST space. In particular, comparing a model-generated distribution to the full population distribution computed from all distinct FRSTs allows us to assess how unbiasedly the model explores the FRST space relative to the total achievable diversity.

    \item \hypertarget{met:hsfrstrep}{\textbf{Height-space representativeness score.}} A model exhibits inference bias if it disproportionately oversamples or undersamples FRSTs from certain regions of the height vector space compared to the true population distribution. For each test polytope, we compare the \hyperlink{met:hsfrstdist}{height-space FRST distribution} generated by the model with the corresponding population distribution. If the two distributions have similar shapes (up to overall scaling), the model is considered representative in its sampling. To quantify this, we compute the cosine similarity between the two distributions, taking advantage of its scale invariance. These similarity scores are then collected across all test polytopes and compiled into a representativeness histogram. A distribution concentrated near one indicates that the model is generally unbiased and capable of producing representative samples of FRSTs.
\end{itemize}

While flip distance could provide a discrete notion of similarity between triangulations, it is not practical in our setting. Flip distance measures the minimal number of bistellar flips needed to transform one triangulation into another. This quantifies the combinatorial proximity of triangulations within the space defined by the flip graph, which connects triangulations via single flips. However, the number of possible flips grows super-exponentially with the number of vertices in any fixed dimension, and in four dimensions this growth becomes prohibitive even for configurations with as few as six vertices. As a result, computing shortest paths in the flip graph becomes intractable and unsuitable for use as a similarity metric in our experiments.

\subsection{Training dynamics}\label{subsec:training_dynamics}

\begin{figure}[htbp]
  \centering
  \includegraphics[height=7.6cm]{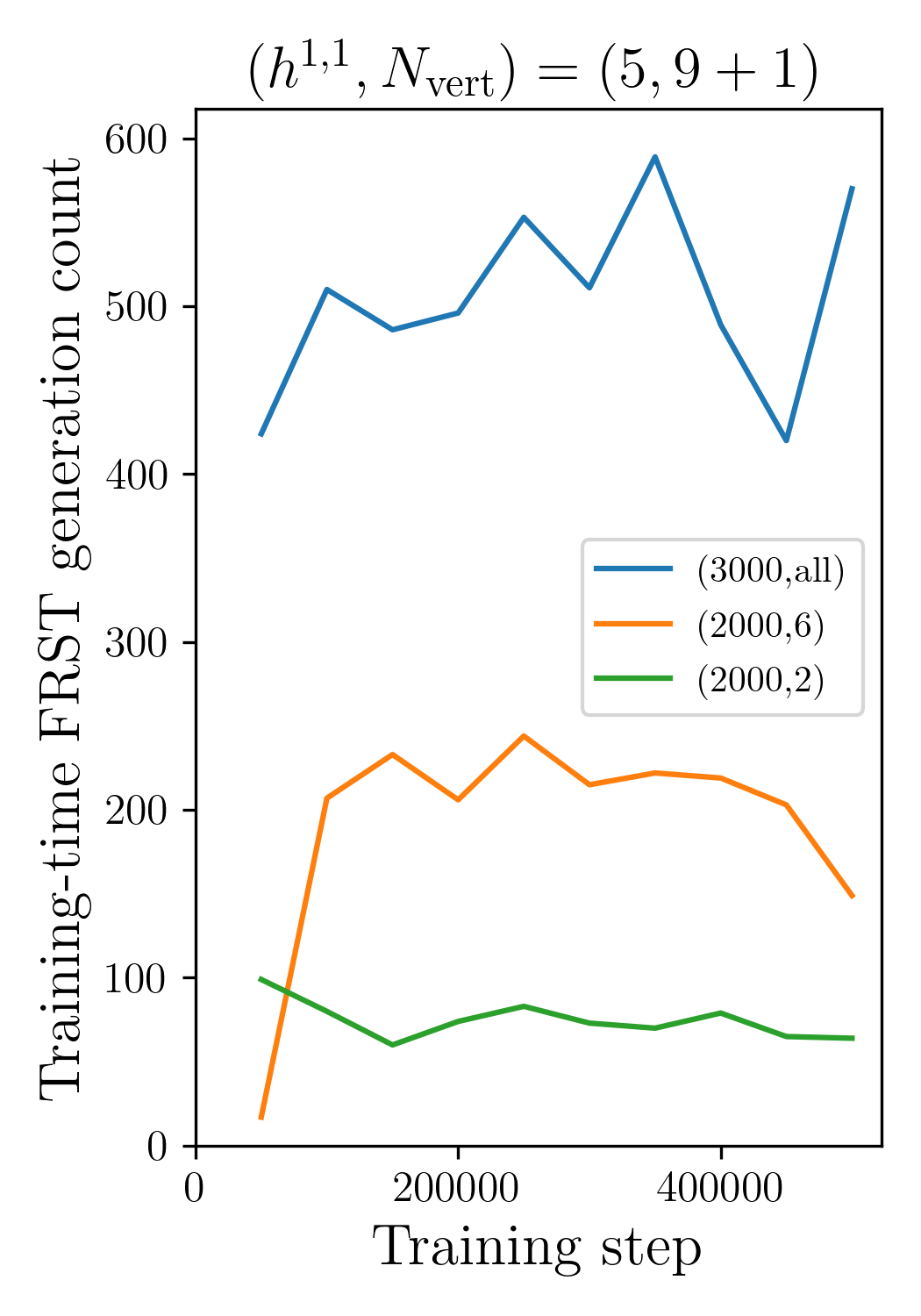}
  \includegraphics[height=7.6cm,trim={1.0cm 0 0 0},clip]{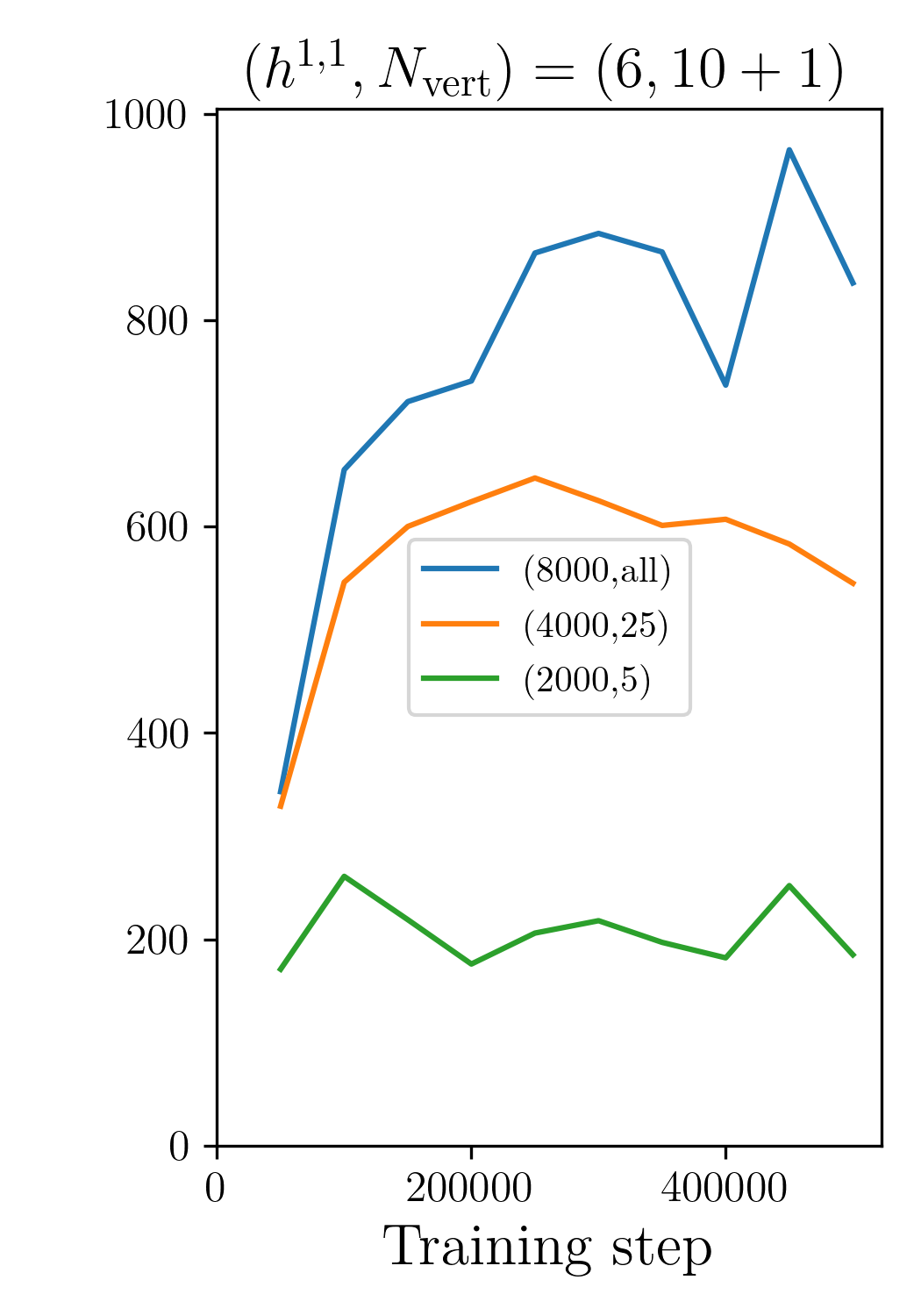}
  \includegraphics[height=7.6cm,trim={1.0cm 0 0 0},clip]{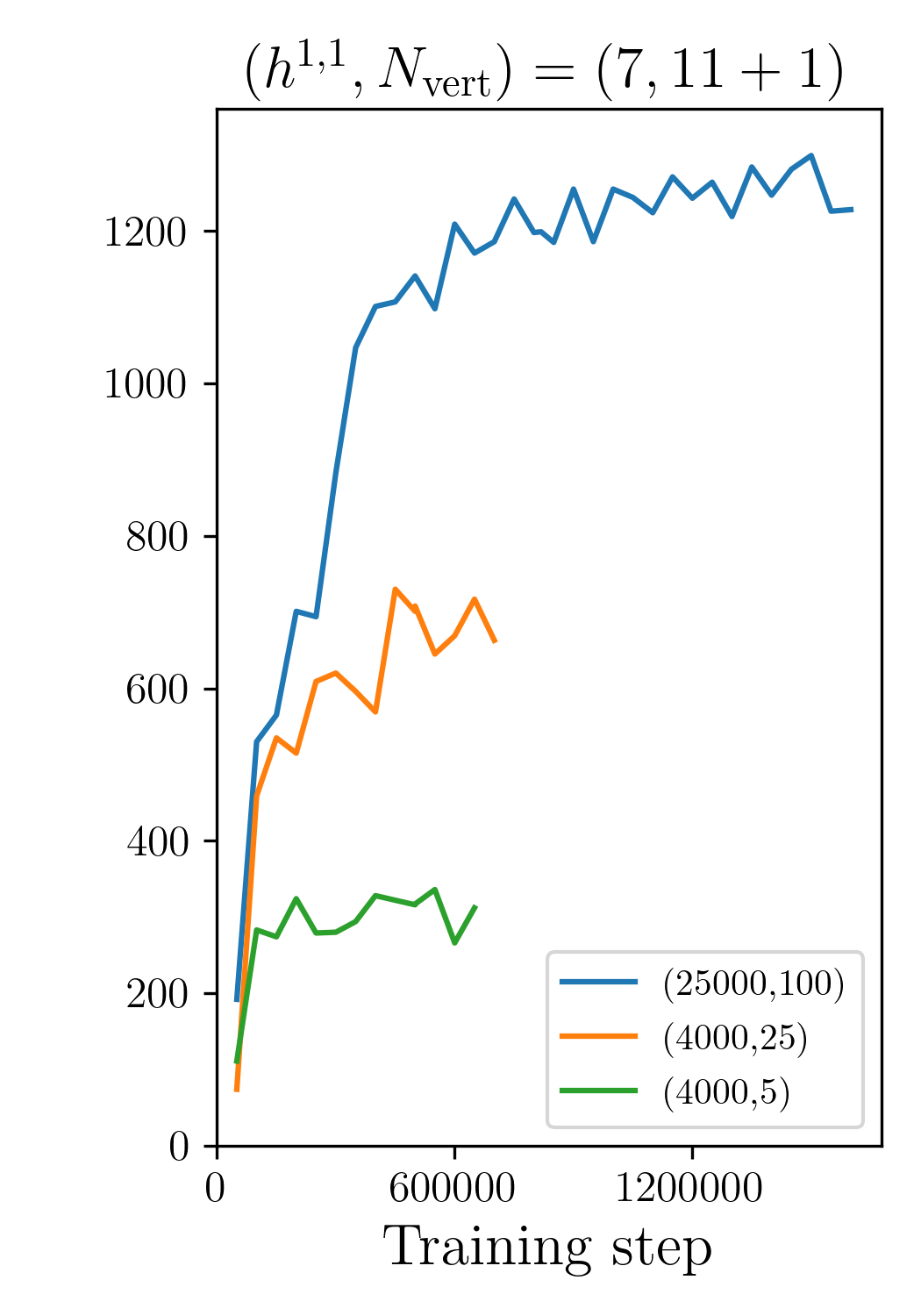}\\
  \includegraphics[height=7.6cm]{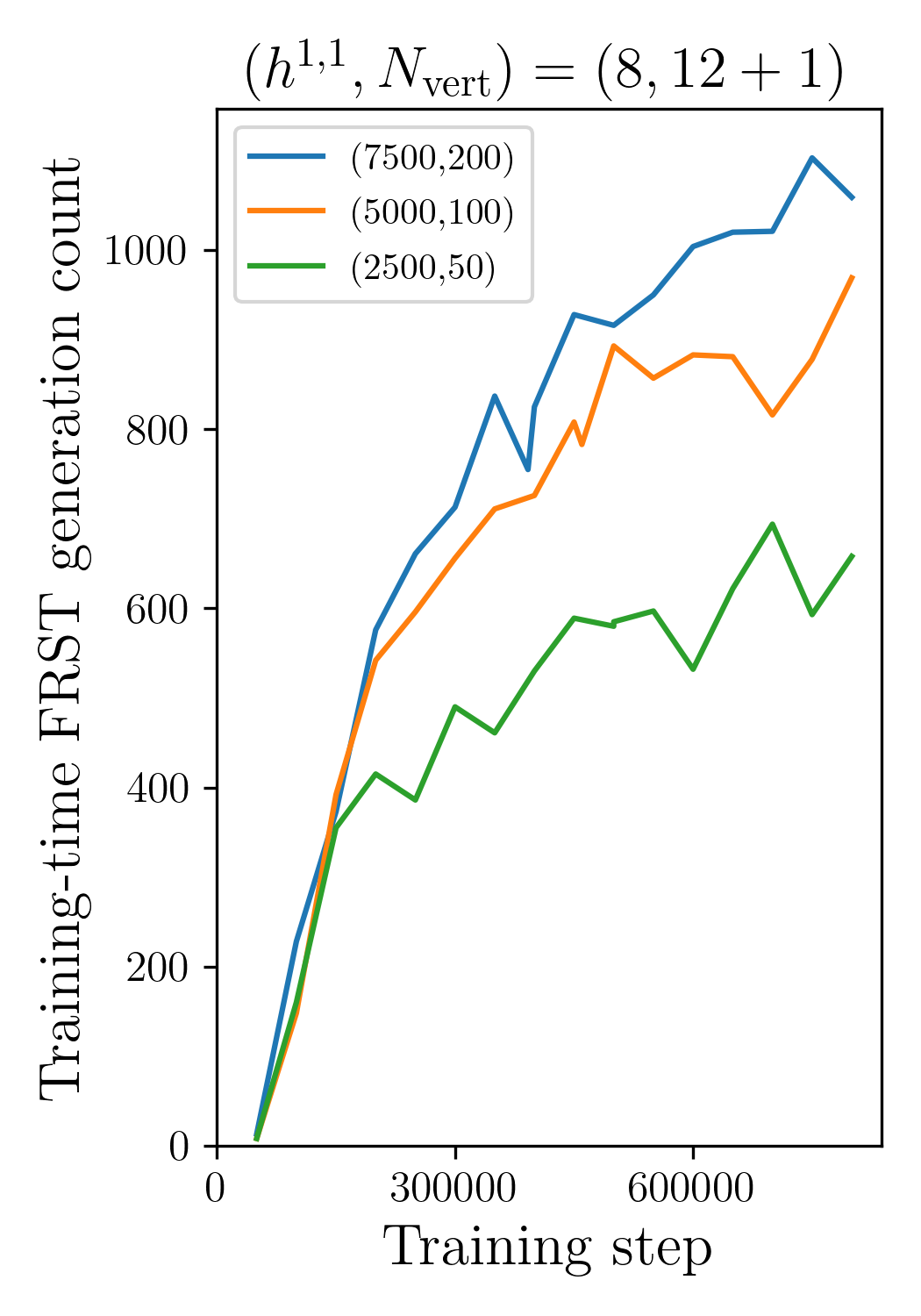}
  \includegraphics[height=7.6cm,trim={1.0cm 0 0 0},clip]{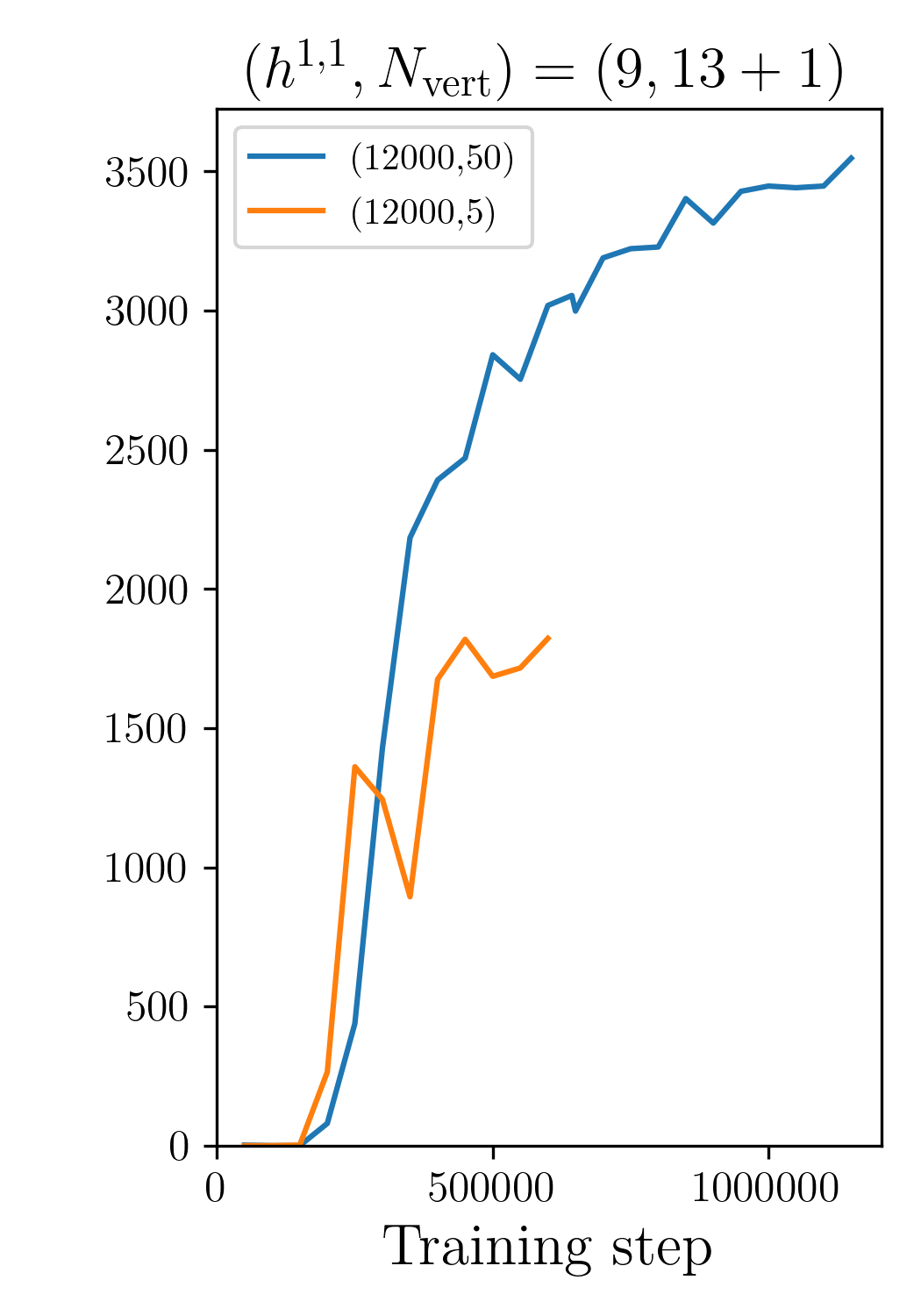}
  \includegraphics[height=7.6cm,trim={1.0cm 0 0 0},clip]{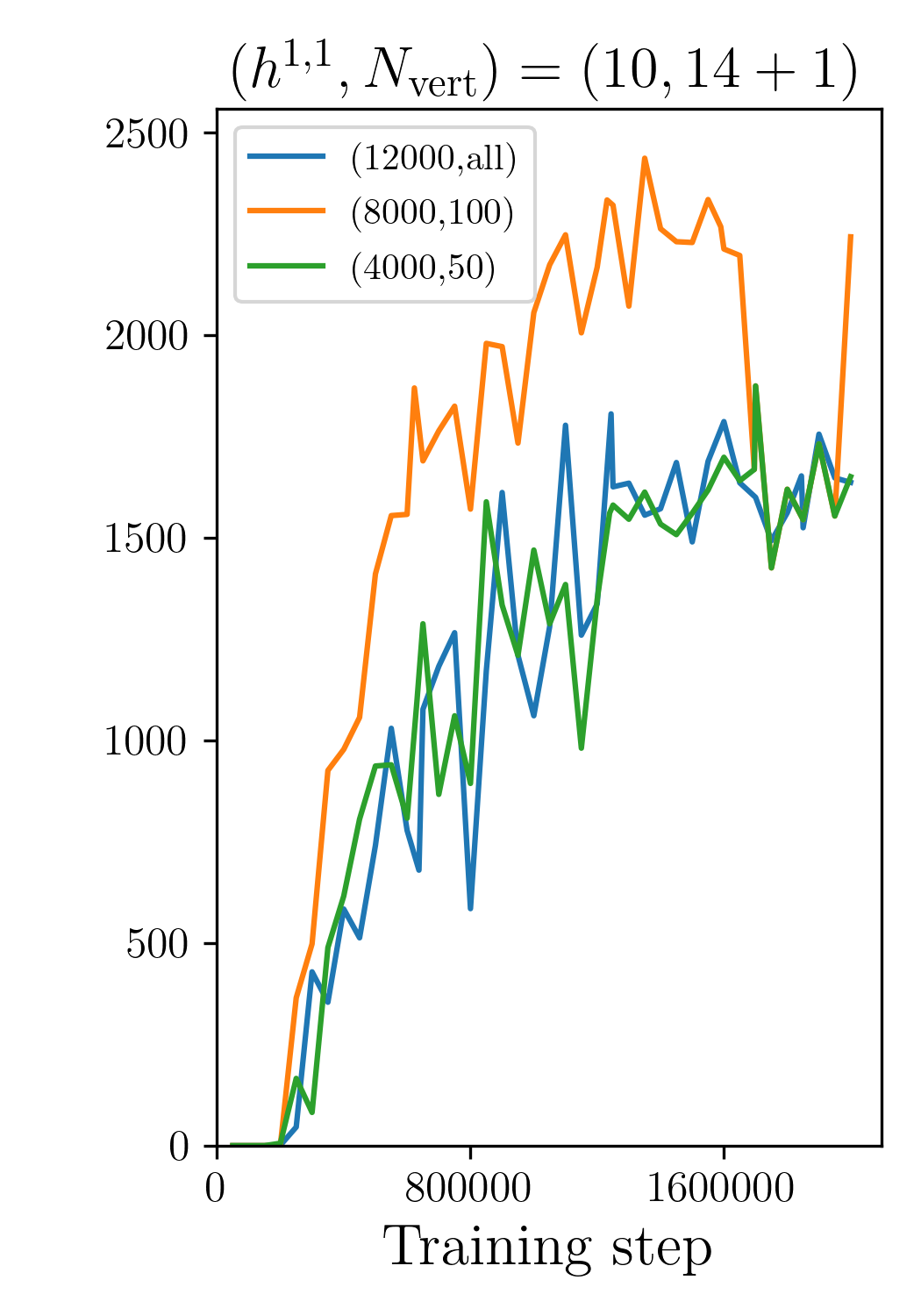}
  \caption{\textbf{CYTransformer \protect\hyperlink{met:ttfrstgencurve}{training-time FRST generation curves}.} Each plot shows the number of distinct FRSTs generated by CYTransformer during training, measured across $1{,}600$ (for $(h^{1,1},N_{\rm vert})=(5,9+1)$ to $(8,12+1)$) or $6{,}400$ (for $(9,13+1)$ and $(10,14+1)$) candidate triangulations, as a function of training step. Each curve within a plot corresponds to a model trained on a different-sized dataset. For example, the label $(2000,6)$ indicates a training set of $2{,}000$ polytopes, each contributing up to $6$ FRSTs (or fewer, if the polytope admits less). The label ``all'' refers to using all available FRSTs for each polytope in the training set, which, depending on the configuration, may either mean the full set of enumerated FRSTs or a capped number generated during data preparation (see sections~\ref{subsec:datasets} and~\ref{subsec:experimental_setup}). As expected, performance generally improves with a larger training set.}
  \label{fig:training_dynamics}
\end{figure}

During training, we periodically evaluate our CYTransformer models by generating $X\times Y$ candidate triangulations every $50{,}000$ training steps (sampling $Y$ candidates for each of $X$ polytopes from the validation set, where $(X,Y)=(40,40)$ for $(h^{1,1},N_{\rm vert})=(5,9+1)$ to $(8,12+1)$, and $(X,Y)=(160,40)$ for $(9,13+1)$ and $(10,14+1)$). We then count the number of distinct FRSTs among these candidates. Figure~\ref{fig:training_dynamics} shows the resulting \hyperlink{met:ttfrstgencurve}{training-time FRST generation curves}. Within each plot, the different curves correspond to models trained on different training sets. We vary both the number of polytopes and the number of FRSTs per polytope to assess how training set composition affects training dynamics. For example, in the case of $(5,9+1)$, we consider three training sets: one with $2{,}000$ polytopes and up to $2$ FRSTs per polytope (fewer if a polytope admits less than $2$), one with $2{,}000$ polytopes and up to $6$ FRSTs per polytope, and one with $3{,}000$ polytopes using all\footnote{Here, ``all'' does not necessarily refer to the full set of possible FRSTs admitted by each polytope, but rather to all FRSTs generated using \texttt{CYTools} in preparation for training, as described in sections~\ref{subsec:datasets} and~\ref{subsec:experimental_setup}. As a reminder, we have access to all possible FRSTs for configurations up to and including $(8,12+1)$, while for $(9,13+1)$ and $(10,14+1)$, we generate $300$ FRSTs per polytope using the fast sampler.} available FRSTs. These are denoted as $(2000,2)$, $(2000,6)$, and $(3000,\text{all})$ in the legend.

Our best models achieve up to $75$\% validity on candidates generated during training. As expected, performance generally improves with a larger training set. We also see that the learning task becomes more challenging with increasing $(h^{1,1},N_{\rm vert})$: for example, models for $(5,9+1)$ and $(6,10+1)$ plateau after $200{,}000$ steps, while those for $(9,13+1)$ and $(10,14+1)$ continue to improve beyond $500{,}000$ steps. Note that the relatively low counts of distinct FRSTs for the $(5,9+1)$ models reflect the limited number of FRSTs available per polytope, not model underperformance. While we do not claim to have fully exploited the potential of our models or reached convergence, we find that the number of polytopes required for training remains on the order of $\sim10{,}000$ even as $(h^{1,1},N_{\rm vert})$ increases. This suggests that the difficulty of the task does not scale significantly with the number of training polytopes needed.

We also monitor (not shown in the figure) the proportion of star triangulations, fine star triangulations and regular star triangulations among generated candidates, and observe that these remain negligible throughout training, indicating that models learn to generate FRSTs directly, rather than progressively satisfying finer constraints.

For each $(h^{1,1},N_{\rm vert})$ configuration, we select the model with the highest generation count (at the training step where this maximum is reached) and refer to it as the trained CYTransformer for that case. The remainder of this section analyzes these models in detail.

\subsection{CYTransformer sampling performance}

This subsection evaluates trained CYTransformers in isolation, focusing on both the efficiency and representativeness of their FRST sampling.

\begin{figure}[htbp]
  \centering
  \includegraphics[height=7.25cm]{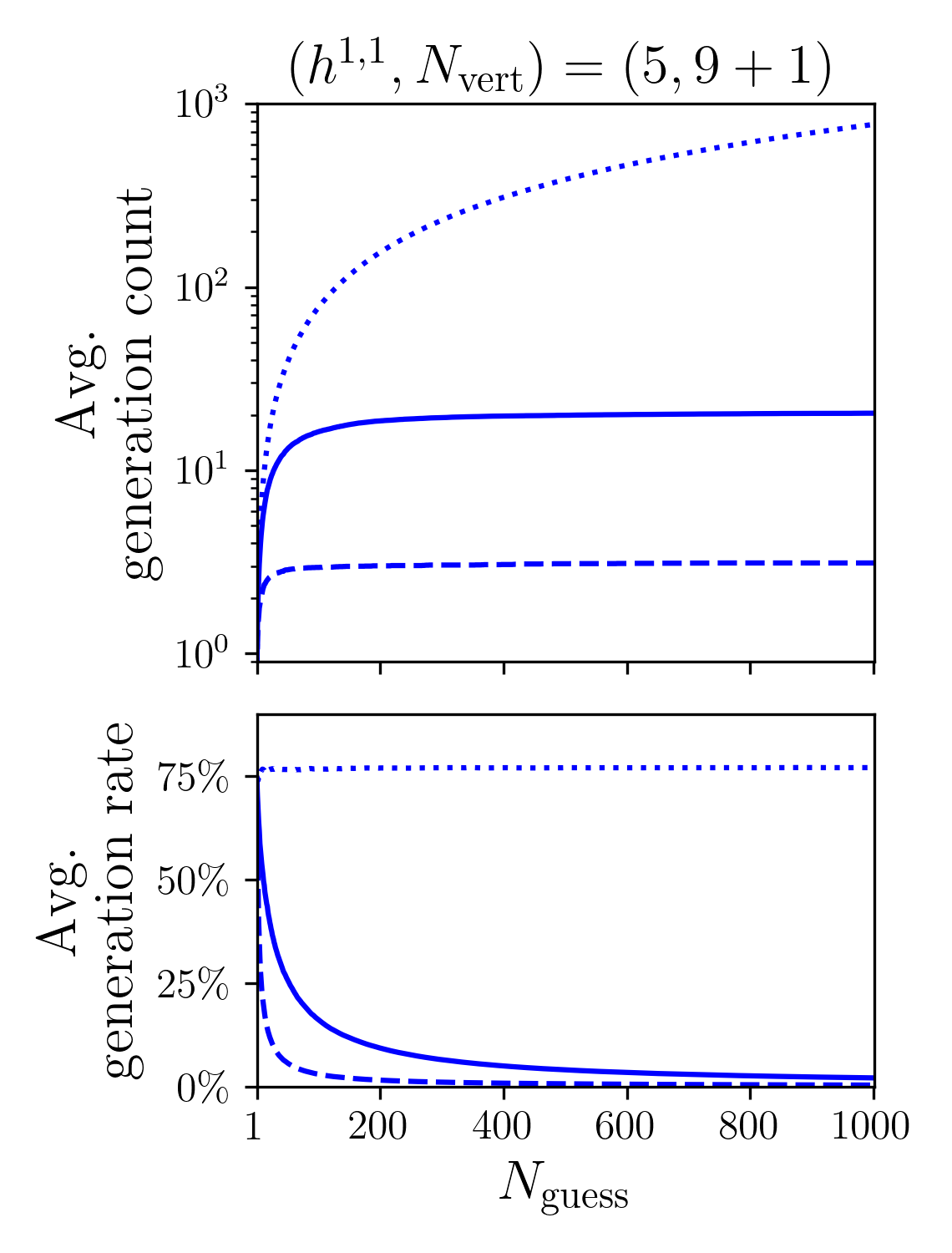}
  \includegraphics[height=7.25cm,trim={1.575cm 0 0 0},clip]{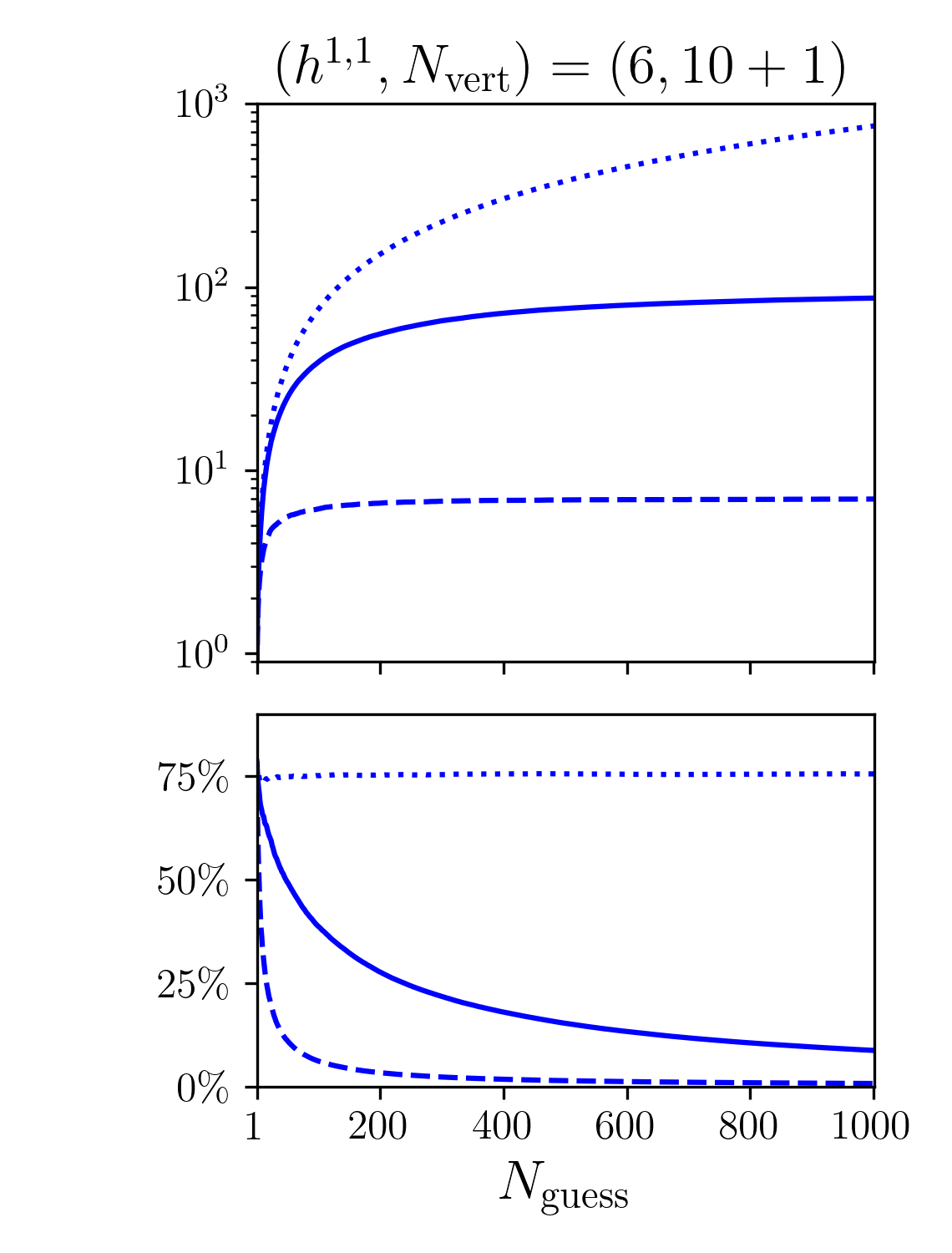}
  \includegraphics[height=7.25cm,trim={1.575cm 0 0 0},clip]{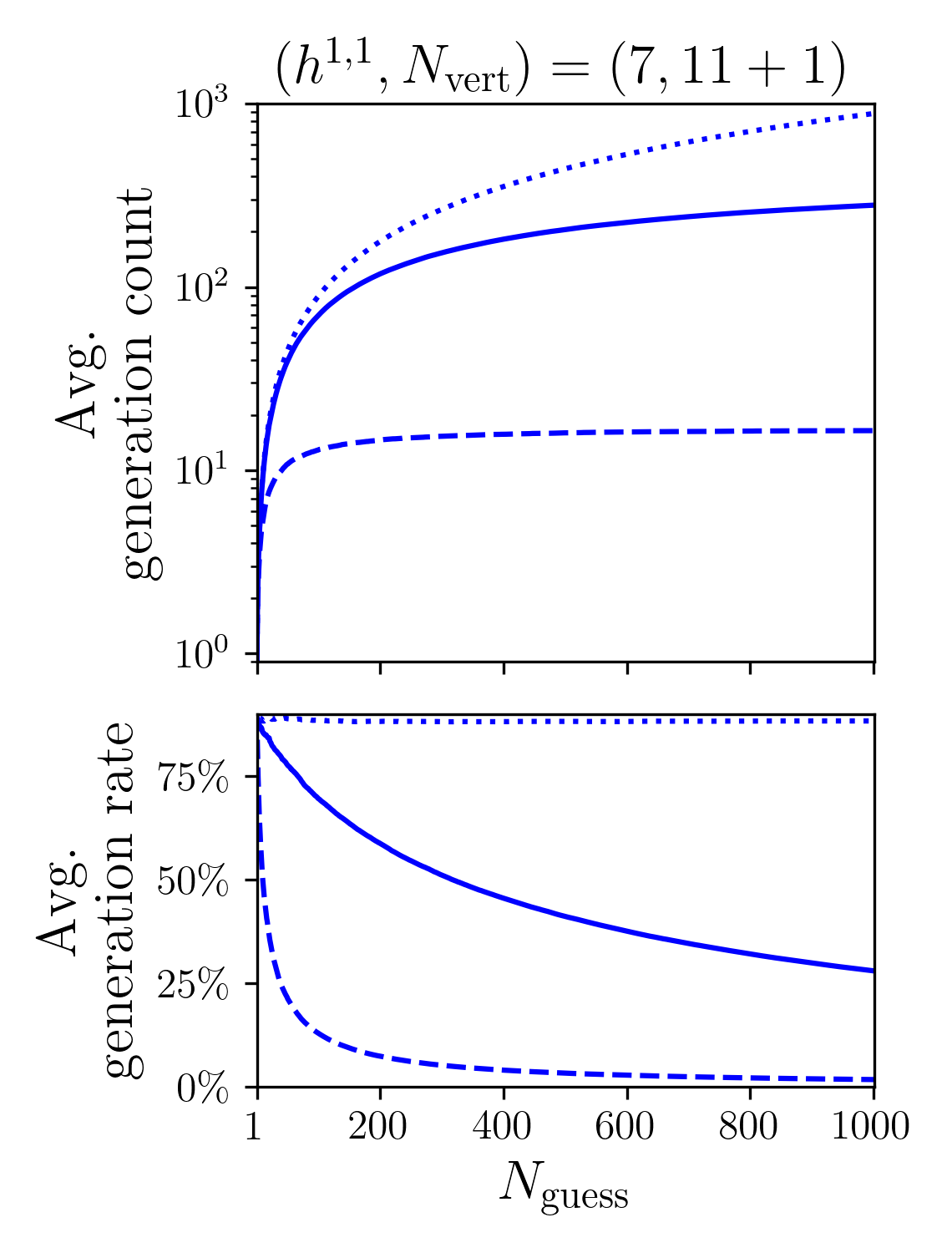}\\
  \includegraphics[height=7.25cm]{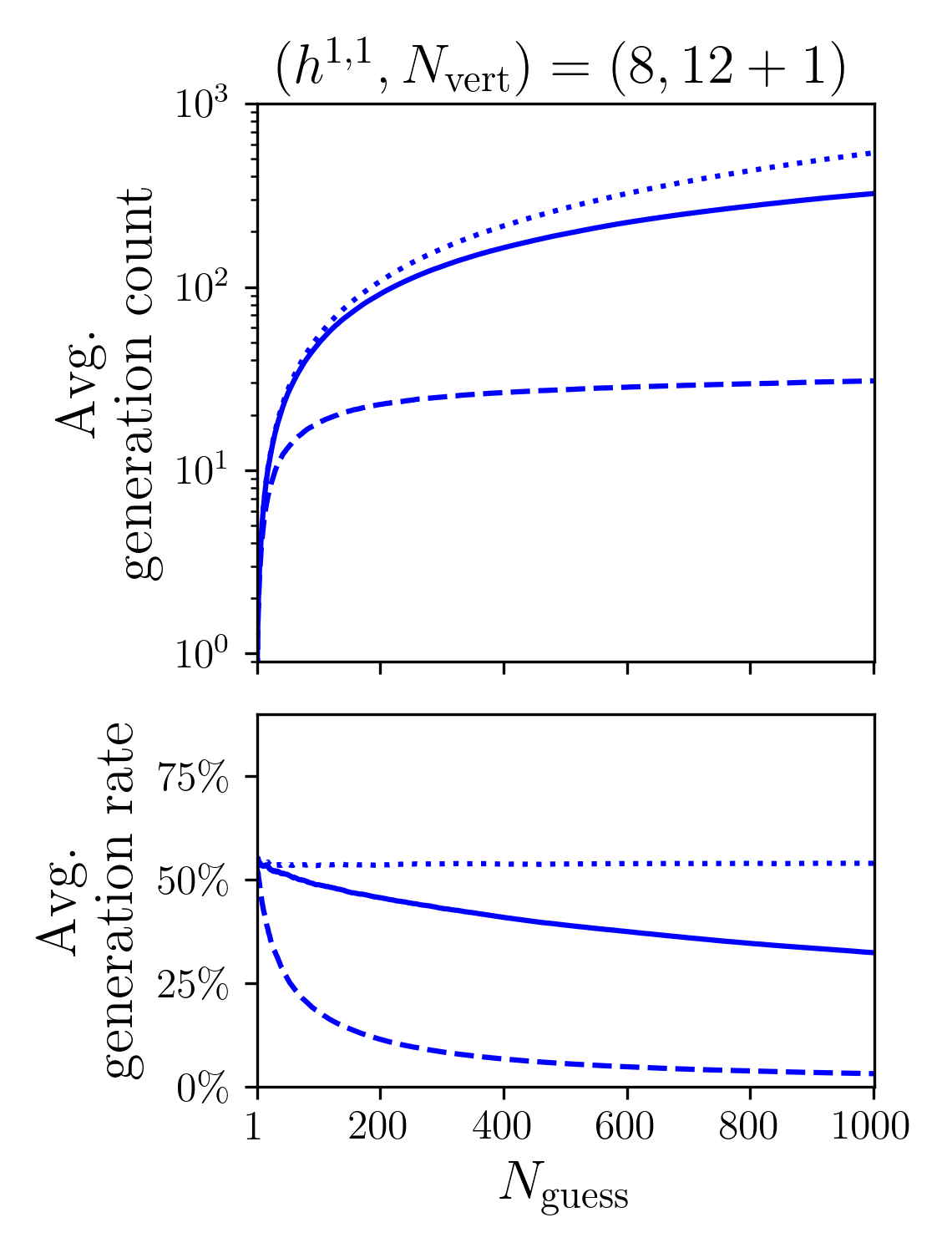}
  \includegraphics[height=7.25cm,trim={1.575cm 0 0 0},clip]{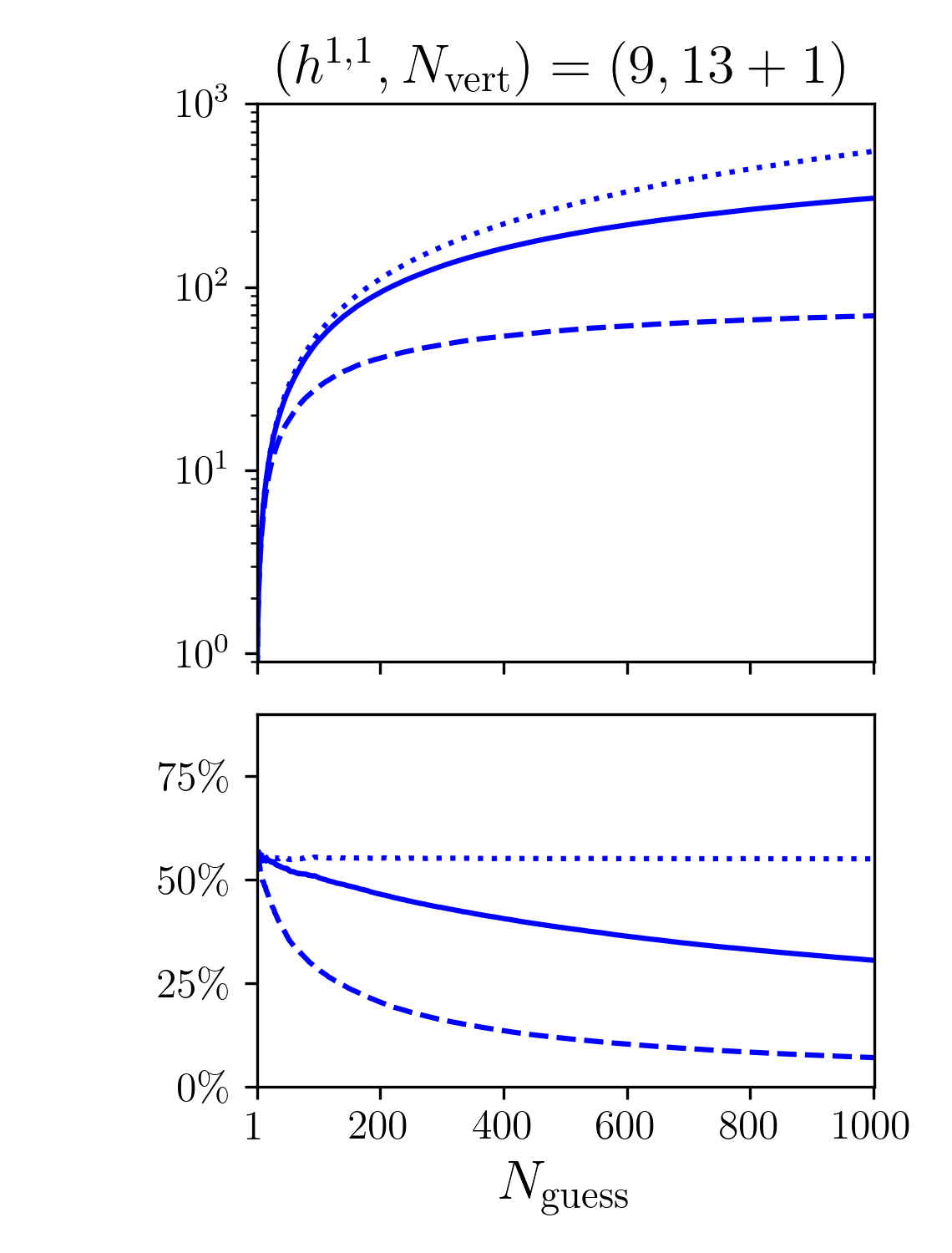}
  \includegraphics[height=7.25cm,trim={1.575cm 0 0 0},clip]{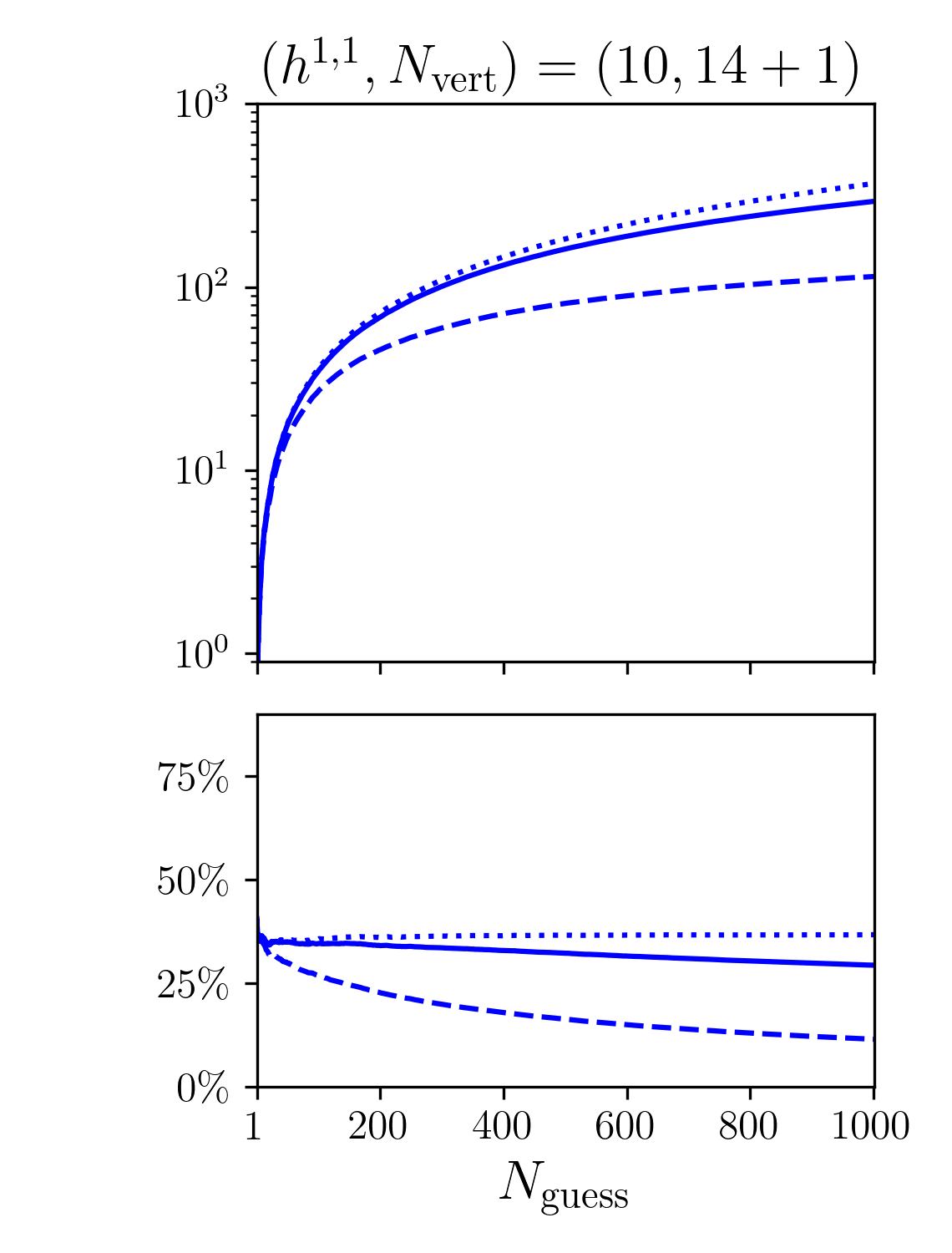}
  \includegraphics[width=1\textwidth]{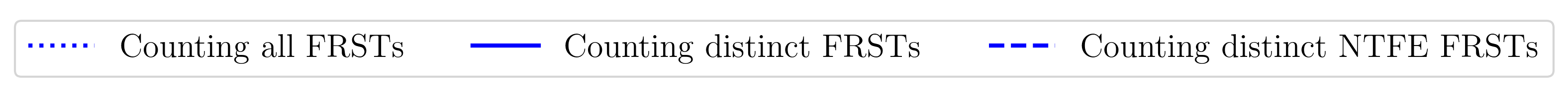}
  \caption{\textbf{Generation efficiency of CYTransformer across polytopes of increasing complexity.}
  Each panel corresponds to a fixed $(h^{1,1},N_{\rm vert})$, showing the \protect\hyperlink{met:frstgencurve}{FRST generation curves} (top row) and the corresponding \protect\hyperlink{met:frstgenrate}{generation rates} (bottom row) as a function of inference calls $N_{\rm guess}$, averaged over $200$ test polytopes. Three curves are shown: all generated FRSTs (dotted), distinct FRSTs (solid), and distinct NTFE FRSTs (dashed). For simpler polytopes (top left), CYTransformer rapidly saturates the FRST space, leading to a steep drop in the distinct generation rate. For more complex polytopes (bottom right), the rate decays more gradually (remains nearly flat for $(10,14+1)$), demonstrating the model’s ability to scale and maintain generative diversity across a vast FRST space. The flatness of the all-FRST rate underscores the model’s stable success probability per candidate triangulation due to the independence of inference calls.}
  \label{fig:solo_ps}
\end{figure}

\textbf{Efficiency.} Figure~\ref{fig:solo_ps} shows the \hyperlink{met:frstgencurve}{FRST generation curves} produced by CYTransformer for $(h^{1,1},N_{\rm vert})$ ranging from $(5,9+1)$ to $(10,14+1)$. Each column presents the average number of FRSTs generated (top row) and the corresponding \hyperlink{met:frstgenrate}{generation rate} (bottom row) as a function of the number of inference calls $N_{\rm guess}$. We show all three variants: counting all FRSTs (dotted), counting distinct FRSTs (solid), and counting distinct NTFE FRSTs (dashed). For clarity, the plots are restricted to the first $1{,}000$ candidate triangulations, where early-time model behavior is most visible.

The generation rate for all FRSTs remains nearly flat across $N_{\rm guess}$, reflecting the independence of inference calls and CYTransformer's stable probability of generating valid FRSTs. In contrast, the generation rate for distinct FRSTs drops as $N_{\rm guess}$ increases. This behavior is expected because once the most common FRSTs are discovered, the model must sample rarer ones, leading to diminishing returns.

For polytopes with smaller $(h^{1,1},N_{\rm vert})$, and hence generally fewer FRSTs, the distinct generation rate drops off quickly, indicating that the model rapidly saturates the available FRST space. As $(h^{1,1},N_{\rm vert})$ increases, the size and complexity of the FRST space grow, and the rate decays more gradually, with only slight decay observed for $(10,14+1)$. This highlights CYTransformer's adaptability: it scales its generative capacity to match and explore the growing FRST space. The models maintain nontrivial generation rates over hundreds of inference calls, suggesting sustained efficiency in discovering new FRSTs. In the small $(h^{1,1},N_{\rm vert})$ cases, the rate does not fall to zero even at $N_{\rm guess}=1{,}000$, indicating continued exploration of more difficult or infrequent regions of the space. In practice, this behavior makes CYTransformer a reliable FRST sampler. Users can expect a steady return of new triangulations across extended sampling runs.

\begin{figure}[htbp]
  \centering
  \includegraphics[height=7.0cm]{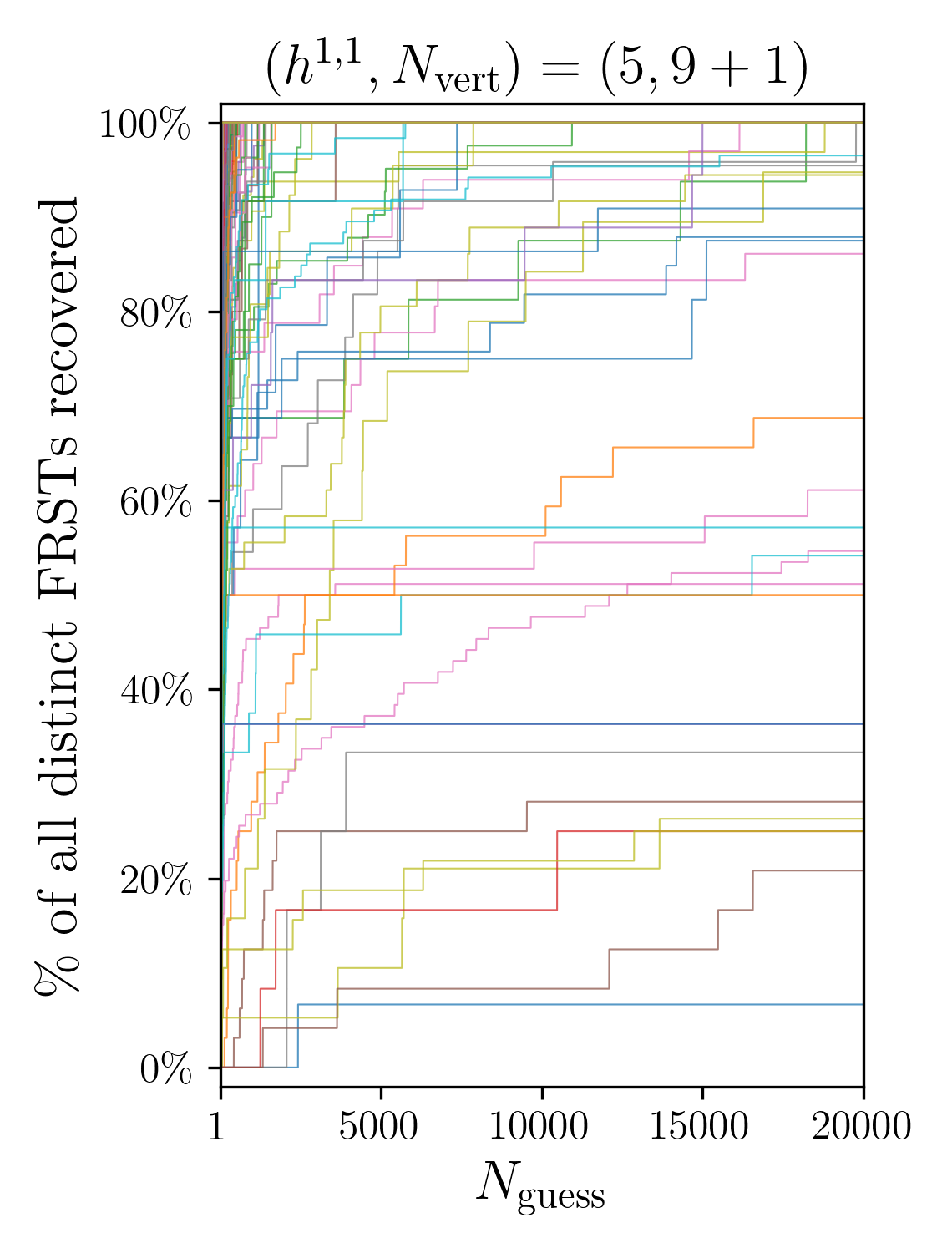}
  \includegraphics[height=7.0cm,trim={1.00cm 0 0 0},clip]{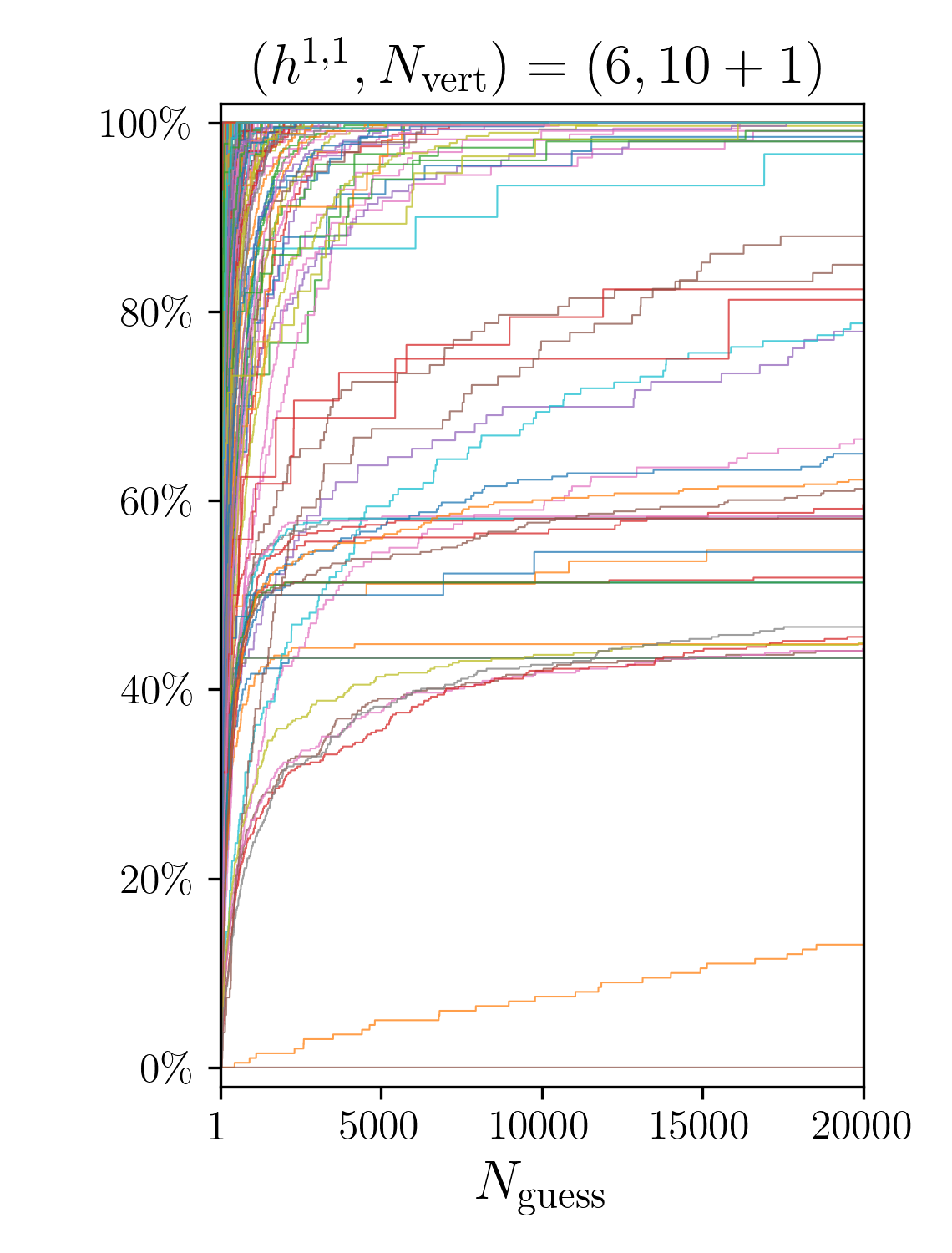}
  \includegraphics[height=7.0cm,trim={1.00cm 0 0 0},clip]{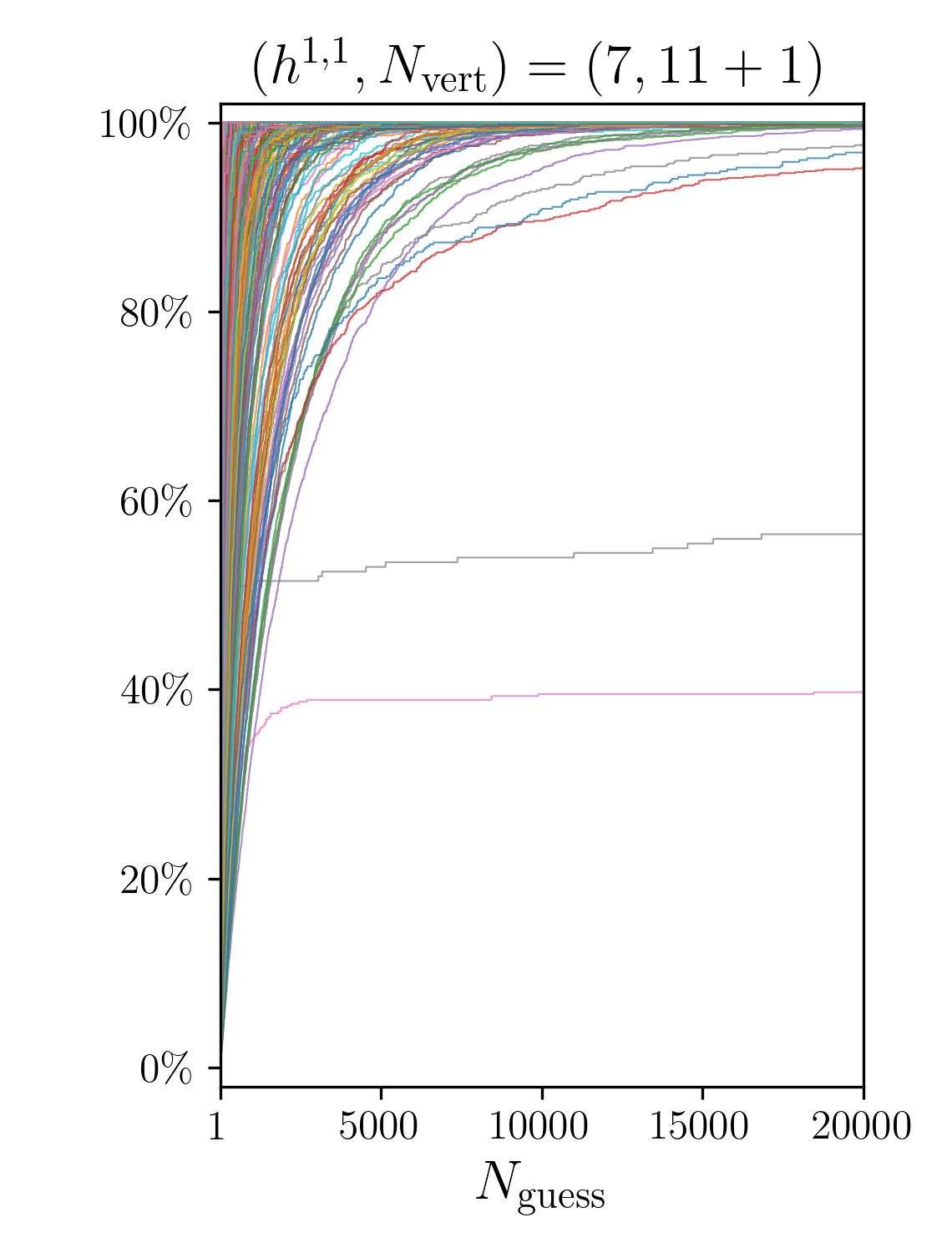} \\
  \includegraphics[height=7.0cm]{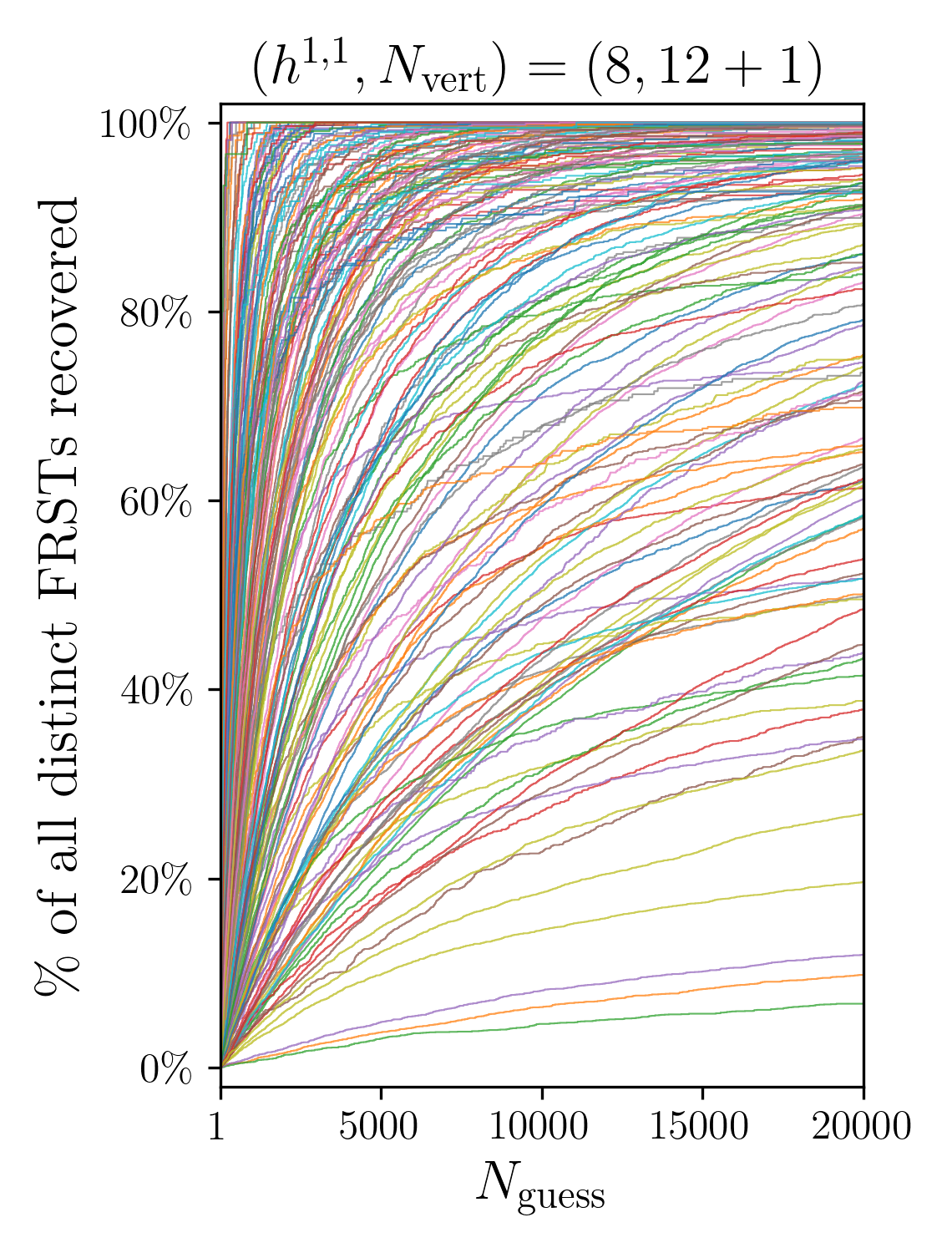}
  \includegraphics[height=7.0cm,trim={1.00cm 0 0 0},clip]{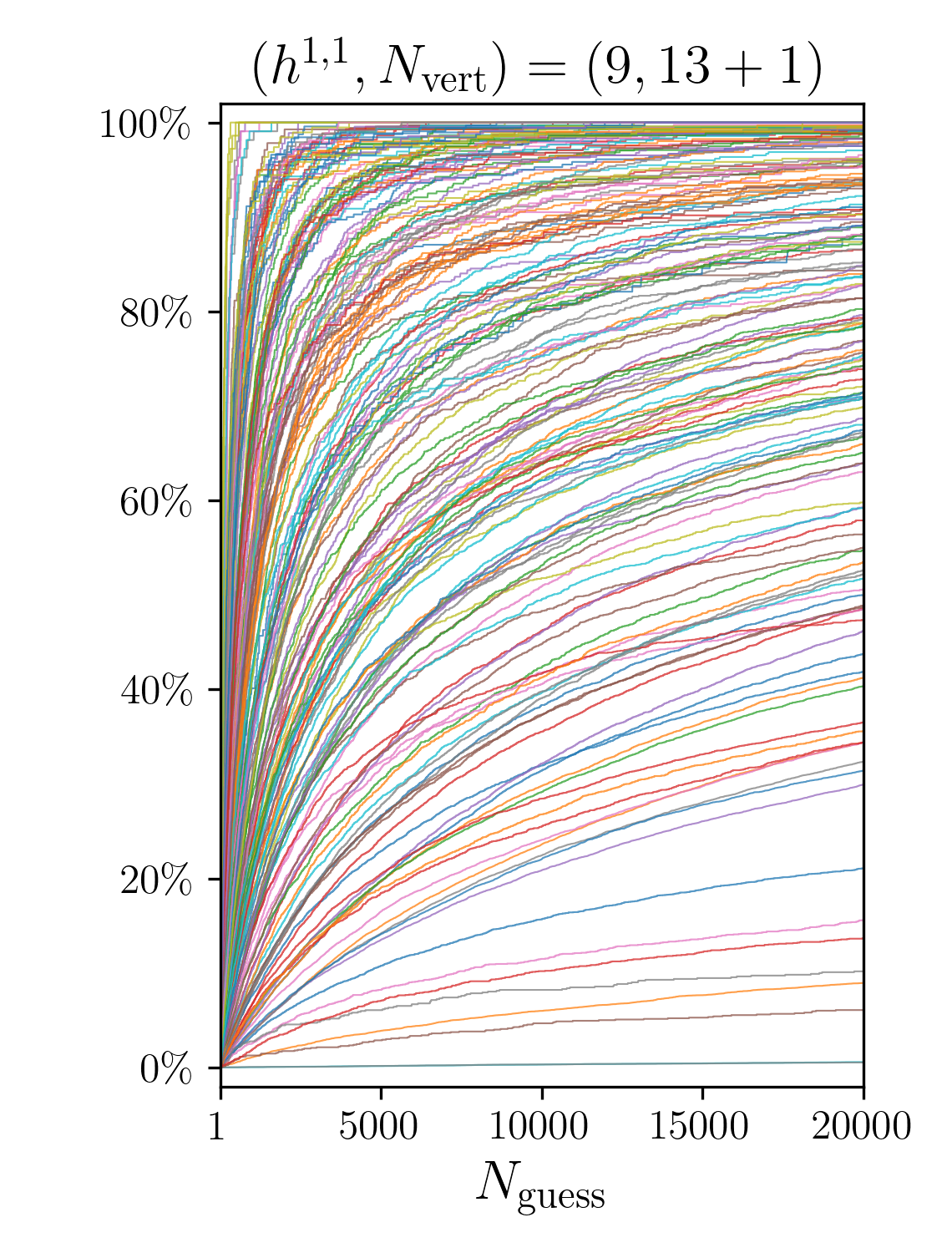}
  \includegraphics[height=7.0cm,trim={1.00cm 0 0 0},clip]{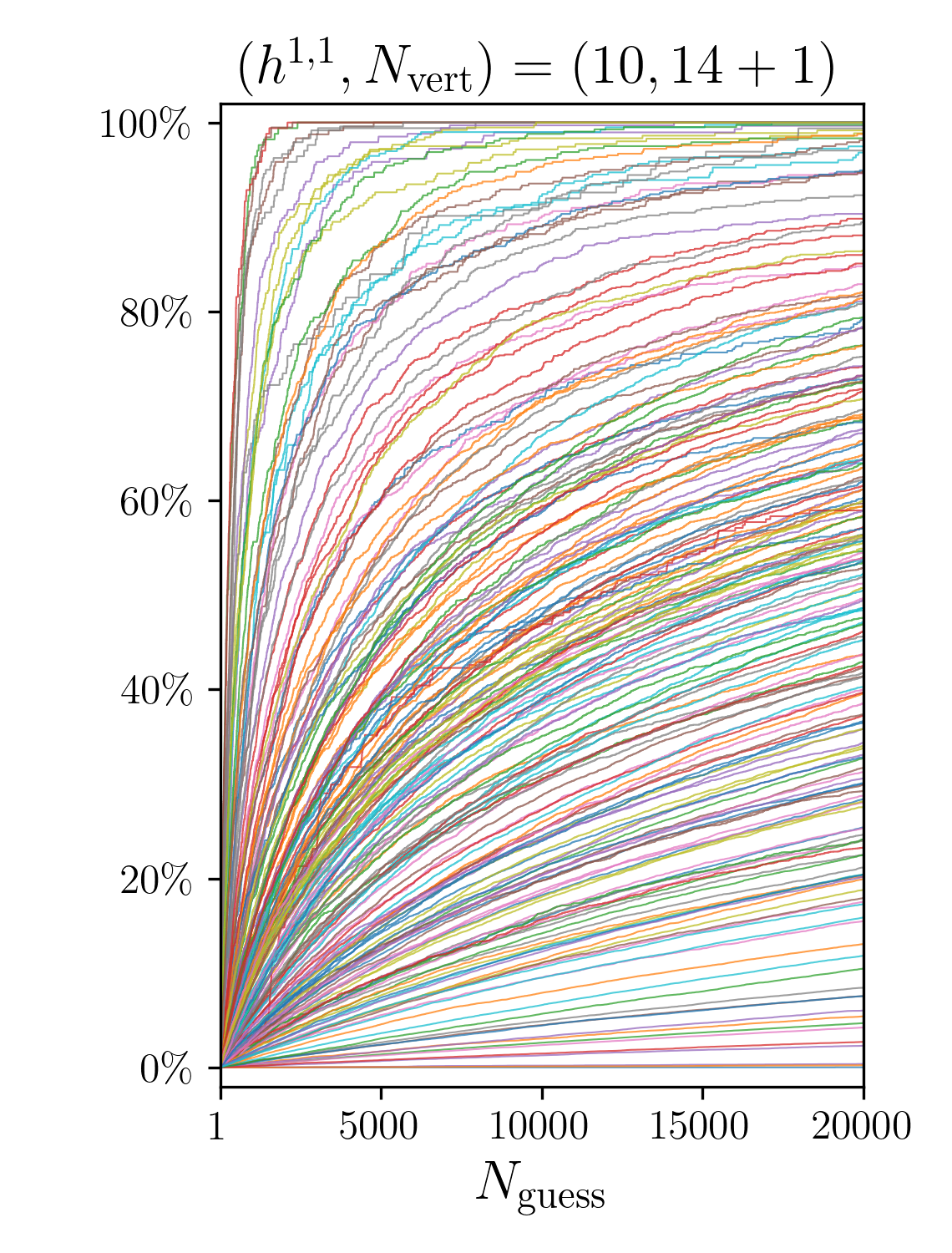}
  \caption{\textbf{Per-polytope \protect\hyperlink{met:frstreccurve}{FRST recovery curves}.} Each panel shows the percentage of all distinct FRSTs recovered as a function of inference calls $N_{\rm guess}$, plotted individually for test polytopes within a fixed $(h^{1,1},N_{\rm vert})$ set. While most polytopes exhibit rapid and high recovery, a noticeable subset show slow or limited recovery, highlighting the influence of polytope geometry on model performance. Some curves plateau early or rise slowly, indicating cases where CYTransformer struggles to learn the full FRST space efficiently. Common in the $(10,14+1)$ case, smooth and gently sloping curves correspond to polytopes with especially large FRST spaces, for which higher sampling budgets are necessary to achieve meaningful recovery.}
  \label{fig:ind_recovery}
\end{figure}

The average generation curves mask the variability that arises from polytope-specific geometry. To uncover this, we turn to individual \hyperlink{met:frstreccurve}{recovery curves}, where each curve represents the number of distinct FRSTs generated, normalized by the total number admitted by the polytope. These plots, shown in figure~\ref{fig:ind_recovery}, illustrate that generation efficiency varies across polytopes, highlighting the impact of polytope geometry on model performance.\footnote{We leave the investigation of how polytope geometry affects learning to future work.} While most test polytopes exhibit recovery curves that quickly reach a high saturation level, a noticeable number show limited or slower recovery. For example, in the $(6,10+1)$ panel, the brown horizontal line at the bottom represents a case where the model fails to recover any FRSTs at all. Just above it, the orange curve shows a steady yet slow increase, suggesting that CYTransformer does learn something, though not very effectively. These rare but poorly performing cases may be attributed to the model’s difficulty in generalizing to uncommon polytope geometries or the inherent complexity of the corresponding FRST spaces. Finally, a smooth and gently sloping curve often corresponds to a polytope that admits a large number of FRSTs; such polytopes are especially prevalent in the $(10,14+1)$ case (curves near the bottom of the plot), where significantly more inference calls are needed to achieve a meaningful recovery rate.

\begin{figure}
    \centering
    \includegraphics[width=1\linewidth]{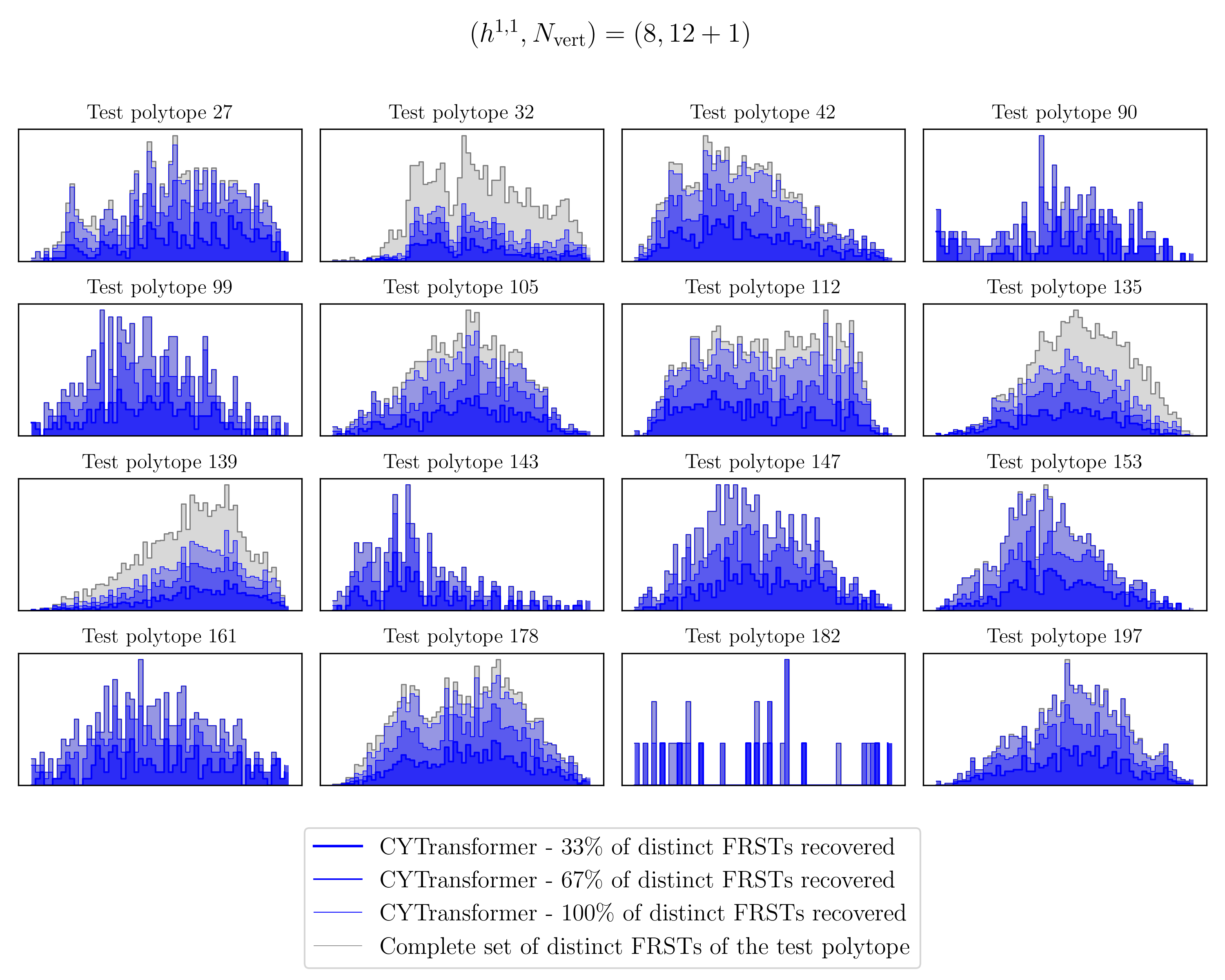}
    \caption{\textbf{CYTransformer-generated \protect\hyperlink{met:hsfrstdist}{height-space FRST distributions} for individual test polytopes.} Each panel shows height-space distributions for CYTransformer-generated FRSTs (blue) compared to the full population (gray) for $16$ randomly selected test polytopes with $(h^{1,1}, N_{\rm vert})=(8,12+1)$. The three shades of blue represent the first $33$\%, $67$\%, and $100$\% of distinct FRSTs recovered by CYTransformer. The close alignment in shape between the partial and full distributions indicates that CYTransformer explores the FRST space in an unbiased, representative manner across all sampling stages, even when full coverage is not yet achieved. The spikiness of the distributions for test polytope $182$ is due to the small number of FRSTs it admits, and does not reflect model failure.}
    \label{fig:solo_percent}
\end{figure}

\textbf{Representativeness.} Figure~\ref{fig:solo_percent} shows the
\hyperlink{met:hsfrstdist}{height-space FRST distributions} for $16$ randomly selected test polytopes for the $(8,12+1)$ configuration. For each polytope, we compare the distribution generated by CYTransformer (blue) to the full FRST population (gray). Despite the limited sampling budget of $20{,}000$ candidate triangulations, CYTransformer recovers nearly all FRSTs for many polytopes, resulting in distributions that closely match the population distribution. This indicates that the model generally explores the FRST space thoroughly, leaving virtually no regions unvisited.

In cases where the population distribution is not fully reconstructed, the mismatch is not due to model bias or failure to learn the polytope geometry. Instead, it reflects the sheer number of FRSTs some polytopes admit, which requires more than $20{,}000$ candidate triangulations to sample adequately. Remarkably, even in these cases, the CYTransformer-generated distributions preserve the overall shape of the population distribution, differing only by scale. This suggests that the model samples the space in an \emph{unbiased} fashion and would recover the full distribution with more inference calls. To make this observation more precise, we additionally plot distributions based on the first $33$\% and $67$\% of distinct FRSTs recovered by the model. Across all polytopes, these partial distributions consistently exhibit the same shape as the full population distribution. This confirms that CYTransformer samples across the entire FRST space in a representative and unbiased manner in all stages of sampling, which is due to the independence of each sample.

\subsection{Performance comparison with the fast sampler}
In this subsection, we evaluate CYTransformer in comparison with the fast sampler (see section~\ref{sec:fast}). To recap, the fast sampler generates candidate triangulations by perturbing the Delaunay height vector with Gaussian noise. We fine-tune the standard deviation $c$ of this noise to optimize performance, and find that $c=20$ yields the best recovery curve across all $(h^{1,1},N_{\rm vert})$ configurations considered.

\begin{figure}[htbp]
  \centering
  \includegraphics[height=4.85cm]{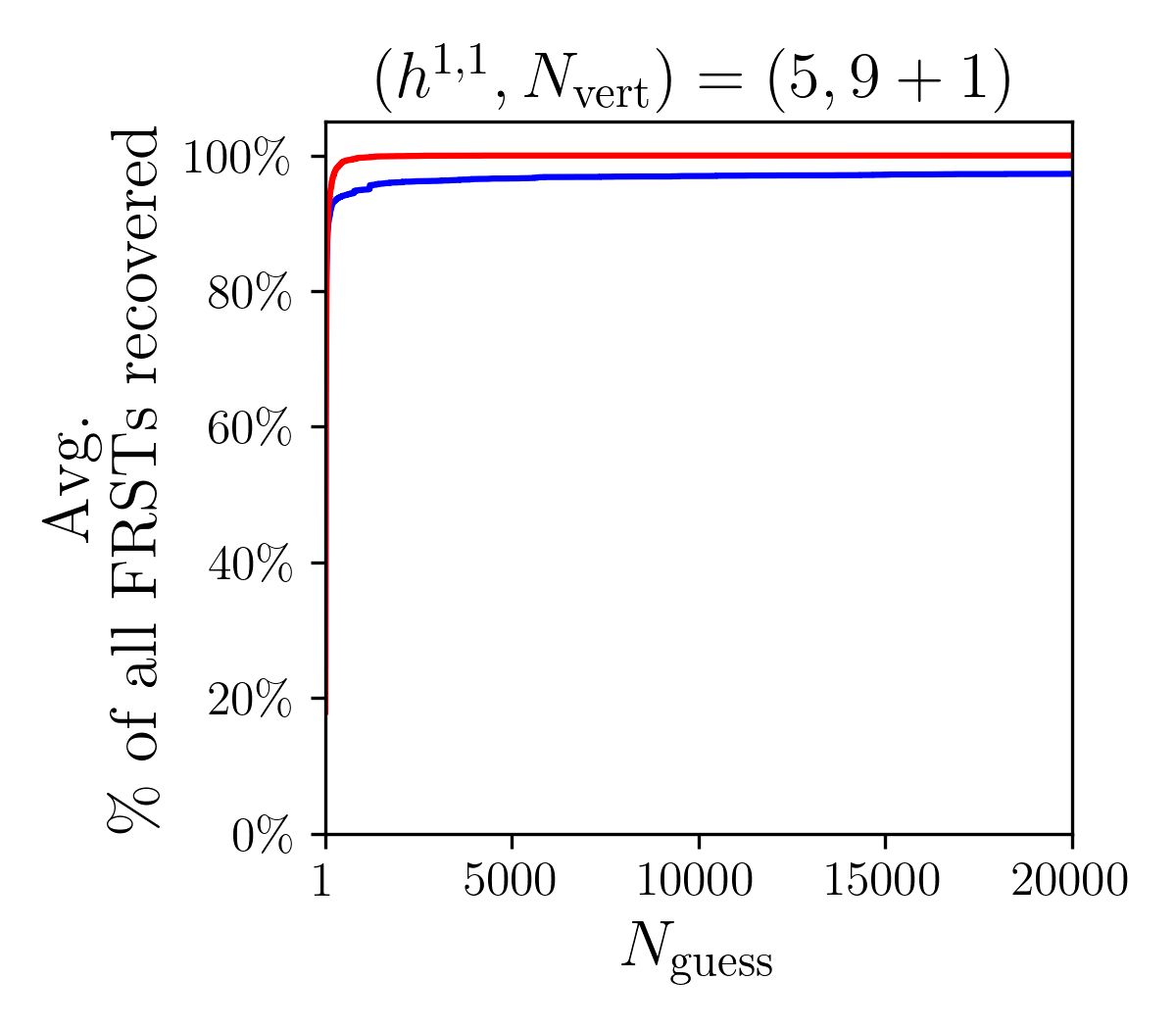}
  \includegraphics[height=4.85cm,trim={1.6cm 0 0 0},clip]{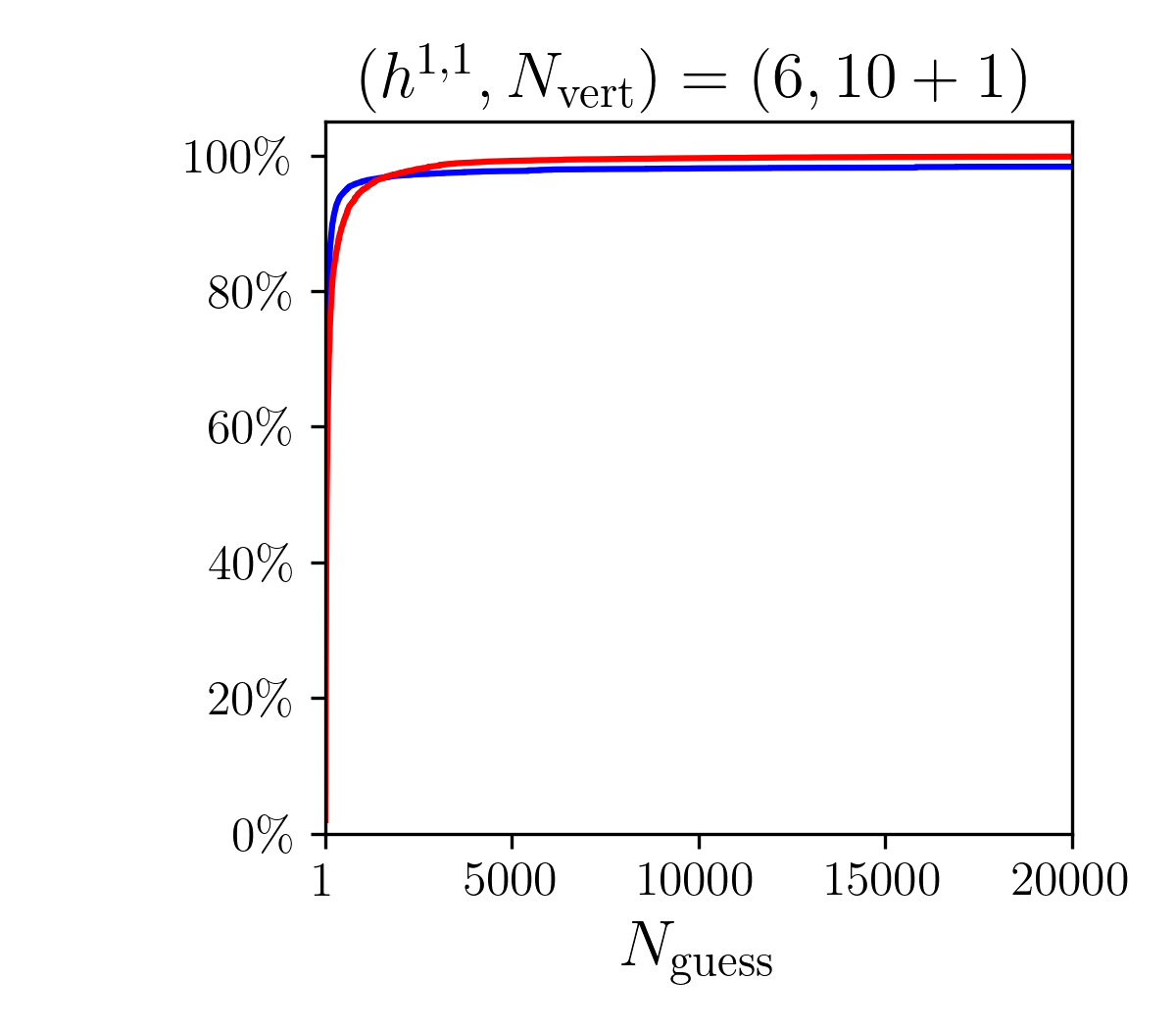}
  \includegraphics[height=4.85cm,trim={1.6cm 0 0 0},clip]{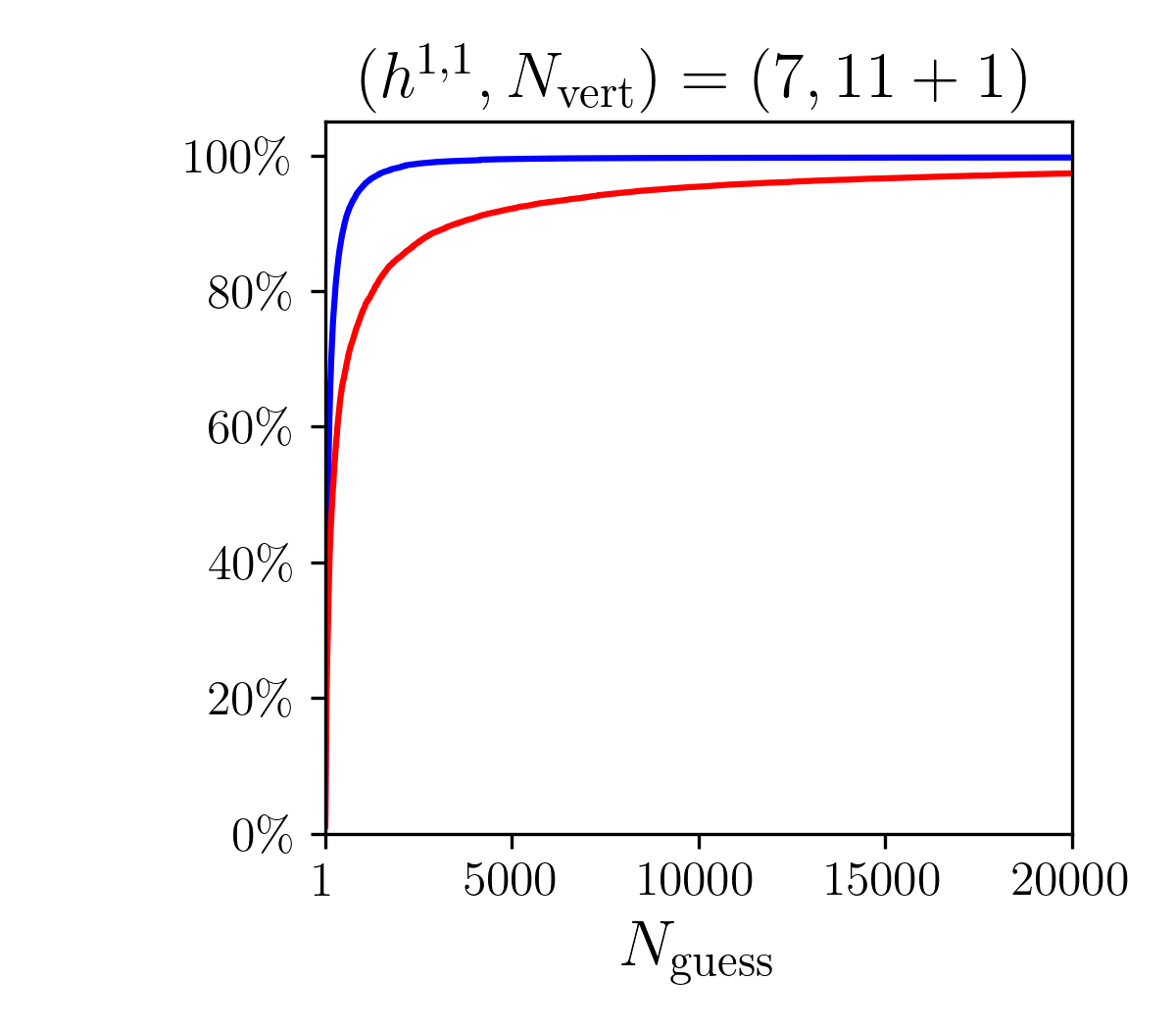} \\
  \includegraphics[height=4.85cm]{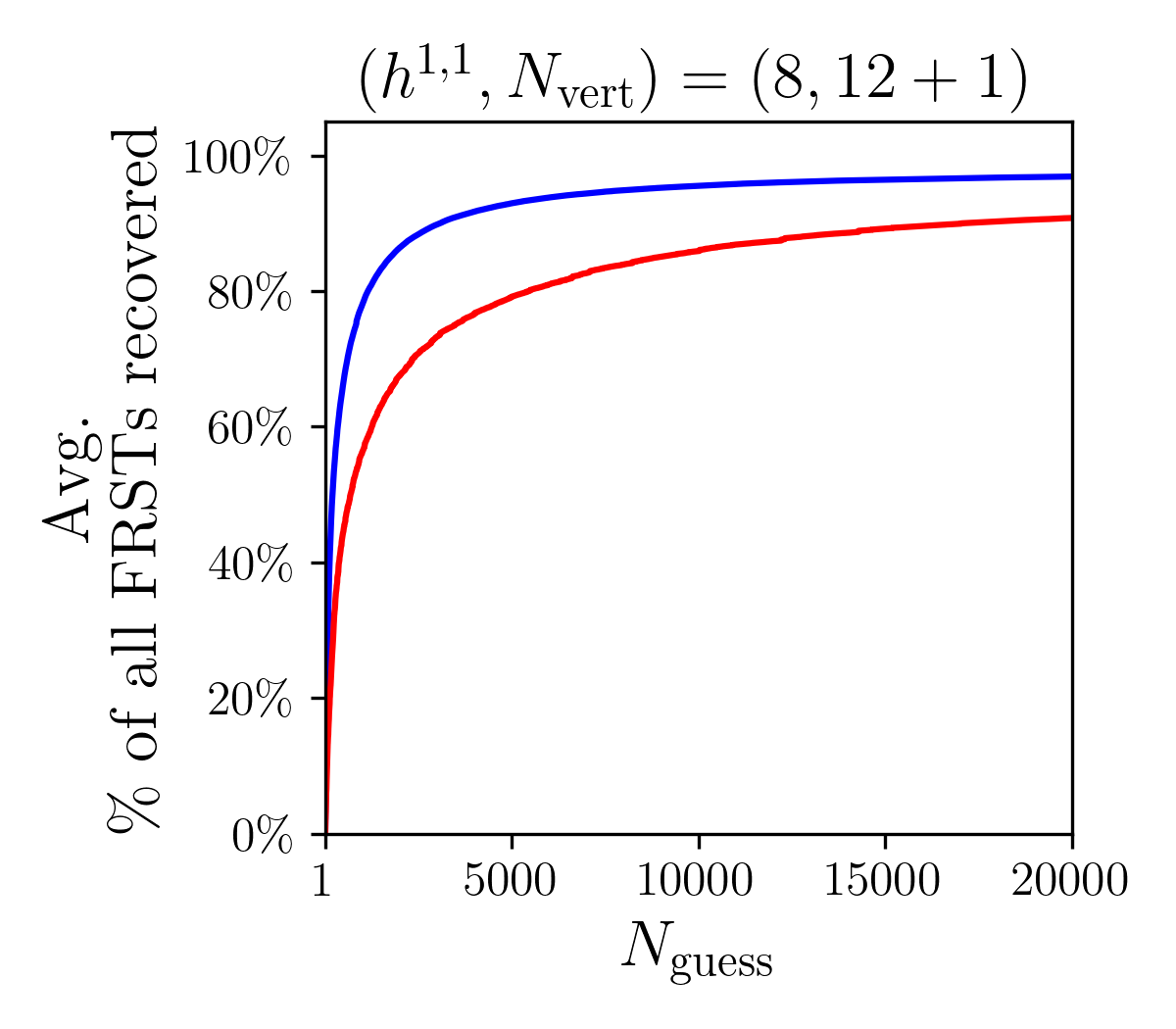}
  \includegraphics[height=4.85cm,trim={1.6cm 0 0 0},clip]{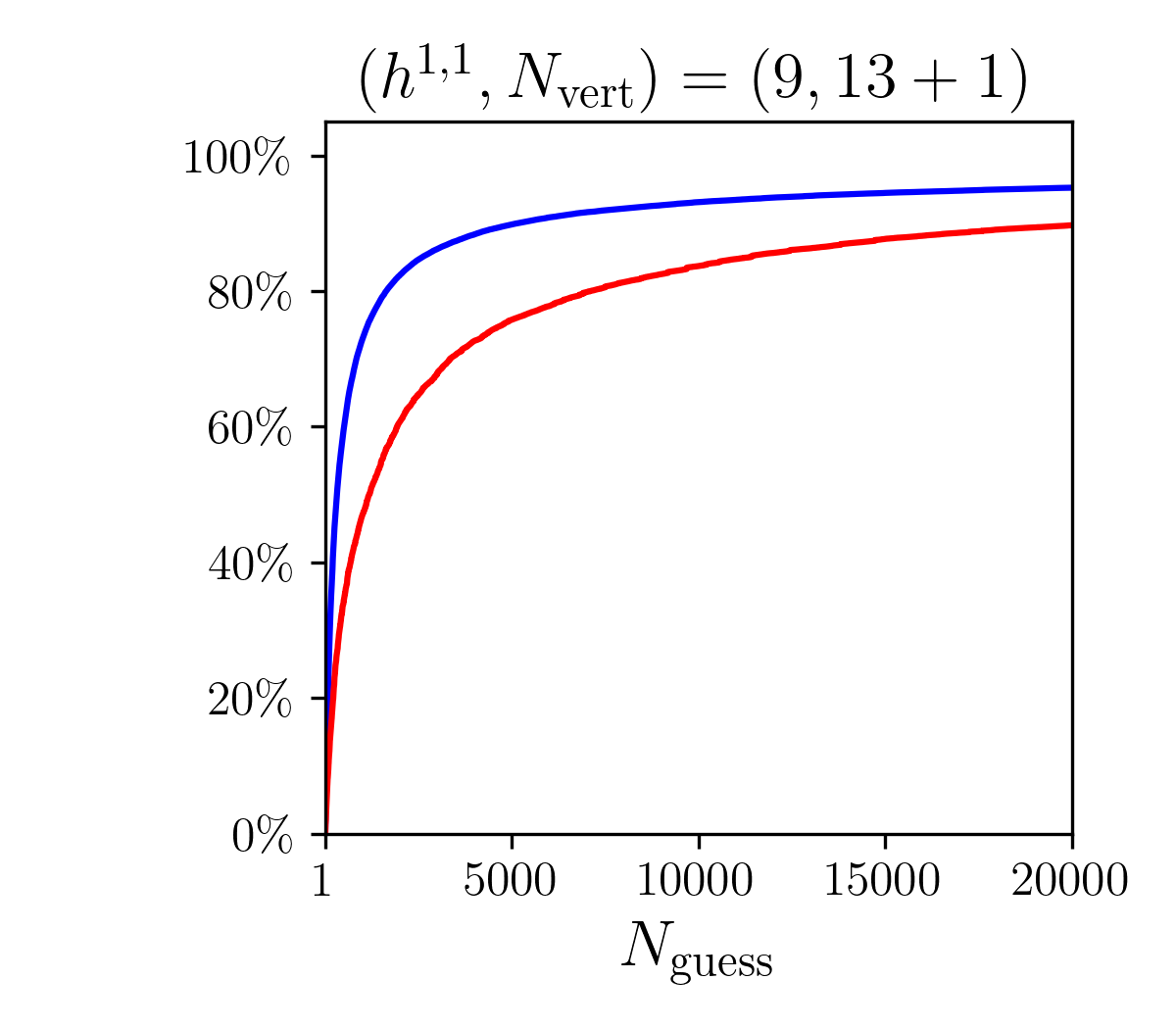}
  \includegraphics[height=4.85cm,trim={1.6cm 0 0 0},clip]{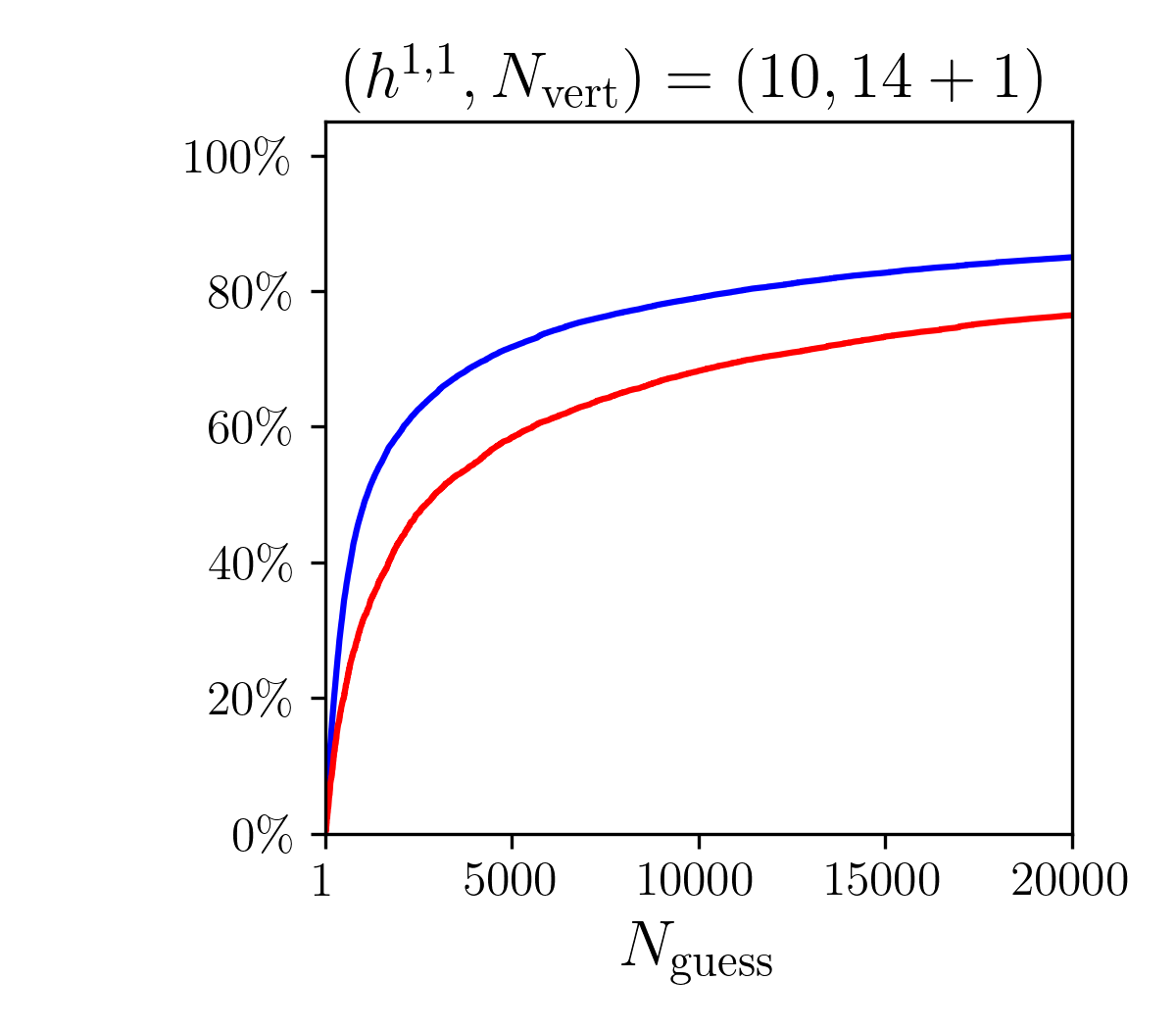}
  \includegraphics[width=1\textwidth]{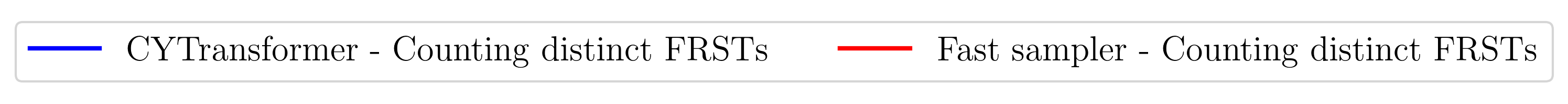}
  \caption{\textbf{Comparison of average \protect\hyperlink{met:frstreccurve}{FRST recovery curves} between CYTransformer and the fast sampler.} Each panel shows the average percentage of distinct FRSTs recovered as a function of inference calls $N_{\rm guess}$, averaged over 200 test polytopes, for increasing $(h^{1,1},N_{\rm vert})$. For the simplest case $(5,9+1)$, the fast sampler (red) outperforms CYTransformer (blue). In the intermediate case $(6,10+1)$, CYTransformer initially leads but is eventually overtaken by the fast sampler at large $N_{\rm guess}$. For more complex polytopes $(7,11+1)$ to $(10,14+1)$, CYTransformer consistently outperforms the fast sampler across the entire inference range. This trend illustrates CYTransformer’s increasing advantage as polytope complexity grows: it learns and samples from a global distribution, enabling diverse and representative FRST generation in large and structured spaces. In contrast, the fast sampler is more effective in smaller, locally scannable ones.}
  \label{fig:compare_recovery}
\end{figure}

We begin by examining the average \hyperlink{met:frstreccurve}{FRST recovery curves} shown in figure~\ref{fig:compare_recovery}. For the simplest case $(5,9+1)$, the fast sampler (red) outperforms CYTransformer (blue) in recovering distinct FRSTs. In the intermediate case $(6,10+1)$, CYTransformer initially performs better, but is eventually overtaken by the fast sampler at large $N_{\rm guess}$. For more complex cases $(7,11+1)$ to $(10,14+1)$, CYTransformer consistently outperforms the fast sampler across the entire range of inference calls. In summary, CYTransformer demonstrates increasingly superior performance compared to the fast sampler as polytope complexity increases.

When restricting to distinct NTFE FRSTs, both methods perform comparably across all configurations when evaluating the weighted average recovery (not shown in the figure). However, if we instead consider the overall percentage of total NTFE FRSTs recovered within the test set, which is a statistic naturally dominated by polytopes with large NTFE FRST counts, CYTransformer outperforms the fast sampler across the full range of $N_{\rm guess}$ for $(8,12+1)$, achieving $87$\% recovery versus $75$\% at $N_{\rm guess}=20{,}000$. This indicates that CYTransformer is not biased toward a narrow subset of TFE triangulations: when applied to a pool of polytopes, especially those admitting many NTFE FRSTs, CYTransformer can serve as a robust generator of topologically distinct Calabi-Yau manifolds. We expect the similar trends for more complex $(h^{1,1},N_{\rm vert})$ configurations, though a clear advantage is not yet visible at our current $N_{\rm guess}$ values, due to the much larger NTFE FRST spaces for those polytopes relative to our limited sampling budget.

We interpret the trend of growing performance advantage as evidence that CYTransformer benefits from the increased complexity of large FRST spaces. For small configurations, the FRST space is small (typically on the order of $10$ FRSTs per polytope for $(5,9+1)$), allowing the fast sampler to efficiently scan the space via random perturbations. In contrast, CYTransformer learns a distribution and tends to prioritize the most representative FRSTs, potentially overlooking rarer ones when the total number is small. However, the FRST space becomes significantly larger and more structured as $(h^{1,1},N_{\rm vert})$ increases. In this regime, CYTransformer captures the \emph{global} distribution more effectively, producing a diverse set of FRSTs that better represent the full range of possibilities. The fast sampler, by contrast, remains a \emph{local} method, confined to the neighborhood of the Delaunay triangulation. In other words, while the fast sampler performs local exploration, CYTransformer learns and samples from a global distribution.

\begin{figure}
    \centering
    \includegraphics[width=1\linewidth]{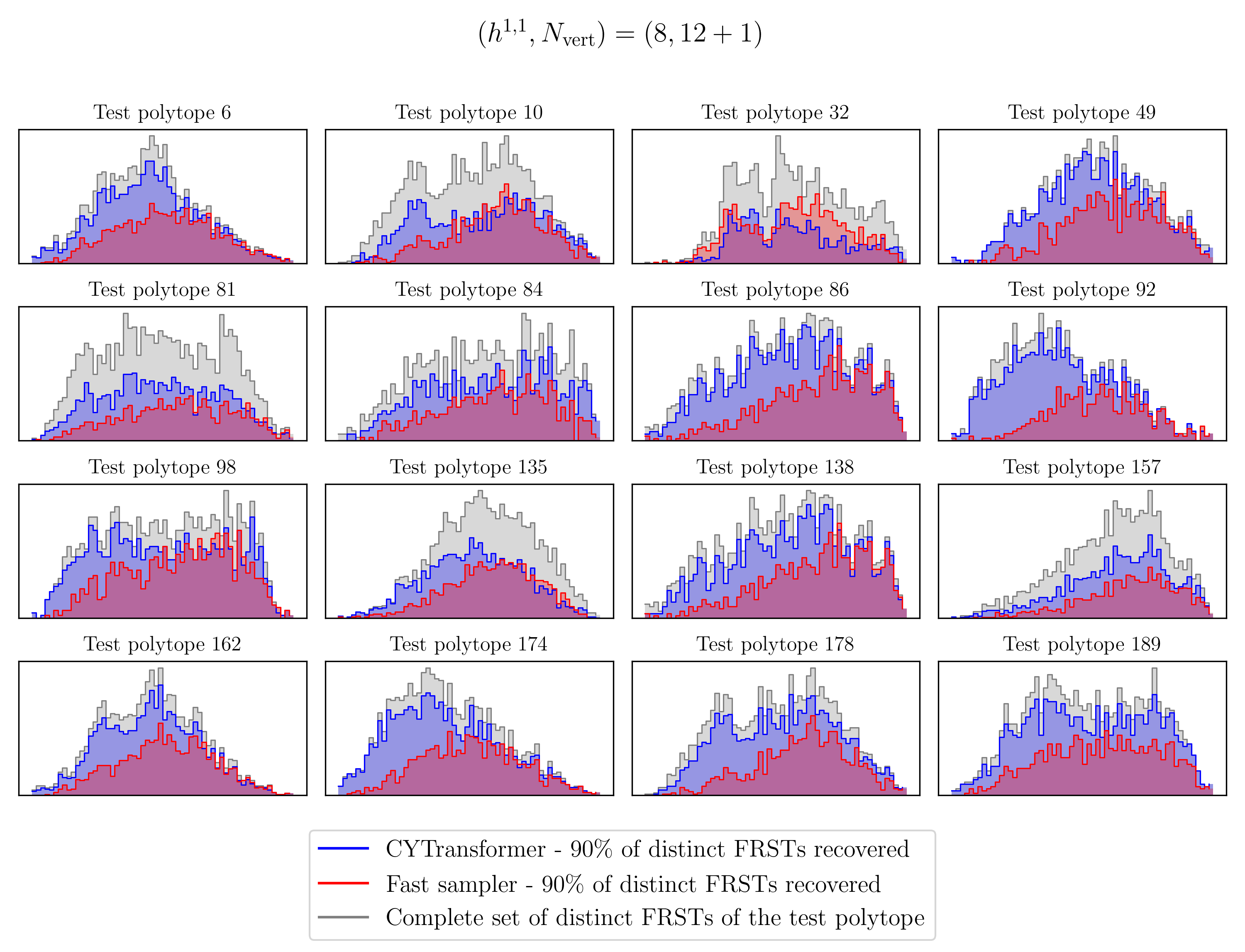}
    \caption{\textbf{Comparison of sampling distributions between CYTransformer and the fast sampler.} \protect\hyperlink{met:hsfrstdist}{Height-space FRST distributions} are shown for 16 test polytopes with $(h^{1,1},N_{\rm vert})=(8,12+1)$. These examples are selected to highlight contrasting sampling behaviors. For each polytope, we compare CYTransformer (blue) and the fast sampler (red) using the first 90\% of distinct FRSTs recovered by each method, shown alongside the full population distribution (gray). The 90\% threshold is used purely for visualization clarity. CYTransformer consistently matches the shape of the population histogram more closely, indicating more representative and unbiased sampling across the FRST space. In contrast, the fast sampler often exhibits skewed distributions, reflecting sampling bias or overconcentration in specific regions.}
    \label{fig:compare_hist}
\end{figure}

To further probe this difference, we compare CYTransformer and the fast sampler in terms of their \hyperlink{met:hsfrstdist}{height-space FRST distributions} for $16$ handpicked test polytopes with $(h^{1,1},N_{\rm vert})=(8,12+1)$, chosen to highlight clear differences in sampling behavior between CYTransformer (blue) and the fast sampler (red); see figure~\ref{fig:compare_hist}. We also plot the population distribution (gray). We find that the CYTransformer's distributions consistently match the shape of the full population distribution, while the fast sampler's distributions often exhibit skewed profiles. This indicates that although both methods can eventually cover much of the FRST space, they do so differently: CYTransformer samples in proportion to the true density of FRSTs, while the fast sampler tends to concentrate on particular regions, resulting in sampling bias.

\begin{figure}
  \centering
  \includegraphics[width=0.495\textwidth]{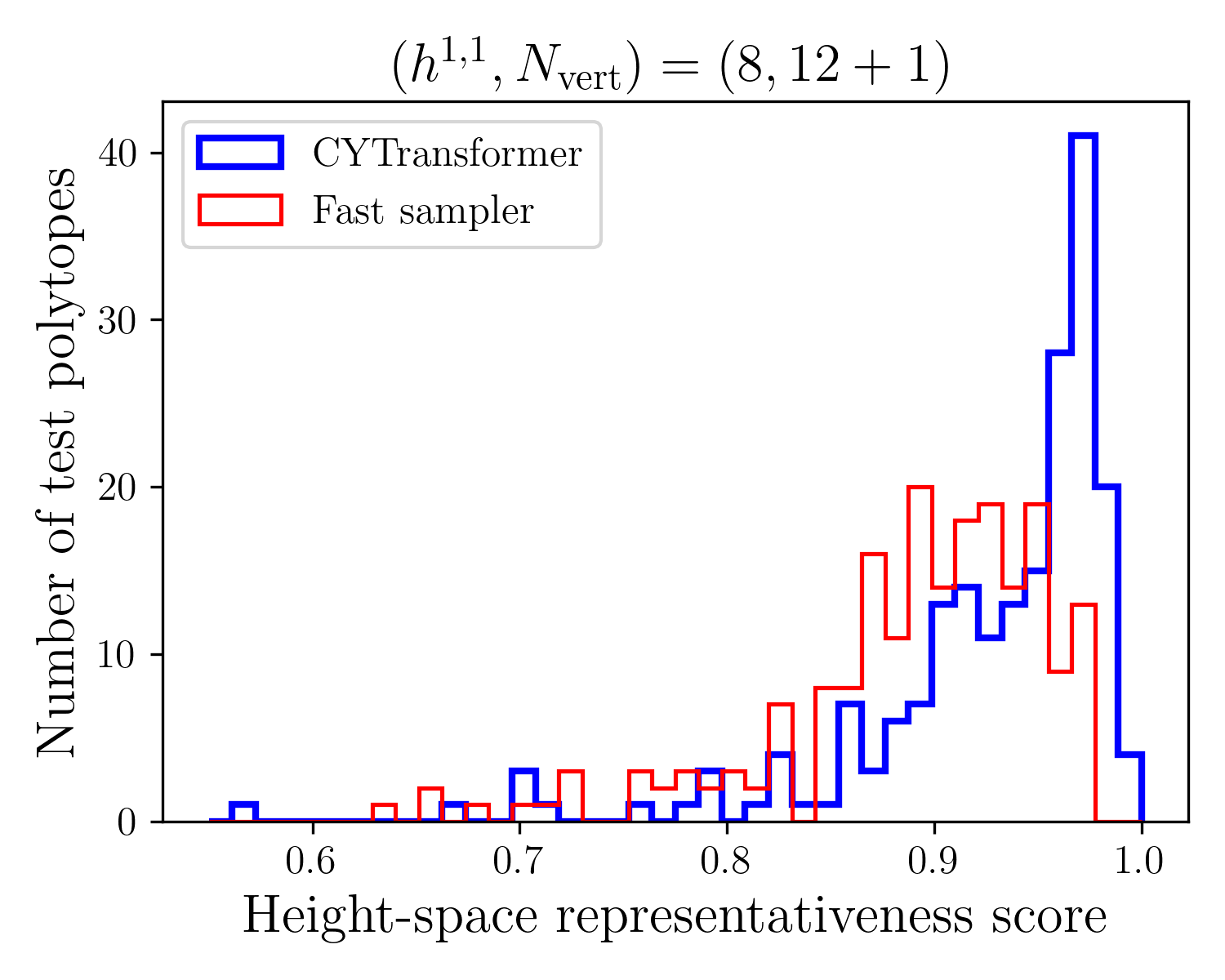}
  \includegraphics[width=0.495\textwidth]{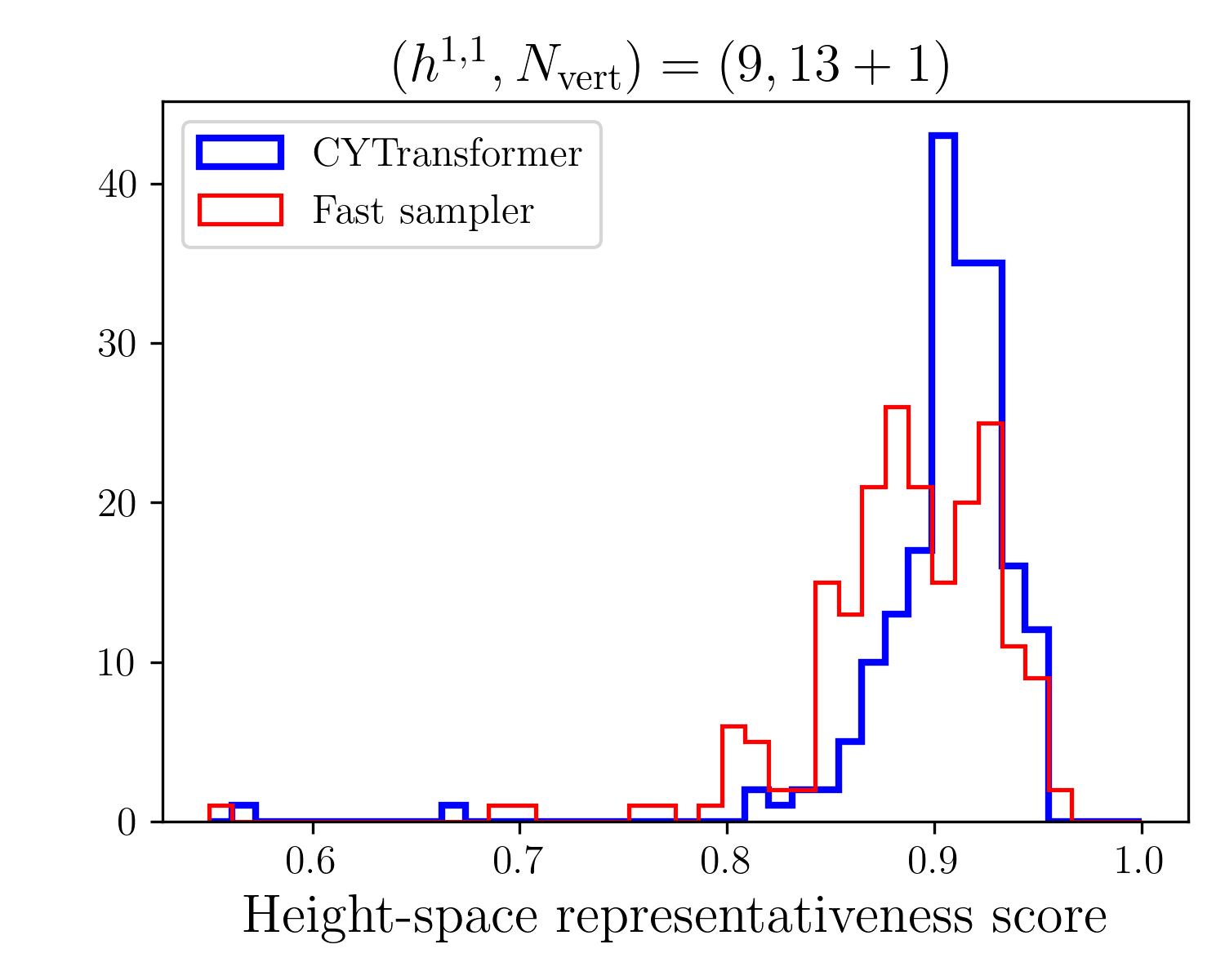} \\
  \includegraphics[width=0.495\textwidth]{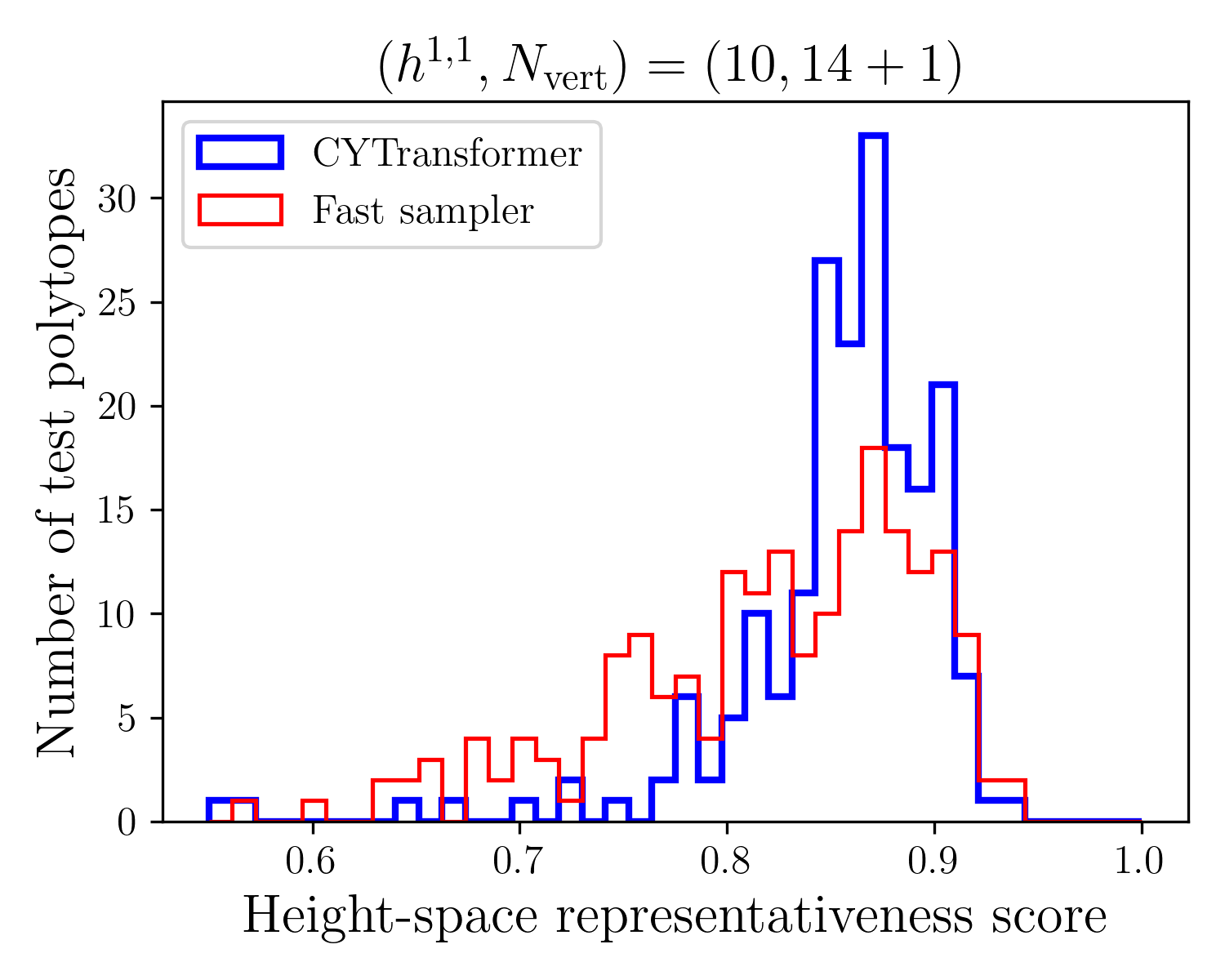}
  \caption{\textbf{Representativeness histograms.} Each of the $200$ \protect\hyperlink{met:hsfrstrep}{height-space representativeness scores} is the cosine similarity between the method's height-space FRST distribution and the full population distribution of the same polytope. Across all $(h^{1,1},N_{\rm vert})$ cases shown, the distributions of scores for CYTransformer (blue) are sharply peaked near unity with low variance, indicating that it consistently and unbiasedly produces samples that reflect the true shape of the FRST distribution. In contrast, the fast sampler (red) exhibits broader and flatter score distributions, confirming that it introduces a stronger sampling bias and tends to favor certain regions of the FRST space over others.}
  \label{fig:compare_rep}
\end{figure}

This conclusion is quantitatively supported by the \hyperlink{met:hsfrstrep}{histograms of height-space representativeness scores} in figure~\ref{fig:compare_rep}, shown for $(8,12+1)$, $(9,13+1)$, and $(10,14+1)$.\footnote{We do not show histograms for $(5,9+1)$, $(6,10+1)$, and $(7,11+1)$, because each polytope in these configurations generally admits relatively few FRSTs, leading to noisy statistics.} Each histogram depicts the distribution of representativeness scores, i.e., the cosine similarity scores between the height-space FRST distribution produced by each method and the corresponding population distribution, computed across $200$ test polytopes. Note that, for polytopes where the model recovers all known FRSTs, the two distributions become identical by definition, resulting in a representativeness score of unity. This artificially inflates the apparent representativeness, even if the model’s intermediate sampling behavior was highly biased. To avoid this issue, we exclude such fully saturated cases and instead compute the representativeness score using only a partial histogram constructed from an early subset of the distinct FRSTs recovered by the model. This allows us to probe the sampling distribution before saturation, providing a more faithful measure of how unbiased the model is during the sampling process.

CYTransformer (blue) achieves consistently higher representativeness scores with lower variance, indicating that its sampling distribution more faithfully captures the true structure of the FRST space. In contrast, the fast sampler (red) shows lower representativeness and broader spread, confirming that it introduces a stronger sampling bias even when the total coverage is similar.

\subsection{A hybrid strategy: CYTransformer-seeded fast sampler}\label{sec:hybrid}
This subsection presents a natural extension of the findings in the previous subsection. Having established that CYTransformer acts as a fair global explorer of the FRST space, while the fast sampler serves as an efficient local scanner, their complementary strengths motivate a hybrid strategy that combines both models: the CYTransformer-seeded fast sampler.

\begin{figure}
    \centering
    \includegraphics[width=1\linewidth]{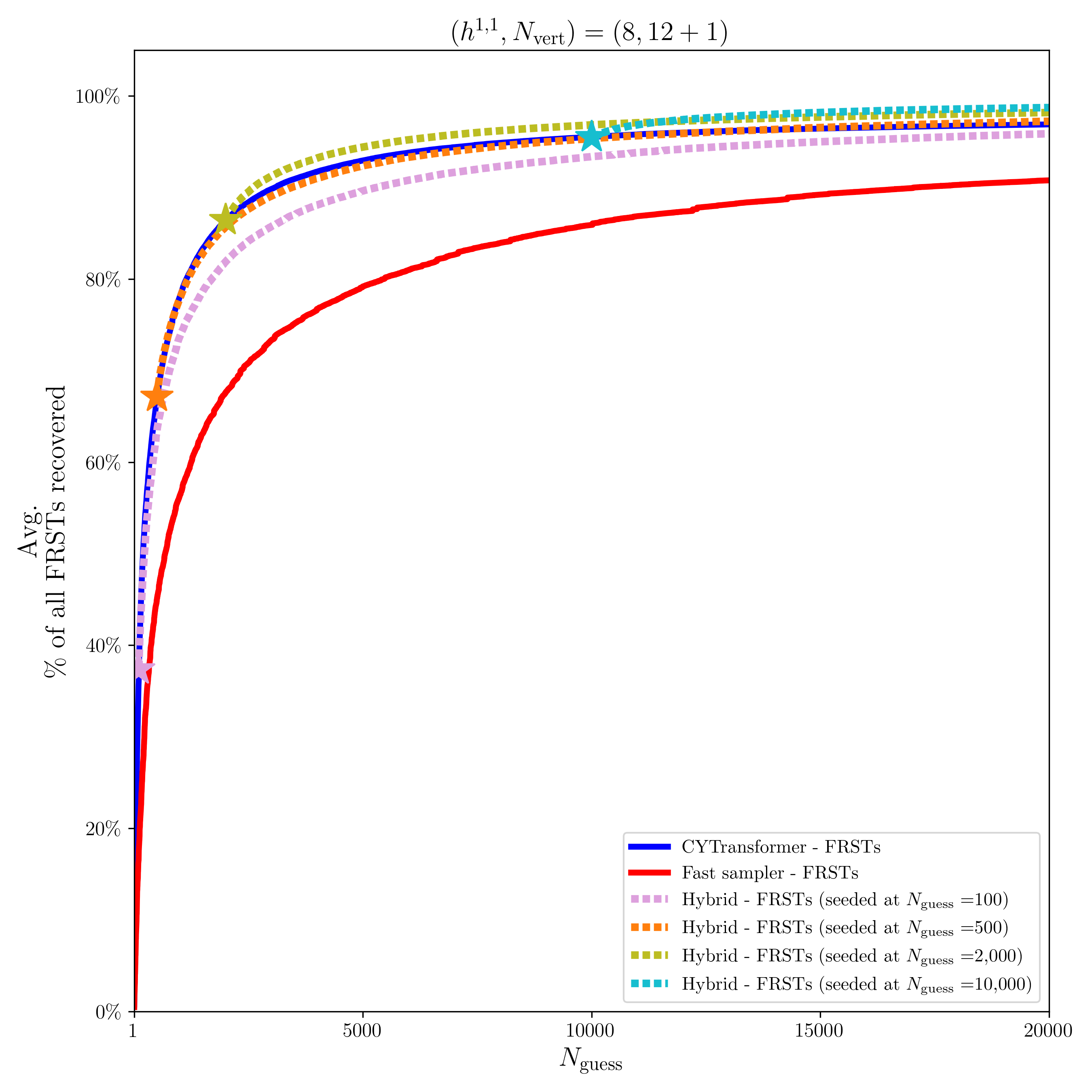}
    \caption{\textbf{\protect\hyperlink{met:frstreccurve}{Recovery curves} for CYTransformer-seeded fast sampler on $(h^{1,1},N_{\rm vert})=(8,12+1)$ polytopes.} Average percentage recovery of distinct FRSTs across $200$ polytopes. We compare pure CYTransformer (blue), pure fast sampler (red), and hybrid strategies (dashed), where CYTransformer is stopped after $N_{\rm cut}$ inference calls (star markers), and the fast sampler continues from random seeds in the resulting pool. Hybrid curves are shown for various $N_{\rm cut}$ values. With $N_{\rm cut}=100$, the hybrid method underperforms due to insufficient seed diversity. As $N_{\rm cut}$ increases, recovery improves, and hybrid performance matches CYTransformer by $N_{\rm cut}=500$, and begins to outperform it for larger values. This indicates that the fast sampler effectively leverages a sufficiently diverse seed pool to enable unbiased exploration. The hybrid method enables early stopping of CYTransformer, offering computational savings without sacrificing recovery quality.}
    \label{fig:combined_recovery12}
\end{figure}

\begin{figure}
    \centering
    \includegraphics[width=1\linewidth]{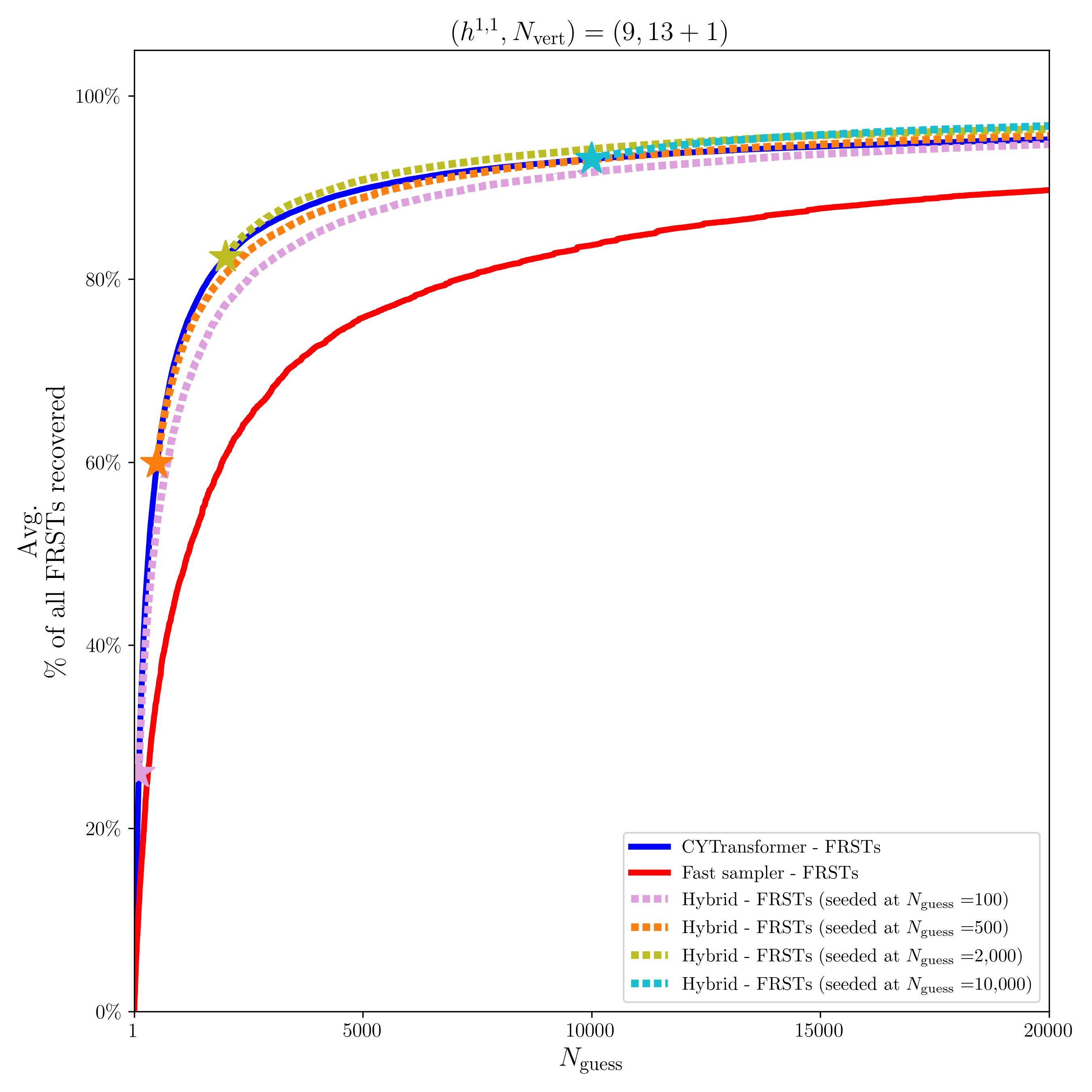}
    \caption{\textbf{\protect\hyperlink{met:frstreccurve}{Recovery curves} for CYTransformer-seeded fast sampler on $(h^{1,1},N_{\rm vert})=(9,13+1)$ polytopes.}}
    \label{fig:combined_recovery13}
\end{figure}

\begin{figure}
    \centering
    \includegraphics[width=1\linewidth]{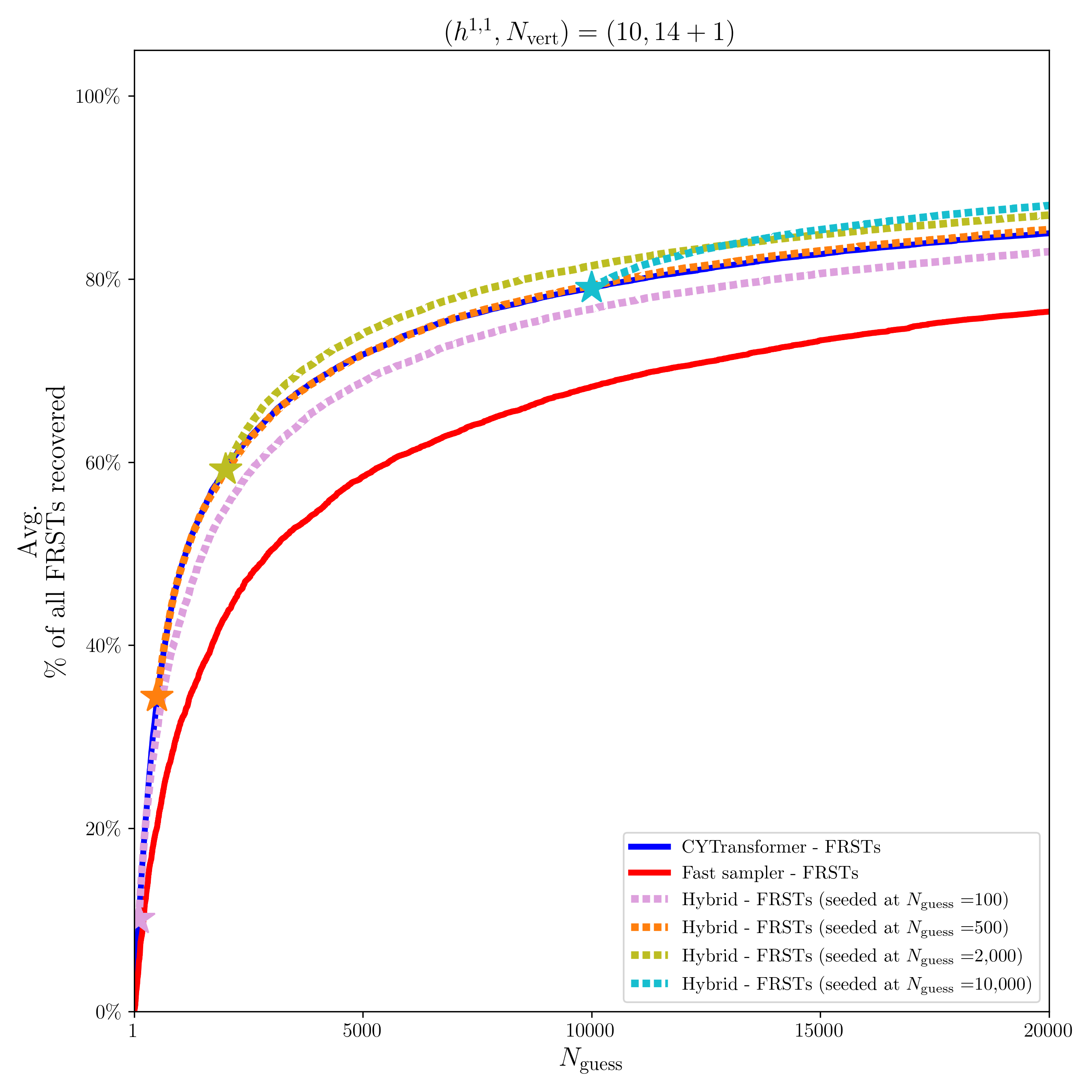}
    \caption{\textbf{\protect\hyperlink{met:frstreccurve}{Recovery curves} for CYTransformer-seeded fast sampler on $(h^{1,1},N_{\rm vert})=(10,14+1)$ polytopes.} For this configuration, increasing $N_{\rm cut}$ to $10{,}000$ yields a hybrid curve that appreciably outperforms pure CYTransformer.}
    \label{fig:combined_recovery14}
\end{figure}

The hybrid method is implemented as follows. We first run CYTransformer for $N_{\rm cut}$ inference calls and collect all distinct FRSTs generated during this phase. These FRSTs form a seed pool. For the remaining $20{,}000-N_{\rm cut}$ candidate triangulations, we sample a seed uniformly at random from this pool to initialize the fast sampler, which then performs a single local perturbation to generate a new candidate triangulation.\footnote{As before, we tune the standard deviation of the Gaussian noise in the fast sampler to optimize the resulting recovery curves.} We apply this hybrid strategy to the $(8,12+1)$, $(9,13+1)$, and $(10,14+1)$ configurations, corresponding to figures~\ref{fig:combined_recovery12},~\ref{fig:combined_recovery13}, and~\ref{fig:combined_recovery14} which show \hyperlink{met:frstreccurve}{recovery curves} for distinct FRSTs across pure CYTransformer (blue), pure fast sampler (red), and several hybrid variants with different $N_{\rm cut}$ values (dashed). Each star marker indicates the point where CYTransformer stops, with the corresponding curve extending from the marker showing the performance of the hybrid method for the remainder of the sampling budget.

For all three cases, we find that when $N_{\rm cut}=100$, the hybrid method underperforms relative to CYTransformer alone, likely because the seed pool is too small to adequately represent the FRST space. As $N_{\rm cut}$ increases to $500$, the hybrid method matches CYTransformer’s performance. Further increases to $2{,}000$, and $10{,}000$ lead to hybrid curves that outperform pure CYTransformer, particularly in the $(10,14+1)$ case. This indicates that the fast sampler can effectively leverage the growing diversity of the seed pool to discover new FRSTs. However, the gain is expected to saturate with increasing $N_{\rm cut}$, suggesting diminishing returns once the seed pool becomes sufficiently representative.

These results reinforce our earlier conclusion that the fast sampler is a highly efficient local scanner. When seeded with CYTransformer-generated samples, which are unbiased across the FRST space, the fast sampler is able to efficiently exhaust nearby triangulations that CYTransformer may overlook. Notably, hybrid strategies with $N_{\rm cut}=500$ performs comparably to full CYTransformer runs, indicating that meaningful computational savings can be achieved without loss in recovery efficiency. Since transformer inference is more expensive than fast sampling, the hybrid method allows early stopping of CYTransformer and continuation with the cheaper fast sampler, making the overall generation process more cost-effective.

\begin{figure}
    \centering
    \includegraphics[width=1\linewidth]{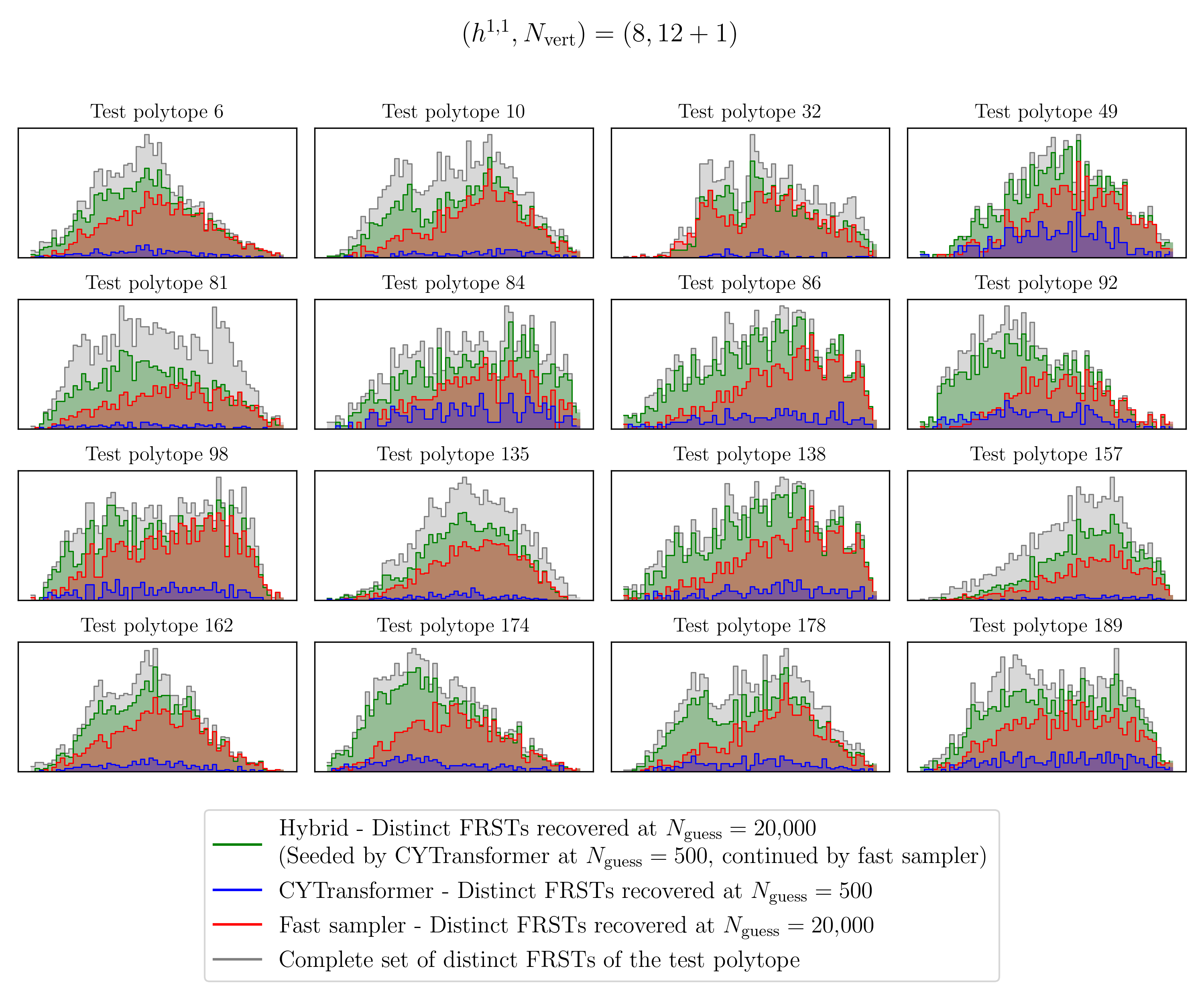}
    \caption{\textbf{Representativeness comparison for hybrid method.} \protect\hyperlink{met:hsfrstdist}{Height-space FRST distributions} for the same $16$ handpicked $(h^{1,1},N_{\rm vert})=(8,12+1)$ test polytopes previously shown in figure~\ref{fig:compare_hist}. For each polytope, we compare the full FRST distribution (gray), CYTransformer-only output at $N_{\rm guess}=500$ (blue), fast-sampler-only output at $N_{\rm guess}=20{,}000$ (red), and the hybrid method using CYTransformer for the first $500$ candidate triangulations followed by the fast sampler for the remaining $19{,}500$ (green). We focus on the $N_{\rm guess}=500$ hybrid setup as its recovery closely matches the pure CYTransformer run, enabling direct comparison with figure~\ref{fig:compare_hist}. Despite its small size, the CYTransformer seed pool already captures key structural features of the true distribution, which the hybrid method preserves and extends. In contrast, the fast sampler alone tends to produce skewed or concentrated distributions, underscoring the importance of CYTransformer's unbiased initialization.}
    \label{fig:combined_hist}
\end{figure}

To confirm that the hybrid method continues to yield representative samples of FRSTs, we examine the \hyperlink{met:hsfrstdist}{height-space FRST distributions} for the same $16$ test polytopes shown earlier in figure~\ref{fig:compare_hist}. In figure~\ref{fig:combined_hist}, each panel shows four distributions: the full set of FRSTs (gray), CYTransformer-only samples at $N_{\rm cut}=500$ (blue), the fast sampler samples at $N_{\rm guess}=20{,}000$ (red), and the hybrid method using $500$ CYTransformer candidate triangulations followed by $19{,}500$ fast samples (green). The purpose of this figure is to highlight that the CYTransformer-only samples, which serve as the seed pool, already reflect key features of the population distribution, despite being small in number. The hybrid method builds on this pool and successfully produces samples that closely match the shape and support of the full FRST distribution, confirming that the hybrid method preserves representativeness. For comparison, the fast sampler alone yields skewed or narrow distributions as previously discussed, underscoring the value of CYTransformer-seeded initialization.

\FloatBarrier
\subsection{Exploring self-improvement}
\label{subsec:selfim_results}

As observed in section~\ref{subsec:training_dynamics} model performance is positively correlated with the training set size. This motivates our exploration of the self-improvement strategy described in section~\ref{subsec:selfim_method}, where we start with a small initial training set and alternate between using the model to generate new training data and and retraining it on the augmented training set. We expect this strategy to be particularly fruitful, and even necessary, for larger values of $N_{\rm vert}$ where acquiring training data is costly. Since this work focuses on smaller $N_{\rm vert}$ for which large datasets are readily generated, our self-improvement experiments serve primarily as a proof of concept for the method.

\begin{figure}[htbp]
  \centering
  \includegraphics[height=4.65cm]{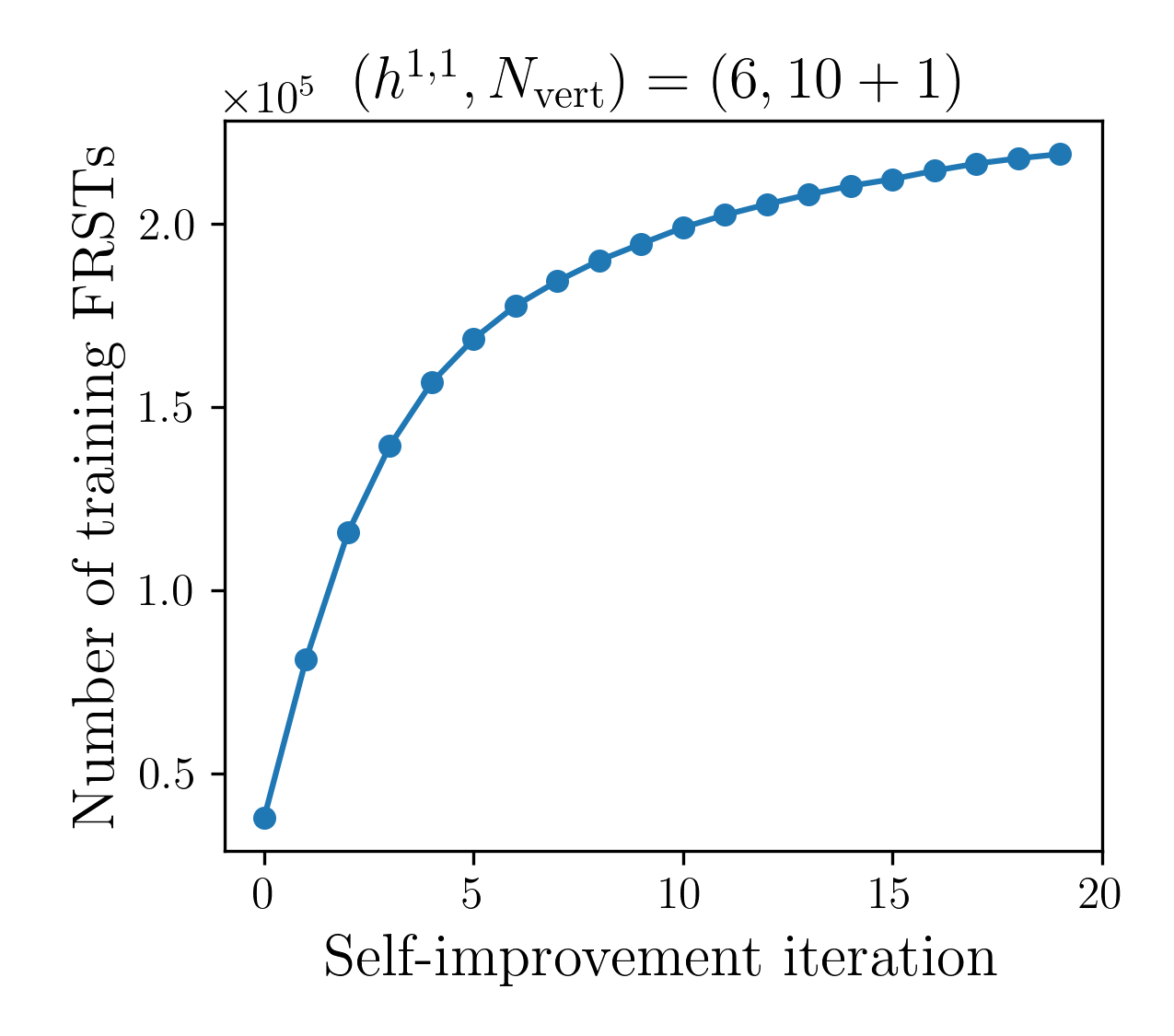}
  \includegraphics[height=4.65cm,trim={1.05cm 0 0 0},clip]{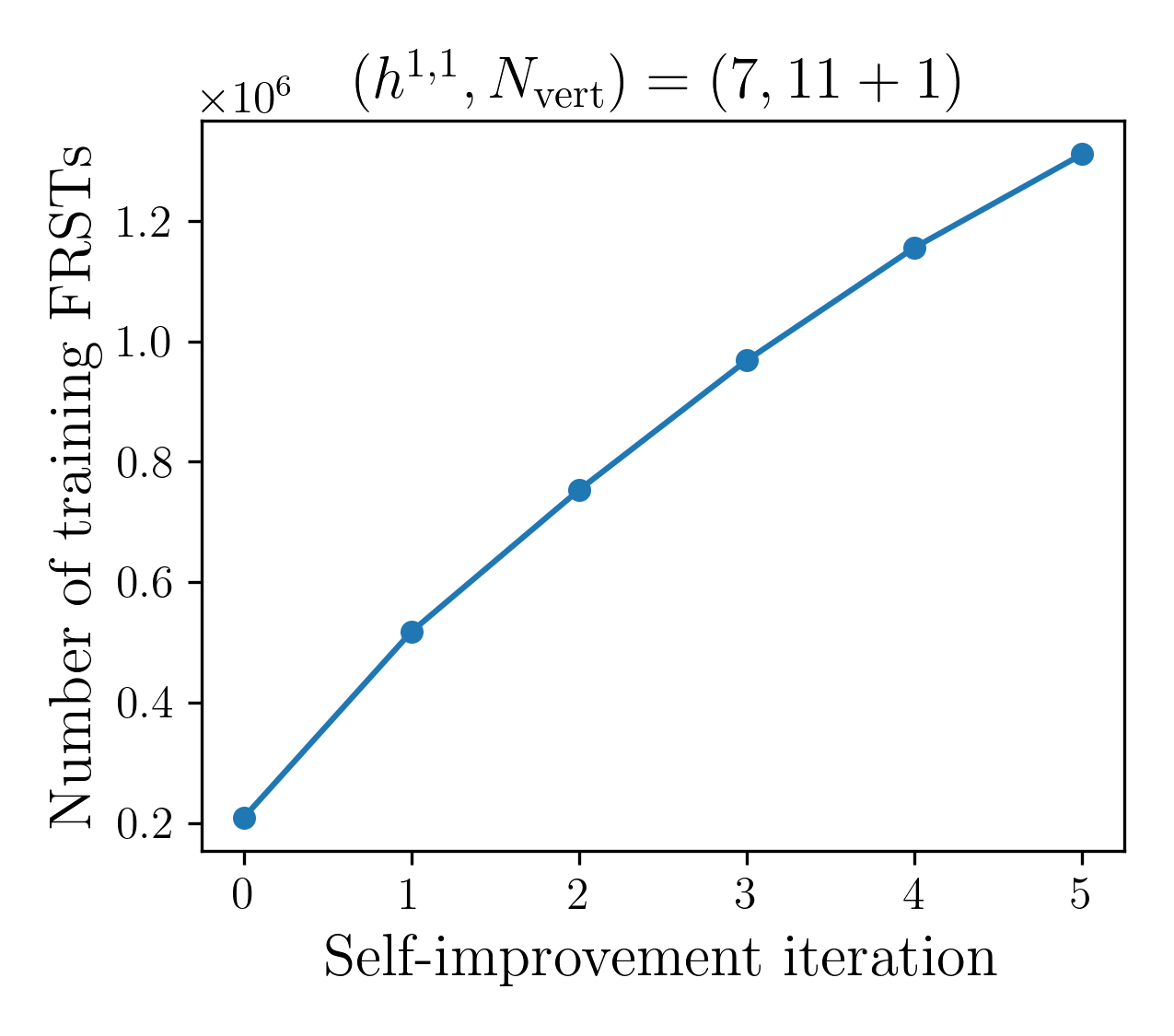}
  \includegraphics[height=4.65cm,trim={1.05cm 0 0 0},clip]{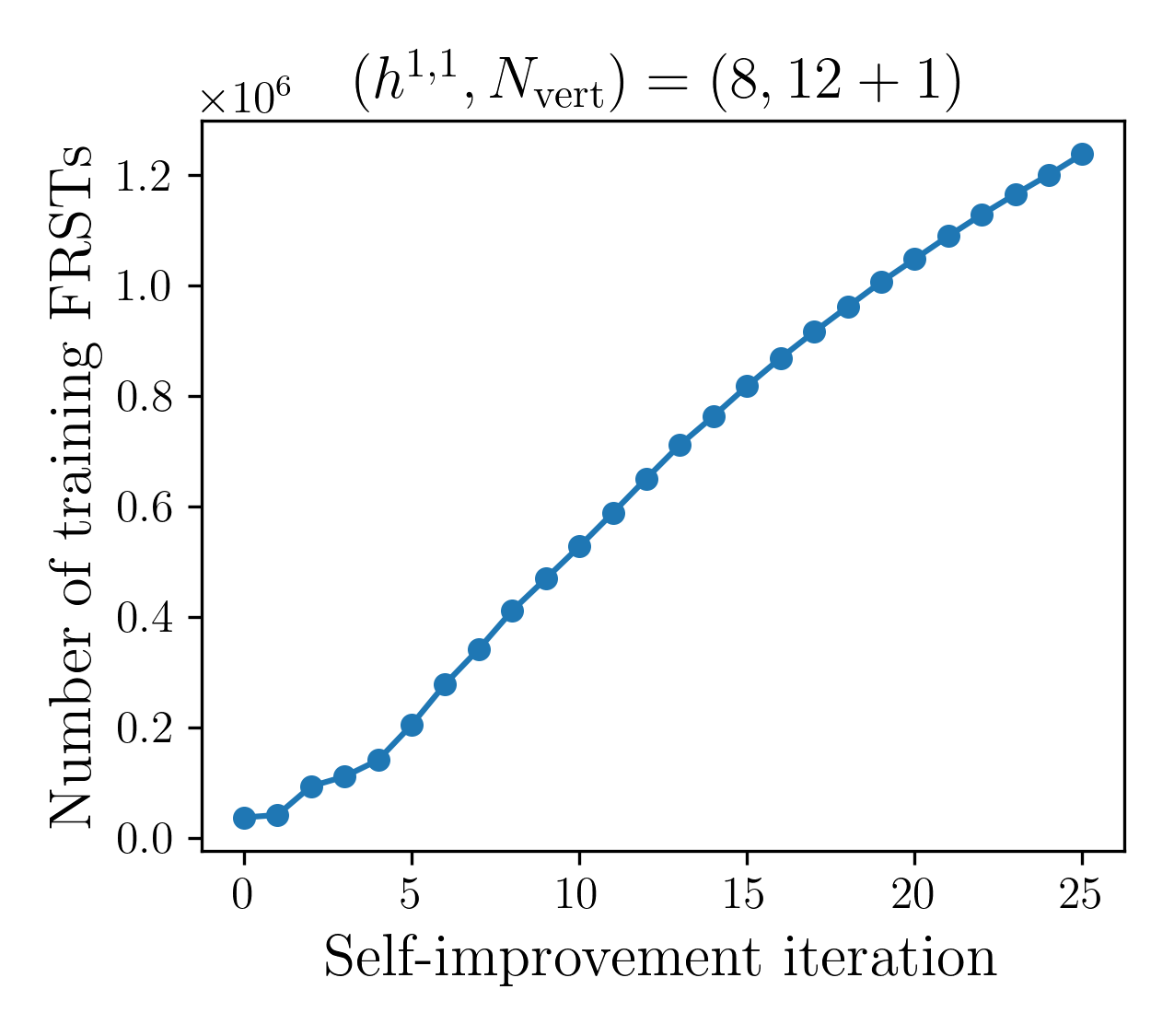} \\
  \includegraphics[height=4.65cm]{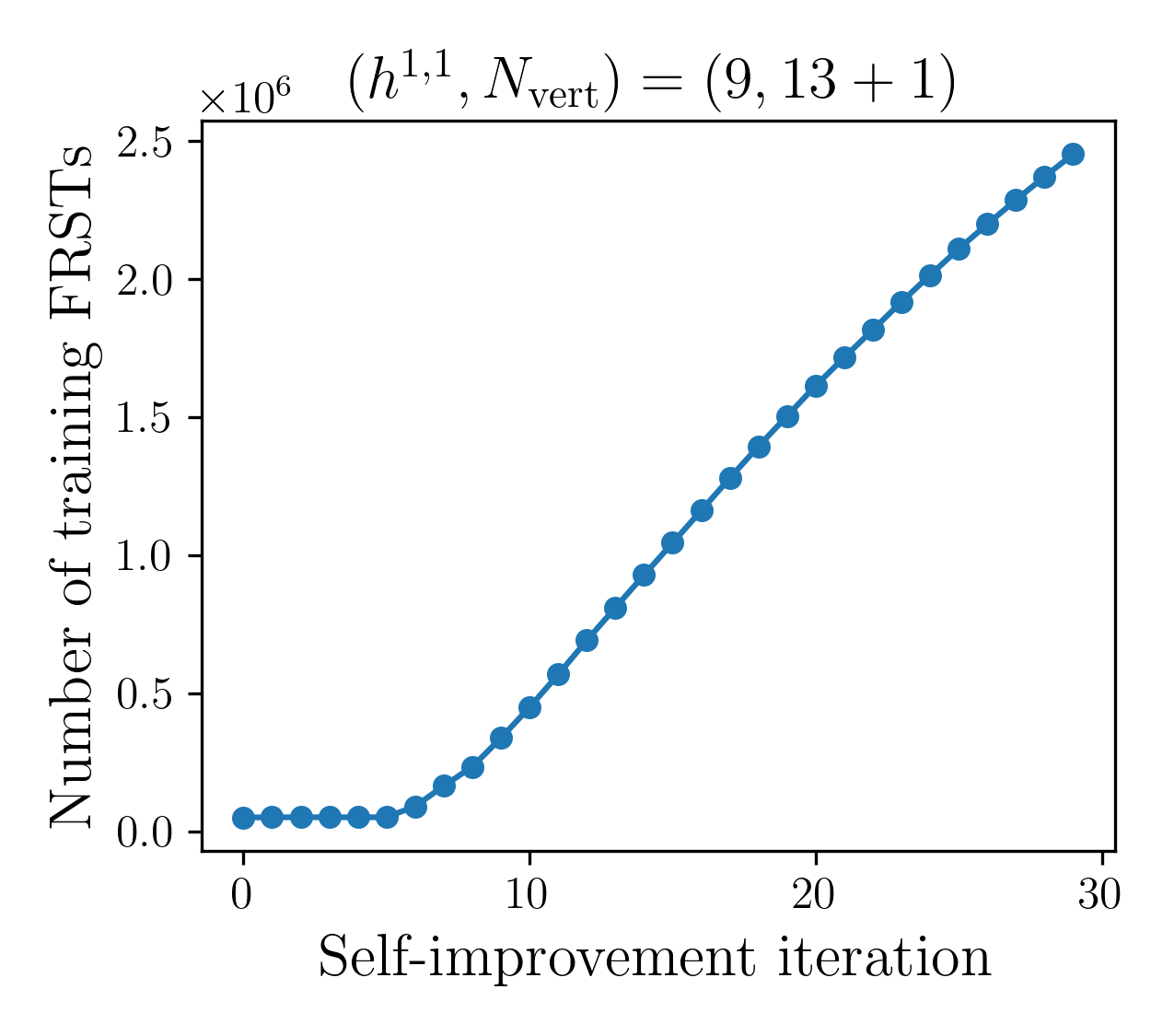}
  \includegraphics[height=4.65cm,trim={1.05cm 0 0 0},clip]{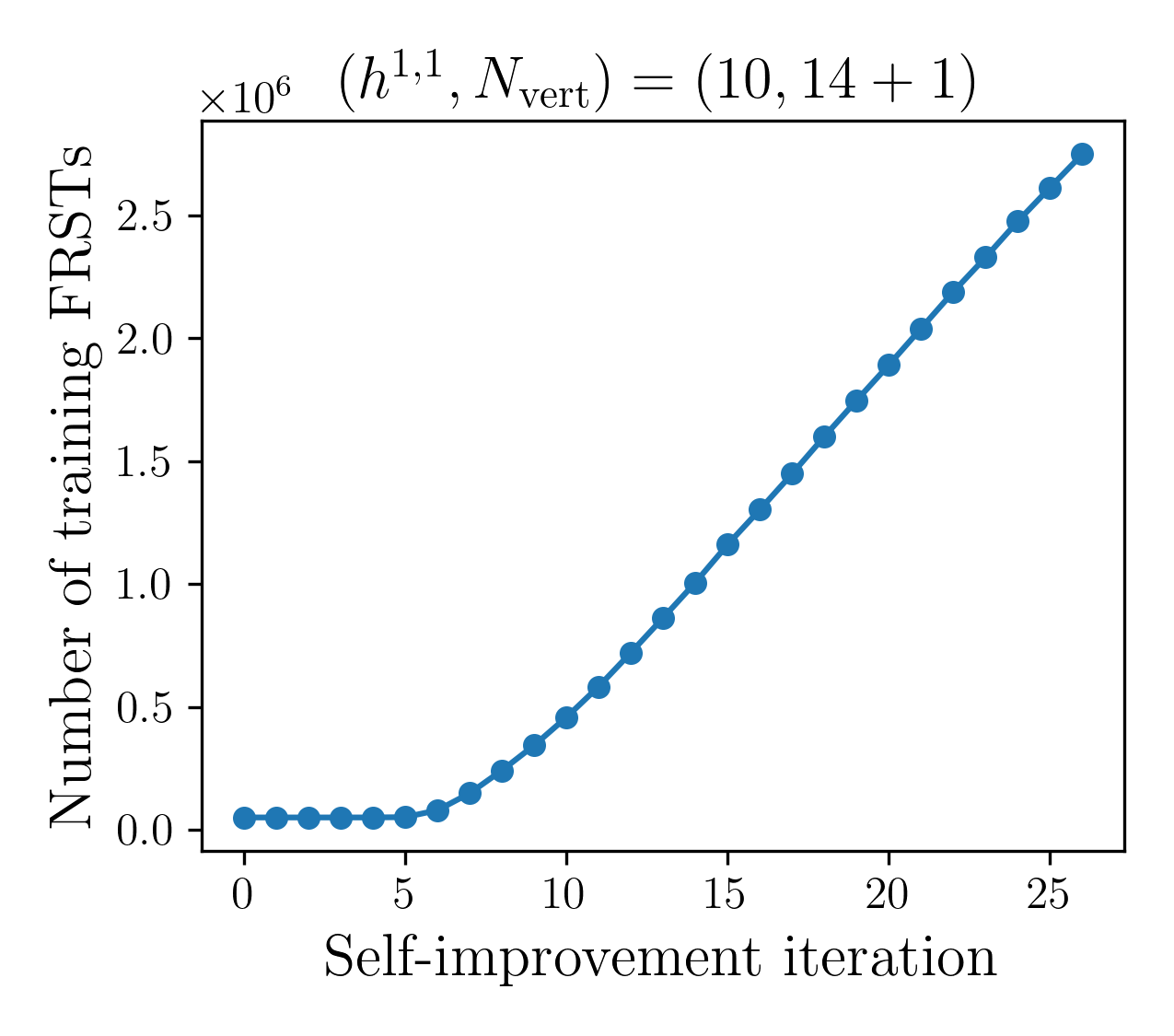}
  \caption{\textbf{Growth of the training set during self-improvement.} For each $(h^{1,1}, N_{\rm vert})$ configuration, we plot the total number of distinct FRSTs in the training set as a function of self-improvement iteration. For smaller configurations (top row), the training set expands fairly quickly during early iterations. In contrast, for larger configurations (bottom row), there is an initial stagnation phase followed by steady growth, reflecting the longer training time needed for the model to begin generating useful new FRSTs for more complex polytopes.}
  \label{fig:self_improvement_data}
\end{figure}

We evaluate self-improvement across $(h^{1,1},N_{\rm vert})$ configurations ranging from $(6,10+1)$ to $(10,14+1)$. As shown in figure~\ref{fig:self_improvement_data}, the size of the training set can grow substantially throughout the self-improvement process, by a factor of $6\times$ for $(6,10+1)$ and up to $55\times$ for $(10,14+1)$. The dynamics of this growth differ by configuration: for smaller polytopes, the training set expands rapidly from the first few iterations, whereas for more complex polytopes, it often stagnates initially before beginning to grow. This behavior is expected because larger polytopes have more complex FRST spaces, so the model needs a few rounds of learning before it can start generating useful new FRSTs.

\begin{figure}[htbp]
  \centering
  \includegraphics[height=4.85cm]{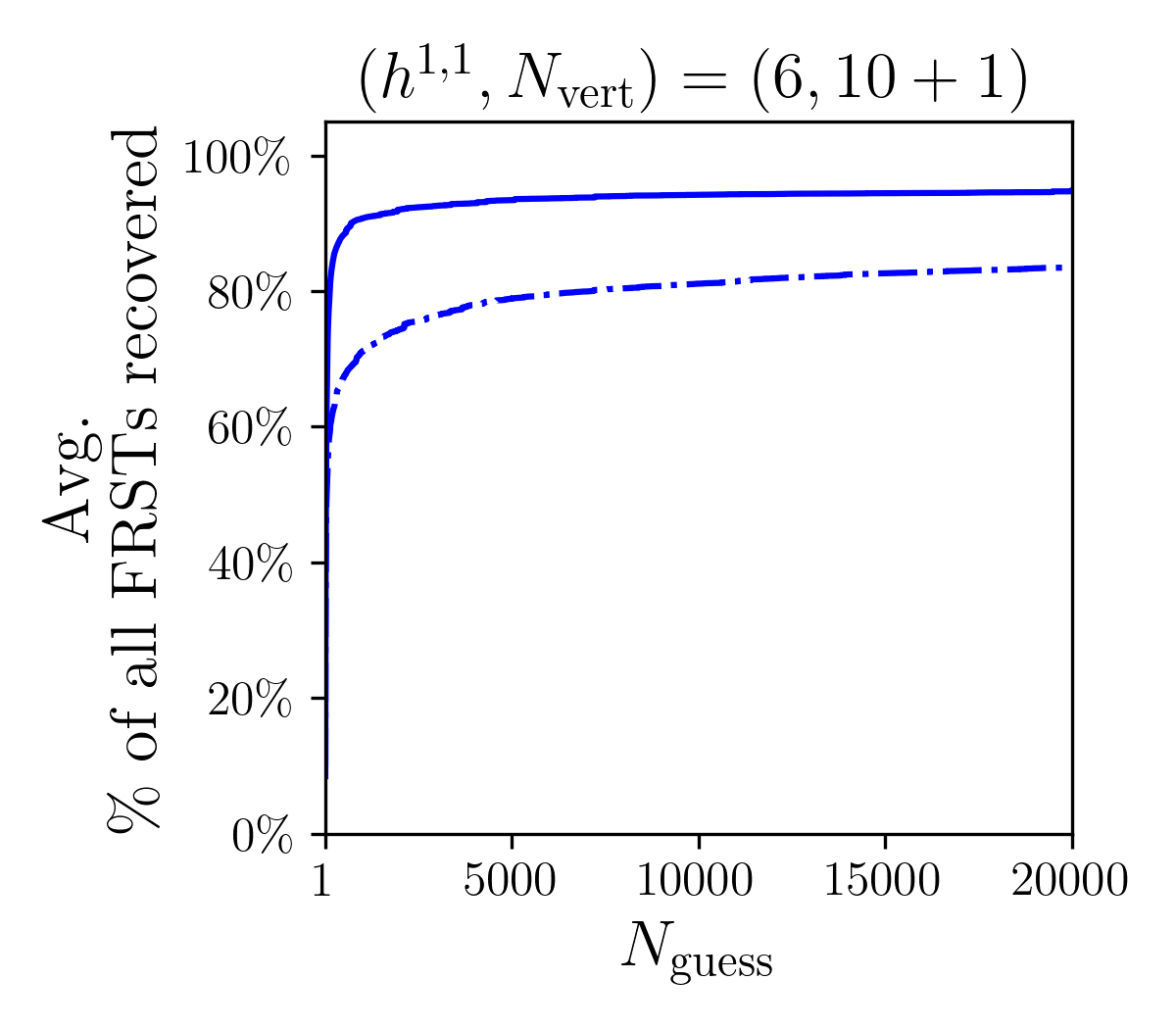}
  \includegraphics[height=4.85cm,trim={1.6cm 0 0 0},clip]{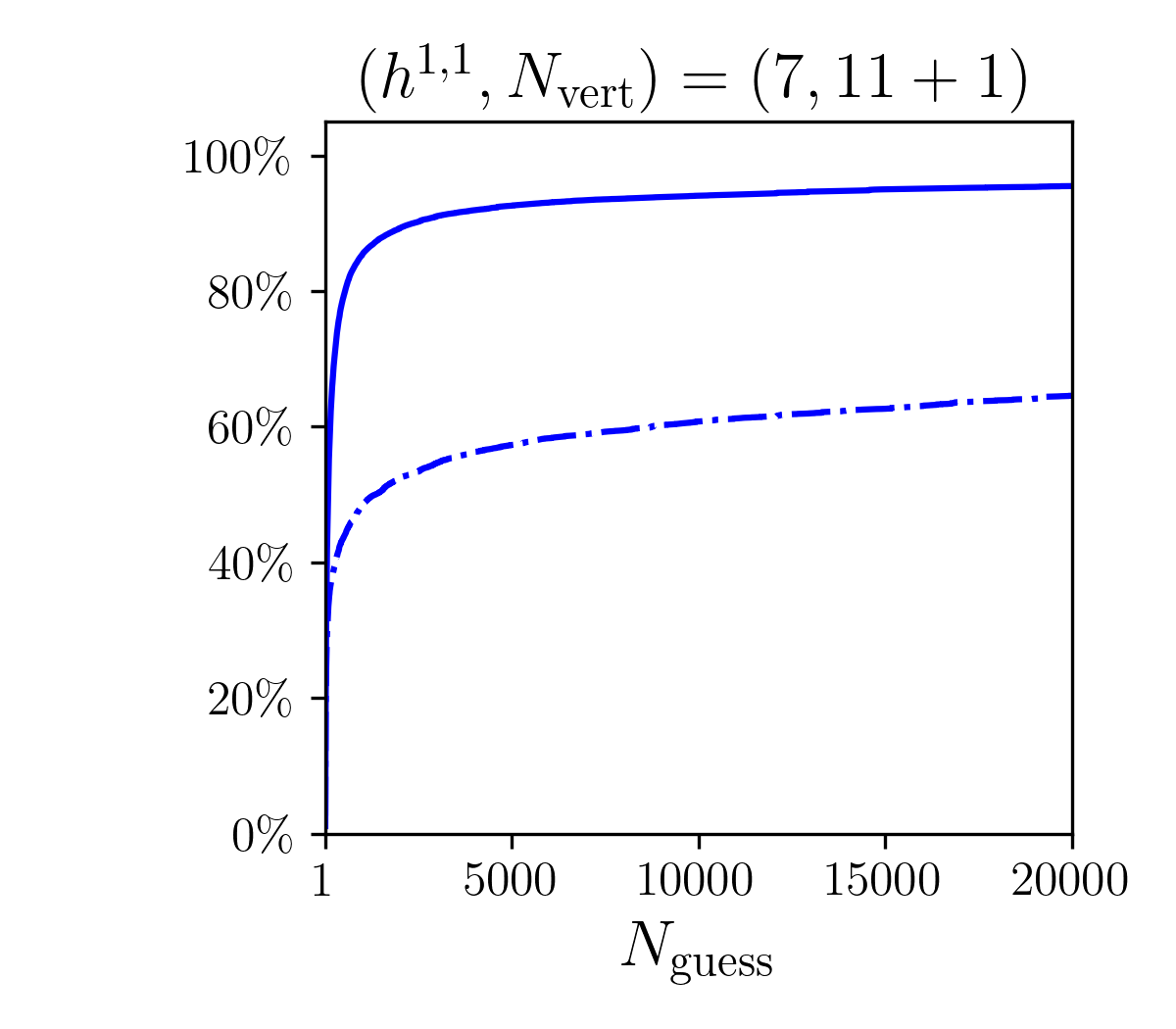}
  \includegraphics[height=4.85cm,trim={1.6cm 0 0 0},clip]{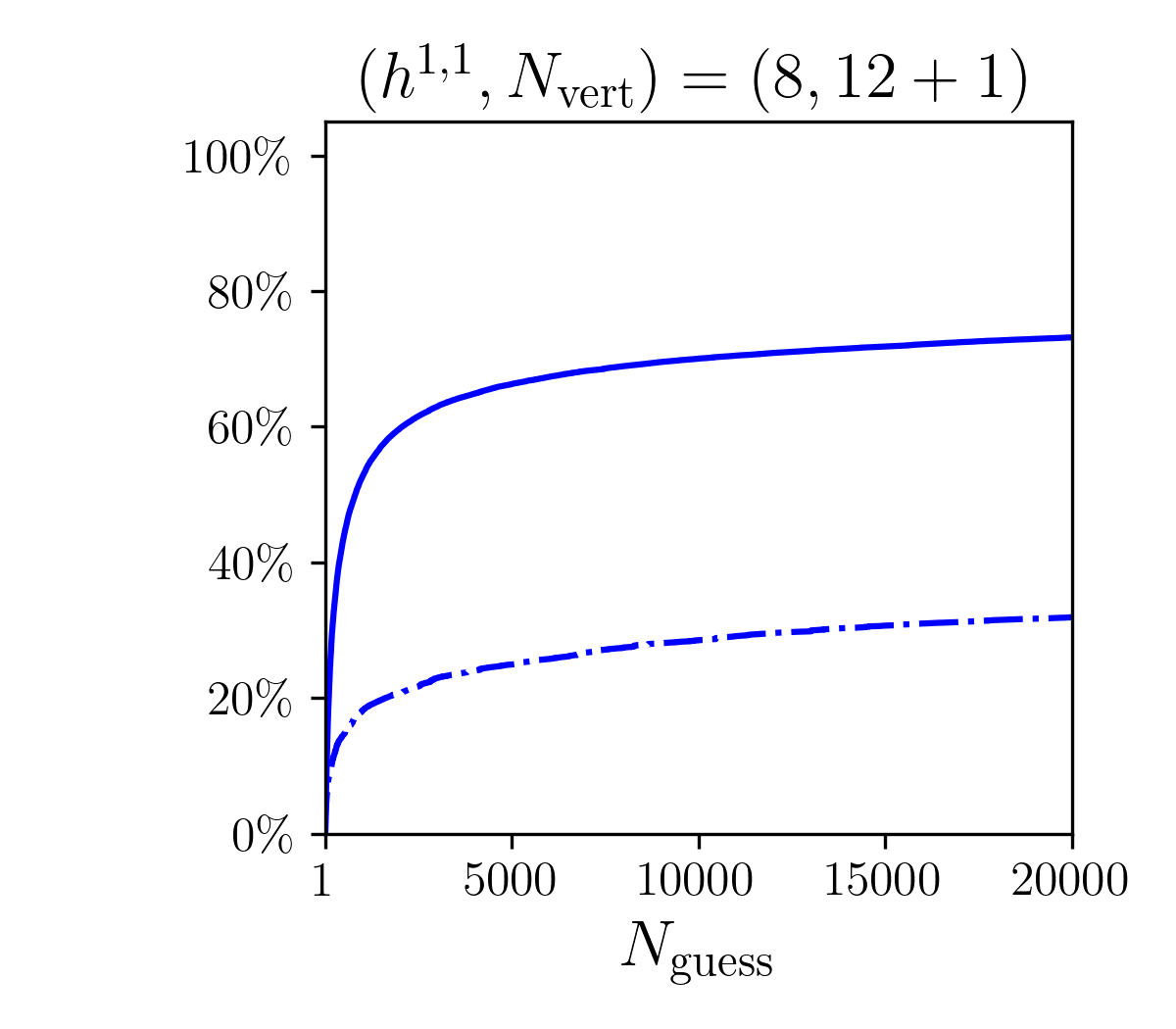} \\
  \includegraphics[height=4.85cm]{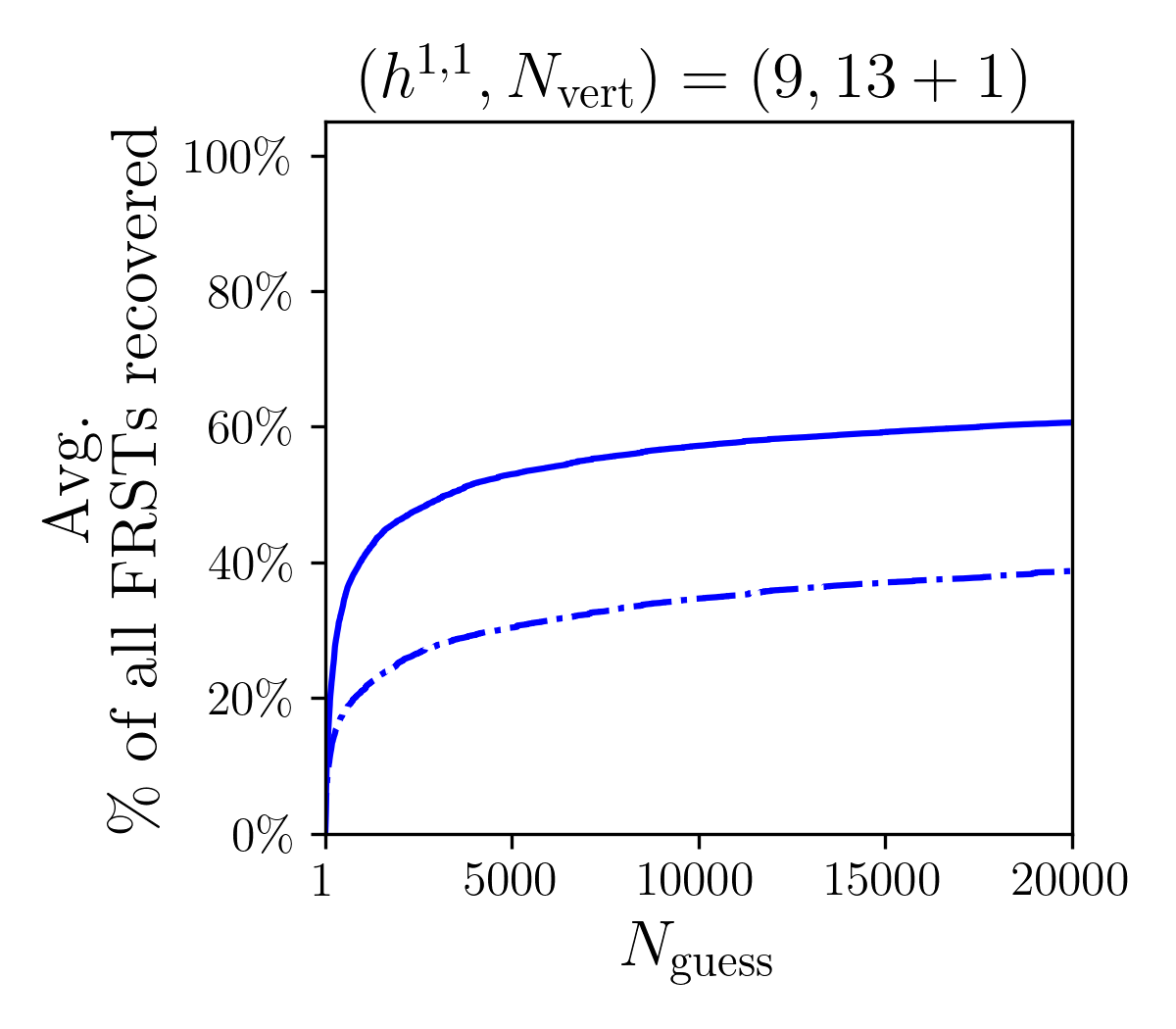}
  \includegraphics[height=4.85cm,trim={1.6cm 0 0 0},clip]{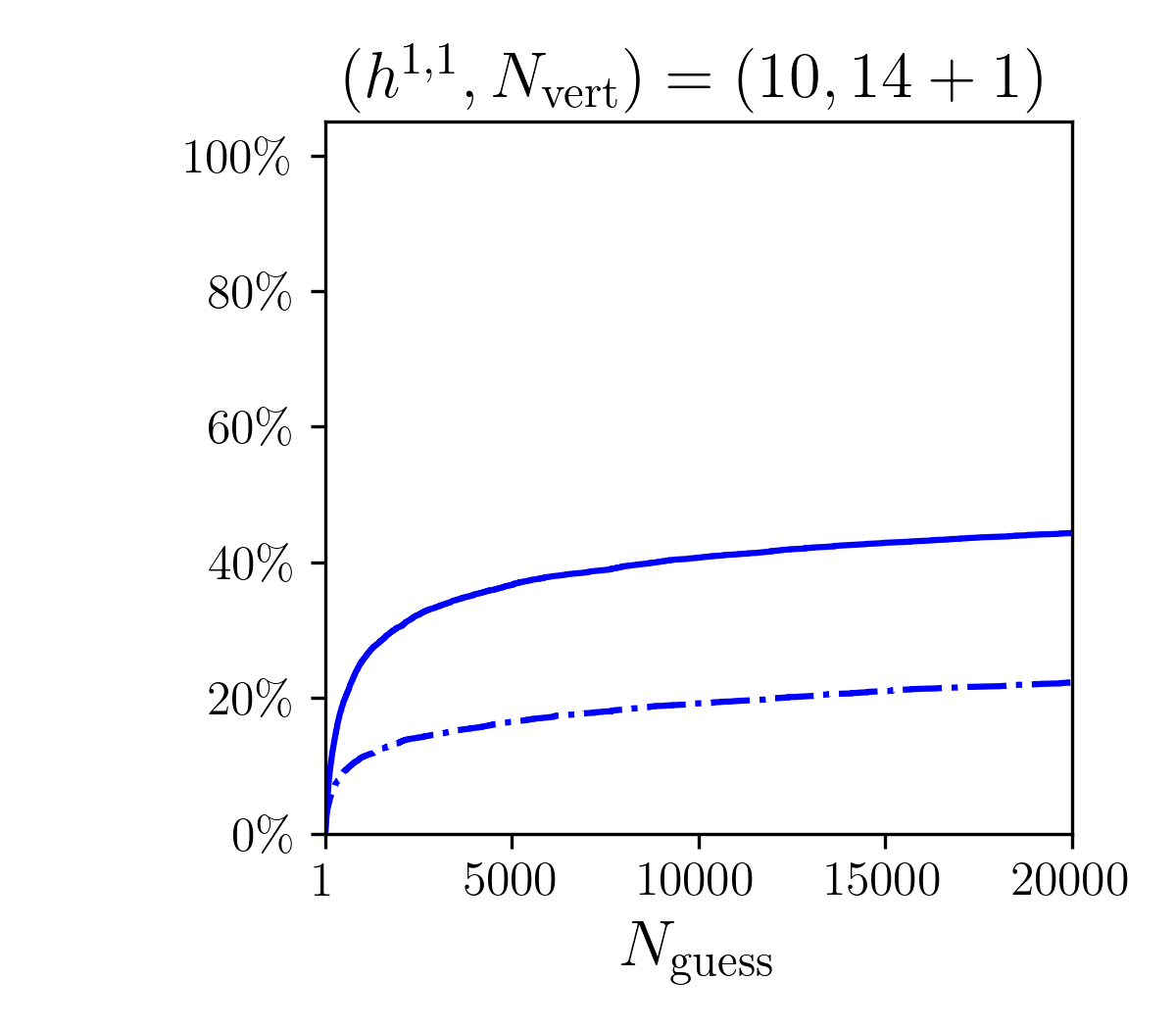}
  \includegraphics[width=0.65\textwidth]{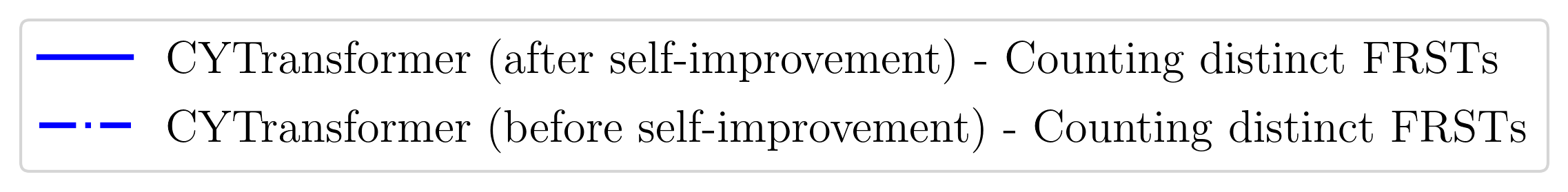}
  \caption{\textbf{Recovery performance before and after self-improvement.} Average \protect\hyperlink{met:frstreccurve}{FRST recovery curves} for CYTransformer after (solid) and before (dash-dotted) self-improvement, across $(h^{1,1}, N_{\rm vert})$ configurations. Self-improvement leads to consistent gains in recovery performance. However, the overall recovery after self-improvement remains lower than that of CYTransformer models trained on large, fully enumerated datasets (cf. figure~\ref{fig:compare_recovery}), suggesting that a minimal amount of initial training data may be necessary for self-improvement to reach full potential.}
  \label{fig:self_improvement_recovery_curves}
\end{figure}

To assess the effectiveness of this approach, we compare self-improved models to baselines trained (to convergence) solely on initial datasets. Figure~\ref{fig:self_improvement_recovery_curves} shows that self-improved CYTransformer models (solid) achieve between $14$\% and $229$\% higher average \hyperlink{met:frstreccurve}{recovery} on the test set than the baselines (dash-dotted). This demonstrates that self-improved models can greatly surpass those trained with straightforward supervised training in data-scarce regimes. However, they still fall short of the performance achieved by models trained on large, fully prepared datasets (cf. figure~\ref{fig:compare_recovery}) in the most complex cases. This gap is primarily due to the self-improved models failing on a fraction of test polytopes, suggesting that for some polytope geometries, there exists a minimal threshold of initial data required for self-improvement to be effective.

\begin{figure}
    \centering
    \includegraphics[width=0.75\linewidth]{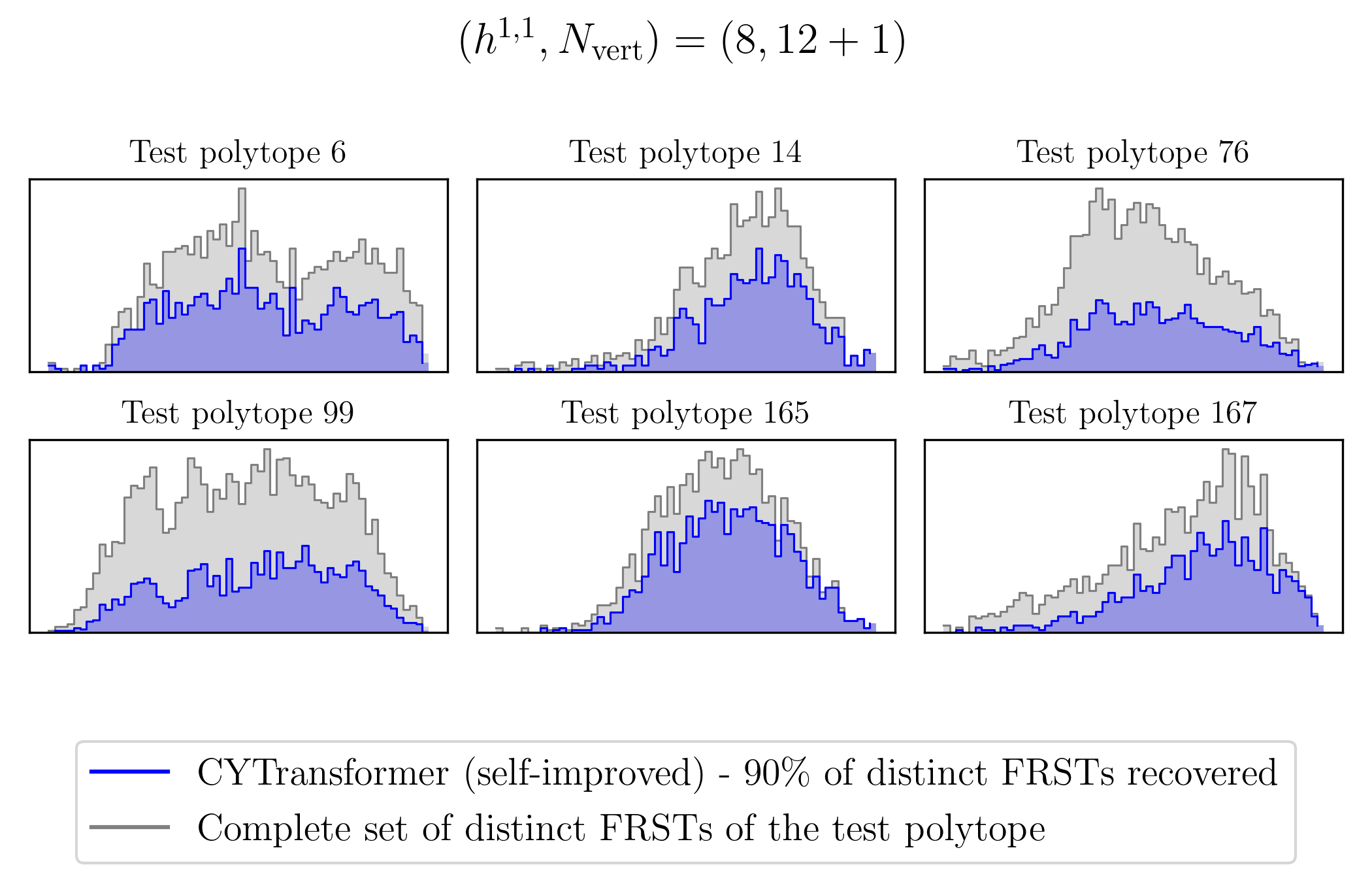}
    \caption{\textbf{Representativeness check of the self-improved CYTransformer.} \protect\hyperlink{met:hsfrstdist}{Height-space FRST distributions} for $6$ test polytopes with $(h^{1,1}, N_{\rm vert})=(8,12+1)$ configuration, selected from the subset where the self-improved CYTransformer achieves high recovery. The recovered distributions (blue) closely match the population distributions (gray) in shape and support, indicating that the self-improved model unbiasedly generate representative samples for these polytope geometries.}
    \label{fig:self_improvement_histogram}
\end{figure}

For those test polytopes on which the self-improved model does succeed, we present the \hyperlink{met:hsfrstdist}{height-space FRST distributions} in figure~\ref{fig:self_improvement_histogram} for the $(8,12+1)$ case. These distributions demonstrate that the self-improved model learns an unbiased representation of the FRST space for those polytope geometries. This result is nontrivial given that the initial training set is extremely sparse compared to the full FRST space. Understanding the minimal information required for successful self-improvement across different polytope geometries remains an interesting direction for future investigation.

As is sometimes done in the context of reinforcement learning (e.g., in~\cite{R1}), we have also attempted to train a new model from scratch on the complete dataset generated over an entire self-improvement run. We have observed that it can sometimes, though not systematically, yield slightly improved performance compared to the model trained via the self-improvement procedure. We hypothesize that this is due to optimization-related phenomena: while trained on a still-small dataset, the self-improved model may converge early to a less favorable local optimum than the model trained directly on the larger, final dataset. We defer a deeper investigation to future work.

\FloatBarrier
\section{\texttt{AICY}: AI-enabled living Calabi-Yau repositories}\label{sec:AICY}

In light of CYTransformer’s outstanding performance and demonstrated capacity for self-improvement, with the goal of extending to larger polytopes, enabling physics-informed targeted searches, and incorporating advanced learning-based strategies, we propose \href{https://aicy.physics.wisc.edu}{\textbf{\texttt{AICY}: AI-enabled living Calabi-Yau repositories}}. \texttt{AICY} is envisioned as a \emph{community-driven} software and data ecosystem for navigating and documenting the landscape of Calabi-Yau manifolds, serving string theorists, geometers, and the broader scientific community.

\begin{figure}
    \centering
    \includegraphics[width=1\linewidth]{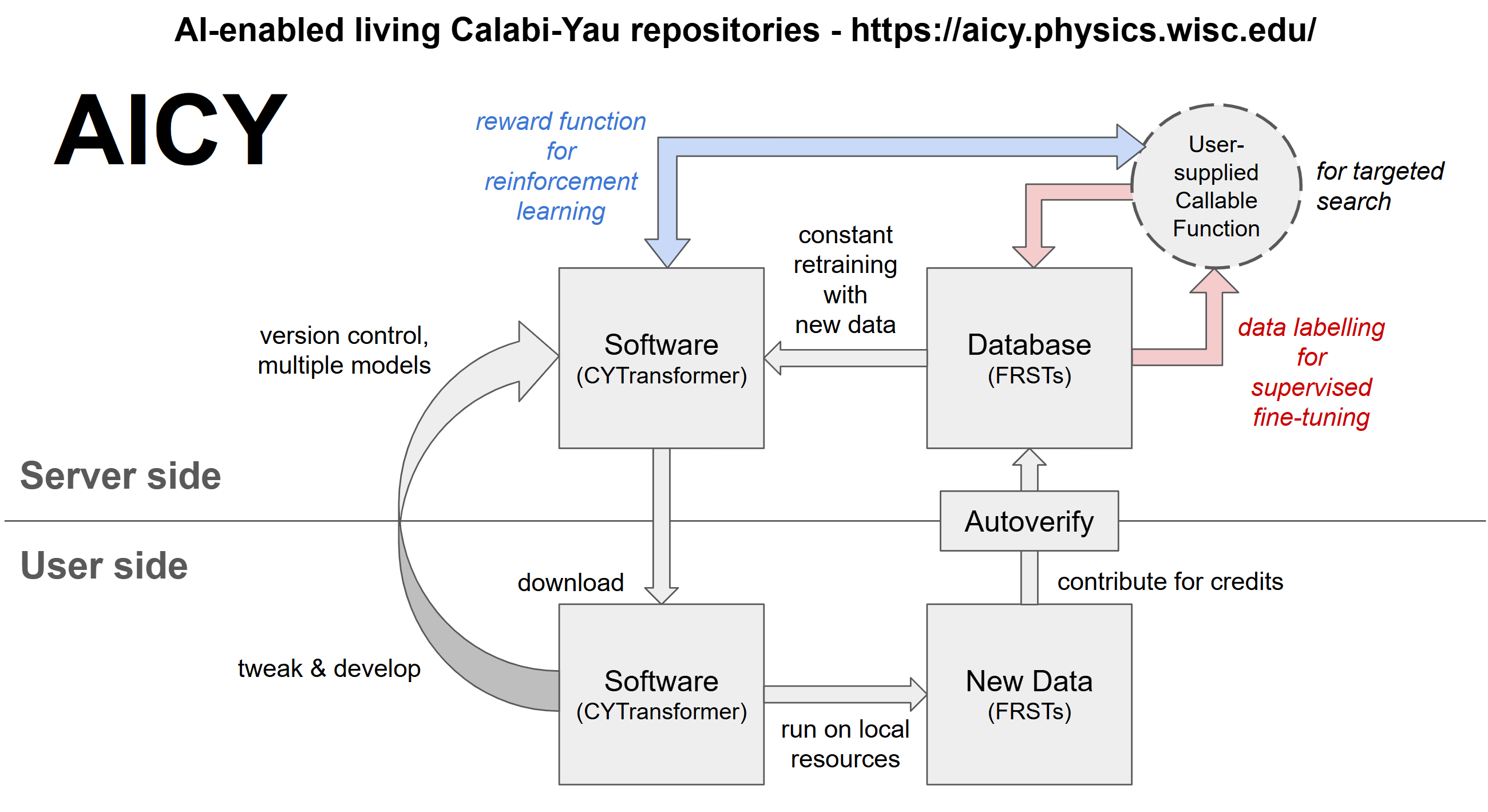}
    \caption{\textbf{Schematic overview of \texttt{AICY}.} The \texttt{AICY} platform integrates a software repository and a data repository into a self-improving, community-driven infrastructure for exploring and documenting the Calabi-Yau landscape. Users can download models such as CYTransformer, run or adapt them locally, and contribute generated data or improved models back to the server. On the server side, learning-based models are continually retrained as the database expands. A key feature is support for targeted search: users may supply a callable function that computes desired Calabi-Yau properties from FRSTs. This function can be used either to label data for supervised fine-tuning or as a reward signal in reinforcement learning. Within this flexible framework, models like CYTransformer learn to favor FRSTs that yield the desired physics or geometry, without requiring changes to the model architecture.}
    \label{fig:aicy}
\end{figure}

As illustrated in figure~\ref{fig:aicy}, \texttt{AICY} consists of two closely interacting components: the \textbf{software repository} and the \textbf{data repository}. These two components \emph{update and reinforce one another continuously}, both on the server side and through user contributions. The software repository includes learning-based tools such as the CYTransformer presented in this work. Users can download pretrained models, adapt or extend them, and contribute enhanced versions back to \texttt{AICY}. These tools can also be utilized locally for practical applications, such as generating FRSTs or Calabi-Yau manifolds for specific research tasks. The data repository is open to community submissions via an automated interface. For example, users can submit candidate triangulations generated using \texttt{AICY}'s tools, which are automatically validated upon upload. All contributors will be properly credited for any software or data they provide.

Through this collaborative pipeline, both repositories remain ``living'', improving continuously with community input. On the server side, we support periodic retraining of the learning-based tools on the growing database, enabling models like CYTransformer to self-improve. A key feature \texttt{AICY} will support is \emph{targeted search}, which we discuss in section~\ref{sec:dis}.

Powered by machine learning and guided by fundamental physics, \texttt{AICY} evolves with the community, for the community. The \texttt{AICY} infrastructure will be launched soon, and we encourage interested readers to visit \href{https://aicy.physics.wisc.edu}{\texttt{https://aicy.physics.wisc.edu}} to sign up for early updates.

\section{Discussion}
\label{sec:dis}

In this work, we applied a transformer architecture to generate triangulations of four-dimensional polytopes, where the triangulations must satisfy nontrivial conditions required for constructing Calabi-Yau manifolds. Our model, CYTransformer, demonstrates promising performance across the range of polytope sizes considered, particularly in its ability to efficiently and unbiasedly sample the FRST space. Nevertheless, several avenues for improvement remain. We outline possible future directions below.

\textbf{Priming.} A key strength of CYTransformer is its ability to sample fairly across the FRST space. This unbiasedness underpins its potential for deriving landscape statistics and motivates its integration with other tools, such as the CYTransformer-seeded fast sampler proposed in this work. We expect this property to become even more beneficial as we move to larger polytopes with increasingly complex FRST spaces. However, practical constraints may limit this advantage. For instance, for polytopes with $N_{\rm vert}\geq13+1$, it becomes effectively infeasible to enumerate all FRSTs of every training polytope. In such cases, we turn to approximate methods like the fast sampler to generate training data. Since these methods are inherently biased, CYTransformer, when trained on such biased data, may also inherit this bias at inference time. While we have shown that no significant bias arises for $N_{\rm vert}=13+1$ and $14+1$ (see figure~\ref{fig:compare_rep}), this highlights a concern for scaling the method to even larger polytopes.

To retain CYTransformer’s strength even when trained on biased data, the model must learn the underlying triangulation rules rather than merely memorizing patterns. We propose using a technique known as priming, originally introduced to support sequence-length generalization in transformers. In our context, priming involves retraining a CYTransformer, initially trained on polytopes of a given size, using a small number of out-of-distribution samples from larger polytopes, gently encouraging generalization. If successful, this would indicate that CYTransformer learns transferrable knowledge, namely, the structural principles of FRSTs, rather than overfitting to specific examples. This represents a major advantage over non-learning methods like the fast sampler, which lack any mechanism for leveraging transferability.

When combined with the self-improvement strategy (see~\ref{subsec:selfim_results}), priming offers a pathway to iterative bootstrapping toward arbitrarily large configurations. Starting from a CYTransformer trained on polytopes with a modest $N^*_{\rm vert}$, one can introduce a priming set with $N^*_{\rm vert}+1$ for retraining. The lack of abundant training data at this size is not a problem, as self-improvement can be applied using CYTransformer-generated FRSTs. This cycle can then repeat for progressively larger $N_{\rm vert}$. \texttt{AICY} is designed to continuously run this iterative process on the server.

\hypertarget{dis:tarsea}{\textbf{Targeted search.}} A key future application of CYTransformer is enabling targeted searches for Calabi-Yau manifolds with desired properties. We envision a general-purpose pipeline where the user simply supplies a callable function that computes the properties of interest, such as volume and intersection numbers, directly from an FRST. This user-supplied function can then be seamlessly integrated into a retraining or fine-tuning process. This flexible design supports optimization for physics-driven use cases without requiring modifications to CYTransformer's underlying architecture or training a new model from scratch.

Two complementary strategies can be implemented for such a targeted search. The first is supervised fine-tuning, where CYTransformer is further trained on a curated subset of FRSTs that yield desirable properties. This subset is labeled by applying the user-supplied function to \texttt{AICY}'s FRST database, and retraining on this labeled dataset biases CYTransformer towards sampling the desired FRSTs. The second strategy is reinforcement learning~\cite{Cole:2019enn,Krippendorf:2021uxu,Cole:2021nnt,Loges:2021hvn,Abel:2021ddu,MacFadden:2024him,Berglund:2024reu}, which removes the need for a pre-labeled dataset. In this framework, the pre-trained CYTransformer serves as an efficient and unbiased navigator of the FRST landscape. It is then further trained via policy gradient methods, using a real-time reward signal built from the user-supplied function and target property values. This guides the model to explore and discover FRSTs that optimize the specified properties.

\textbf{Specialized tokenization.}
We adopted a general-purpose tokenization scheme capable of representing arbitrary triangulations, enabling broad applicability. However, more specialized tokenizations may simplify specific tasks. For example, using the height vector representation guarantees regularity by construction, with tokens corresponding directly to height values. Another possibility would be to design token sequences that encode specific structural features, such as NTFE constraints, by explicitly including two-face configurations in the input. While specialized schemes can improve performance on targeted tasks, we prioritized generality in this work to demonstrate the versatility of CYTransformer.

\textbf{Code and hyperparameters optimization.} 
While our models already achieve a satisfying level of performance, this work is chiefly exploratory, and our code, hyperparameters selection, and training procedures could be further optimized. In particular, larger models are expected to yield higher performance, while the implementation of advanced machine learning techniques, such as key-value caching, could considerably speed up training.

\textbf{Inference algorithm.} While CYTransformer exhibits high sampling efficiency in terms of $N_{\rm guess}$, each inference step remains computationally expensive compared to lightweight methods like the fast sampler. This cost could be reduced by making inference steps stateful rather than independent. For example, the model could track previously generated candidate triangulations and dynamically adjust token sampling probabilities to avoid duplications and promote diversity. Such an approach could significantly enhance inference efficiency without retraining.

\textbf{CYTransformer as a database.} Given the vast number of FRSTs across all four-dimensional reflexive polytopes, storing them in a static database is infeasible. Instead, a well-trained CYTransformer that samples fairly from the FRST space offers a dynamic alternative: rather than retrieving from a finite database, one queries the model to generate FRSTs on demand.


\acknowledgments
We thank Alessandro Mininno and Jakob Moritz for valuable discussions. We are grateful to Moritz Münchmeyer for providing access to his research group's computing cluster, which was essential for the initial exploratory stage of this work. The work of G.S. and J.H.T.Y. is supported by the U.S. Department of Energy, Office of Science, Office of High Energy Physics under Award Numbers DE-SC-0023719 and DE-SC-0017647. F.C. was affiliated with Meta FAIR during the period when the main part of this research was conducted.

\bibliographystyle{JHEP}
\bibliography{biblio.bib}

@BOOK{Fulton1993-hj,
  title     = "Introduction to toric varieties. ({AM-131)}, volume 131",
  author    = "Fulton, William",
  publisher = "Princeton University Press",
  series    = "Annals of Mathematics Studies",
  month     =  jul,
  year      =  1993,
  address   = "Princeton, NJ",
  language  = "en"
}

@BOOK{Tadao2012-bj,
  title     = "Convex bodies and algebraic geometry",
  author    = "Tadao, Oda",
  publisher = "Springer",
  series    = "Ergebnisse der Mathematik und ihrer Grenzgebiete. 3. Folge / A
               Series of Modern Surveys in Mathematics",
  month     =  feb,
  year      =  2012,
  address   = "Berlin, Germany",
  language  = "en"
}

@BOOK{Cox2024-vp,
  title     = "Toric varieties",
  author    = "Cox, David A and Little, John B and Schenck, Henry K",
  publisher = "American Mathematical Society",
  month     =  jun,
  year      =  2024,
  address   = "Providence, RI",
  language  = "en"
}

@inproceedings{Cole:2021nnt,
    author = "Cole, Alex and Krippendorf, Sven and Schachner, Andreas and Shiu, Gary",
    title = "{Probing the Structure of String Theory Vacua with Genetic Algorithms and Reinforcement Learning}",
    booktitle = "{35th Conference on Neural Information Processing Systems}",
    eprint = "2111.11466",
    archivePrefix = "arXiv",
    primaryClass = "hep-th",
    month = "11",
    year = "2021"
}

@article{Cole:2019enn,
    author = "Cole, Alex and Schachner, Andreas and Shiu, Gary",
    title = "{Searching the Landscape of Flux Vacua with Genetic Algorithms}",
    eprint = "1907.10072",
    archivePrefix = "arXiv",
    primaryClass = "hep-th",
    reportNumber = "MAD-TH-19-05",
    doi = "10.1007/JHEP11(2019)045",
    journal = "JHEP",
    volume = "11",
    pages = "045",
    year = "2019"
}

@article{Loges:2021hvn,
    author = "Loges, Gregory J. and Shiu, Gary",
    title = "{Breeding Realistic D-Brane Models}",
    eprint = "2112.08391",
    archivePrefix = "arXiv",
    primaryClass = "hep-th",
    doi = "10.1002/prop.202200038",
    journal = "Fortsch. Phys.",
    volume = "70",
    number = "5",
    pages = "2200038",
    year = "2022"
}

@article{Krippendorf:2021uxu,
    author = "Krippendorf, Sven and Kroepsch, Rene and Syvaeri, Marc",
    title = "{Revealing systematics in phenomenologically viable flux vacua with reinforcement learning}",
    eprint = "2107.04039",
    archivePrefix = "arXiv",
    primaryClass = "hep-th",
    reportNumber = "LMU-ASC 20/21, MPP-2021-108",
    month = "7",
    year = "2021"
}

@inproceedings{Abel:2021ddu,
    author = "Abel, Steven and Constantin, Andrei and Harvey, Thomas R. and Lukas, Andre",
    title = "{String Model Building, Reinforcement Learning and Genetic Algorithms}",
    booktitle = "{Nankai Symposium on Mathematical Dialogues}: {In celebration of S.S.Chern's 110th anniversary}",
    eprint = "2111.07333",
    archivePrefix = "arXiv",
    primaryClass = "hep-th",
    month = "11",
    year = "2021"
}

@article{WALL1966,
    author = {Wall, C.T.C.},
    title = {Classification Problems in Differential Topology. V. On Certain 6-Manifolds},
    journal = {Inventiones Mathematicae},
    volume = {1},
    number = {},
    pages = {355-374},
    year = {1966},
    doi = {10.1007/BF01389738},
    url = {http://eudml.org/doc/141839},
    publisher = {Springer}
}

@article{Vafa:2005ui,
    author = "Vafa, Cumrun",
    title = "{The String landscape and the swampland}",
    eprint = "hep-th/0509212",
    archivePrefix = "arXiv",
    reportNumber = "HUTP-05-A043",
    month = "9",
    year = "2005"
}

@article{Acharya:2006zw,
    author = "Acharya, Bobby Samir and Douglas, Michael R",
    title = "{A Finite landscape?}",
    eprint = "hep-th/0606212",
    archivePrefix = "arXiv",
    reportNumber = "IC-2006-42",
    month = "6",
    year = "2006"
}

@article{R1,
author = {DeepSeek-AI},
year = {2025},
title = {DeepSeek-R1: Incentivizing Reasoning Capability in LLMs via Reinforcement Learning},
journal = {arXiv.2501.12948}
}

@article{Douglas:2003um,
    author = "Douglas, Michael R.",
    title = "{The Statistics of string / M theory vacua}",
    eprint = "hep-th/0303194",
    archivePrefix = "arXiv",
    reportNumber = "RUNHETC-2003-09",
    doi = "10.1088/1126-6708/2003/05/046",
    journal = "JHEP",
    volume = "05",
    pages = "046",
    year = "2003"
}

@article{Douglas:2006xy,
    author = "Douglas, Michael R. and Taylor, Washington",
    title = "{The Landscape of intersecting brane models}",
    eprint = "hep-th/0606109",
    archivePrefix = "arXiv",
    reportNumber = "MIT-CTP-3748, SU-ITP-06-15, RU-NHETC-06-04",
    doi = "10.1088/1126-6708/2007/01/031",
    journal = "JHEP",
    volume = "01",
    pages = "031",
    year = "2007"
}

@article{Loges:2022mao,
    author = "Loges, Gregory J. and Shiu, Gary",
    title = "{134 billion intersecting brane models}",
    eprint = "2206.03506",
    archivePrefix = "arXiv",
    primaryClass = "hep-th",
    doi = "10.1007/JHEP12(2022)097",
    journal = "JHEP",
    volume = "12",
    pages = "097",
    year = "2022"
}

@article{Candelas:1985en,
    author = "Candelas, P. and Horowitz, Gary T. and Strominger, Andrew and Witten, Edward",
    title = "{Vacuum configurations for superstrings}",
    reportNumber = "NSF-ITP-84-170",
    doi = "10.1016/0550-3213(85)90602-9",
    journal = "Nucl. Phys. B",
    volume = "258",
    pages = "46--74",
    year = "1985"
}

@article{Marchesano:2024gul,
    author = "Marchesano, Fernando and Shiu, Gary and Weigand, Timo",
    title = "{The Standard Model from String Theory: What Have We Learned?}",
    eprint = "2401.01939",
    archivePrefix = "arXiv",
    primaryClass = "hep-th",
    reportNumber = "IFT-UAM/CSIC-24-01, ZMP-HH/24-01",
    doi = "10.1146/annurev-nucl-102622-012235",
    journal = "Ann. Rev. Nucl. Part. Sci.",
    volume = "74",
    pages = "113--140",
    year = "2024"
}

@online{KS-Database,
author = "Kreuzer, Maximilian and Skarke, Harald",
  title = {{Calabi-Yau} Data},
 howpublished = {\url{http://hep.itp.tuwien.ac.at/~kreuzer/CY/}}
}

@online{Altman-Database,
author = "Altman, Ross",
  title = {{Toric Calabi-Yau} Database},
 howpublished = {\url{http://www.rossealtman.com/toriccy/}}
}

@article{Llama3,
author = {Dubey, Abhimanyu and Jauhri, Abhinav and Pandey, Abhinav and Kadian, Abhishek and Al-Dahle, Ahmad and Letman, Aiesha and Mathur, Akhil and Schelten, Alan and Yang, Amy and Fan, Angela and Goyal, Anirudh and Hartshorn, Anthony and Yang, Aobo and Mitra, Archi and Sravankumar, Archie and Korenev, Artem and Hinsvark, Arthur and Rao, Arun and Zhang, Aston and Zhao, Zhiwei},
year = {2024},
journal={ArXiv},
title = {The Llama 3 Herd of Models},
volume={arXiv.2407.21783},
}

@article{Smatrix,
author = {Dersy, Aurélien and Schwartz, Matthew and Zhiboedov, Alexander},
year = {2024},
month = {05},
pages = {},
title = {Reconstructing S-matrix Phases with Machine Learning},
volume = {2024},
journal = {Journal of High Energy Physics},
doi = {10.1007/JHEP05(2024)200}
}

@article{Funsearch,
author = {Romera-Paredes, Bernardino and Barekatain, Mohammadamin and Novikov, Alexander and Balog, Matej and Kumar, M. and Dupont, Emilien and Ruiz, Francisco and Ellenberg, Jordan and Wang, Pengming and Fawzi, Omar and Kohli, Pushmeet and Fawzi, Alhussein},
year = {2023},
month = {12},
pages = {},
title = {Mathematical discoveries from program search with large language models},
volume = {625},
journal = {Nature},
doi = {10.1038/s41586-023-06924-6}
}

@article{Cai:2024znx,
    author = "Cai, Tianji and Merz, Garrett W. and Charton, Fran{\c{c}}ois and Nolte, Niklas and Wilhelm, Matthias and Cranmer, Kyle and Dixon, Lance J.",
    title = "{Transforming the bootstrap: using transformers to compute scattering amplitudes in planar $\mathcal{N} = 4$ super Yang{\textendash}Mills theory}",
    eprint = "2405.06107",
    archivePrefix = "arXiv",
    primaryClass = "cs.LG",
    reportNumber = "SLAC-PUB-17774",
    doi = "10.1088/2632-2153/ad743e",
    journal = "Mach. Learn. Sci. Tech.",
    volume = "5",
    number = "3",
    pages = "035073",
    year = "2024"
}

@article{Alfarano,
author = {Alfarano, Alberto and Charton, Francois and Hayat, Amaury},
year = {2024},
title = {Global Lyapunov functions: a long-standing open problem in mathematics, with symbolic transformers},
  journal= {10.48550/arXiv.2410.08304}
}

@article{Inductionheads,
  title={In-context learning and induction heads},
  author={Olsson, Catherine and Elhage, Nelson and Nanda, Neel and Joseph, Nicholas and DasSarma, Nova and Henighan, Tom and Mann, Ben and Askell, Amanda and Bai, Yuntao and Chen, Anna and others},
  journal={arXiv preprint arXiv:2209.11895},
  year={2022}
}

@misc{encoderdecodersource,
  author       = { Godoy, Daniel},
  title        = {Deep Learning Visuals},
  year         = {2025},
  howpublished = {\url{https://github.com/dvgodoy/dl-visuals/}},
}

@article{FunctionVI,
  title={Function Vectors in Large Language Models},
  author={Eric Todd and Millicent Li and Arnab Sen Sharma and Aaron Mueller and Byron C. Wallace and David Bau},
  journal={ArXiv},
  year={2023},
  volume={abs/2310.15213},
  url={https://api.semanticscholar.org/CorpusID:264439657}
}

@inproceedings{Iterationhead,
author = {Cabannes, Vivien and Arnal, Charles and Bouaziz, Wassim and Yang, Alice and Charton, Francois and Kempe, Julia},
title = {Iteration head: a mechanistic study of chain-of-thought},
year = {2025},
isbn = {9798331314385},
publisher = {Curran Associates Inc.},
address = {Red Hook, NY, USA},
booktitle = {Proceedings of the 38th International Conference on Neural Information Processing Systems},
articleno = {3463},
numpages = {22},
location = {Vancouver, BC, Canada},
series = {NIPS '24}
}

@article{charton2024patternboost,
  title={PatternBoost: Constructions in mathematics with a little help from AI},
  author={Charton, Fran{\c{c}}ois and Ellenberg, Jordan S and Wagner, Adam Zsolt and Williamson, Geordie},
  journal={arXiv preprint arXiv:2411.00566},
  year={2024}
}

@article{Geneva2020TransformersFM,
  title={Transformers for Modeling Physical Systems},
  author={Nick De Geneva and Nicholas Zabaras},
  journal={Neural networks : the official journal of the International Neural Network Society},
  year={2020},
  volume={146},
  pages={
          272-289
        },
  url={https://api.semanticscholar.org/CorpusID:222208767}
}

@article{Janny2023EagleLL,
  title={Eagle: Large-Scale Learning of Turbulent Fluid Dynamics with Mesh Transformers},
  author={Steeven Janny and Aur'elien B'eneteau and Nicolas Thome and Madiha Nadri Wolf and Julie Digne and Christian Wolf},
  journal={ArXiv},
  year={2023},
  volume={abs/2302.10803},
  url={https://api.semanticscholar.org/CorpusID:257050214}
}

@article{jumper2021highly,
  title={Highly accurate protein structure prediction with AlphaFold},
  author={Jumper, John and Evans, Richard and Pritzel, Alexander and Green, Tim and Figurnov, Michael and Ronneberger, Olaf and Tunyasuvunakool, Kathryn and Bates, Russ and {\v{Z}}{\'\i}dek, Augustin and Potapenko, Anna and others},
  journal={nature},
  volume={596},
  number={7873},
  pages={583--589},
  year={2021},
  publisher={Nature Publishing Group}
}

@misc{arnal2025asymmetricreinforceoffpolicyreinforcement,
      title={Asymmetric REINFORCE for off-Policy Reinforcement Learning: Balancing positive and negative rewards}, 
      author={Charles Arnal and Gaëtan Narozniak and Vivien Cabannes and Yunhao Tang and Julia Kempe and Remi Munos},
      year={2025},
      eprint={2506.20520},
      archivePrefix={arXiv},
      primaryClass={cs.LG},
      url={https://arxiv.org/abs/2506.20520}, 
}

@misc{faircodegenteam2025cwmopenweightsllmresearch,
      title={CWM: An Open-Weights LLM for Research on Code Generation with World Models}, 
      author={FAIR CodeGen team and Jade Copet and Quentin Carbonneaux and Gal Cohen and Jonas Gehring and Jacob Kahn and Jannik Kossen and Felix Kreuk and Emily McMilin and Michel Meyer and Yuxiang Wei and David Zhang and Kunhao Zheng and Jordi Armengol-Estapé and Pedram Bashiri and Maximilian Beck and Pierre Chambon and Abhishek Charnalia and Chris Cummins and Juliette Decugis and Zacharias V. Fisches and François Fleuret and Fabian Gloeckle and Alex Gu and Michael Hassid and Daniel Haziza and Badr Youbi Idrissi and Christian Keller and Rahul Kindi and Hugh Leather and Gallil Maimon and Aram Markosyan and Francisco Massa and Pierre-Emmanuel Mazaré and Vegard Mella and Naila Murray and Keyur Muzumdar and Peter O'Hearn and Matteo Pagliardini and Dmitrii Pedchenko and Tal Remez and Volker Seeker and Marco Selvi and Oren Sultan and Sida Wang and Luca Wehrstedt and Ori Yoran and Lingming Zhang and Taco Cohen and Yossi Adi and Gabriel Synnaeve},
      year={2025},
      eprint={2510.02387},
      archivePrefix={arXiv},
      primaryClass={cs.SE},
      url={https://arxiv.org/abs/2510.02387}, 
}

@inproceedings{Achiam2023GPT4TR,
  title={GPT-4 Technical Report},
  author={OpenAI Josh Achiam and Steven Adler and Sandhini Agarwal and Lama Ahmad and Ilge Akkaya and Florencia Leoni Aleman and Diogo Almeida and Janko Altenschmidt and Sam Altman and Shyamal Anadkat and Red Avila and Igor Babuschkin and Suchir Balaji and Valerie Balcom and Paul Baltescu and Haim-ing Bao and Mo Bavarian and Jeff Belgum and Irwan Bello and Jake Berdine and Gabriel Bernadett-Shapiro and Christopher Berner and Lenny Bogdonoff and Oleg Boiko and Made-laine Boyd and Anna-Luisa Brakman and Greg Brockman and Tim Brooks and Miles Brundage and Kevin Button and Trevor Cai and Rosie Campbell and Andrew Cann and Brittany Carey and Chelsea Carlson and Rory Carmichael and Brooke Chan and Che Chang and Fotis Chantzis and Derek Chen and Sully Chen and Ruby Chen and Jason Chen and Mark Chen and Benjamin Chess and Chester Cho and Casey Chu and Hyung Won Chung and Dave Cummings and Jeremiah Currier and Yunxing Dai and Cory Decareaux and Thomas Degry and Noah Deutsch and Damien Deville and Arka Dhar and David Dohan and Steve Dowling and Sheila Dunning and Adrien Ecoffet and Atty Eleti and Tyna Eloundou and David Farhi and Liam Fedus and Niko Felix and Sim'on Posada Fishman and Juston Forte and Is-abella Fulford and Leo Gao and Elie Georges and Christian Gibson and Vik Goel and Tarun Gogineni and Gabriel Goh and Raphael Gontijo-Lopes and Jonathan Gordon and Morgan Grafstein and Scott Gray and Ryan Greene and Joshua Gross and Shixiang Shane Gu and Yufei Guo and Chris Hallacy and Jesse Han and Jeff Harris and Yuchen He and Mike Heaton and Johannes Heidecke and Chris Hesse and Alan Hickey and Wade Hickey and Peter Hoeschele and Brandon Houghton and Kenny Hsu and Shengli Hu and Xin Hu and Joost Huizinga and Shantanu Jain and Shawn Jain and Joanne Jang and Angela Jiang and Roger Jiang and Haozhun Jin and Denny Jin and Shino Jomoto and Billie Jonn and Heewoo Jun and Tomer Kaftan and Lukasz Kaiser and Ali Kamali and Ingmar Kanitscheider and Nitish Shirish Keskar and Tabarak Khan and Logan Kilpatrick and Jong Wook Kim and Christina Kim and Yongjik Kim and Hendrik Kirchner and Jamie Ryan Kiros and Matthew Knight and Daniel Kokotajlo and Lukasz Kondraciuk and Andrew Kondrich and Aris Konstantinidis and Kyle Kosic and Gretchen Krueger and Vishal Kuo and Michael Lampe and Ikai Lan and Teddy Lee and Jan Leike and Jade Leung and Daniel Levy and Chak Li and Rachel Lim and Molly Lin and Stephanie Lin and Ma-teusz Litwin and Theresa Lopez and Ryan Lowe and Patricia Lue and Anna Makanju and Kim Malfacini and Sam Manning and Todor Markov and Yaniv Markovski and Bianca Martin and Katie Mayer and Andrew Mayne and Bob McGrew and Scott Mayer McKinney and Christine McLeavey and Paul McMillan and Jake McNeil and David Medina and Aalok Mehta and Jacob Menick and Luke Metz and An-drey Mishchenko and Pamela Mishkin and Vinnie Monaco and Evan Morikawa and Daniel P. Mossing and Tong Mu and Mira Murati and Oleg Murk and David M'ely and Ashvin Nair and Reiichiro Nakano and Rajeev Nayak and Arvind Neelakantan and Richard Ngo and Hyeonwoo Noh and Ouyang Long and Cullen O'Keefe and Jakub W. Pachocki and Alex Paino and Joe Palermo and Ashley Pantuliano and Giambattista Parascandolo and Joel Parish and Emy Parparita and Alexandre Passos and Mikhail Pavlov and Andrew Peng and Adam Perelman and Filipe de Avila Belbute Peres and Michael Petrov and Henrique Pond{\'e} de Oliveira Pinto and Michael Pokorny and Michelle Pokrass and Vitchyr H. Pong and Tolly Powell and Alethea Power and Boris Power and Elizabeth Proehl and Raul Puri and Alec Radford and Jack W. Rae and Aditya Ramesh and Cameron Raymond and Francis Real and Kendra Rimbach and Carl Ross and Bob Rotsted and Henri Roussez and Nick Ryder and Mario D. Saltarelli and Ted Sanders and Shibani Santurkar and Girish Sastry and Heather Schmidt and David Schnurr and John Schulman and Daniel Selsam and Kyla Sheppard and Toki Sherbakov and Jessica Shieh and Sarah Shoker and Pranav Shyam and Szymon Sidor and Eric Sigler and Maddie Simens and Jordan Sitkin and Katarina Slama and Ian Sohl and Benjamin Sokolowsky and Yang Song and Natalie Staudacher and Felipe Petroski Such and Natalie Summers and Ilya Sutskever and Jie Tang and Nikolas A. Tezak and Madeleine Thompson and Phil Tillet and Amin Tootoonchian and Elizabeth Tseng and Preston Tuggle and Nick Turley and Jerry Tworek and Juan Felipe Cer'on Uribe and Andrea Vallone and Arun Vijayvergiya and Chelsea Voss and Carroll L. Wainwright and Justin Jay Wang and Alvin Wang and Ben Wang and Jonathan Ward and Jason Wei and CJ Weinmann and Akila Welihinda and Peter Welinder and Jiayi Weng and Lilian Weng and Matt Wiethoff and Dave Willner and Clemens Winter and Samuel Wolrich and Hannah Wong and Lauren Workman and Sherwin Wu and Jeff Wu and Michael Wu and Kai Xiao and Tao Xu and Sarah Yoo and Kevin Yu and Qim-ing Yuan and Wojciech Zaremba and Rowan Zellers and Chong Zhang and Marvin Zhang and Shengjia Zhao and Tianhao Zheng and Juntang Zhuang and William Zhuk and Barret Zoph},
  year={2023},
  url={https://api.semanticscholar.org/CorpusID:257532815}
}

@article{Jiang2023Mistral7,
  title={Mistral 7B},
  author={Albert Qiaochu Jiang and Alexandre Sablayrolles and Arthur Mensch and Chris Bamford and Devendra Singh Chaplot and Diego de Las Casas and Florian Bressand and Gianna Lengyel and Guillaume Lample and Lucile Saulnier and L'elio Renard Lavaud and Marie-Anne Lachaux and Pierre Stock and Teven Le Scao and Thibaut Lavril and Thomas Wang and Timoth{\'e}e Lacroix and William El Sayed},
  journal={ArXiv},
  year={2023},
  volume={abs/2310.06825},
  url={https://api.semanticscholar.org/CorpusID:263830494}
}

@article{Touvron2023LLaMAOA,
  title={LLaMA: Open and Efficient Foundation Language Models},
  author={Hugo Touvron and Thibaut Lavril and Gautier Izacard and Xavier Martinet and Marie-Anne Lachaux and Timoth{\'e}e Lacroix and Baptiste Rozi{\`e}re and Naman Goyal and Eric Hambro and Faisal Azhar and Aur'elien Rodriguez and Armand Joulin and Edouard Grave and Guillaume Lample},
  journal={ArXiv},
  year={2023},
  volume={abs/2302.13971},
  url={https://api.semanticscholar.org/CorpusID:257219404}
}

@misc{hashemi2025transformersenumerativegeometry,
      title={Can Transformers Do Enumerative Geometry?}, 
      author={Baran Hashemi and Roderic G. Corominas and Alessandro Giacchetto},
      year={2025},
      eprint={2408.14915},
      archivePrefix={arXiv},
      primaryClass={cs.LG},
      url={https://arxiv.org/abs/2408.14915}, 
}

@inbook{Topcom,
author = {Rambau, Jörg},
title = {TOPCOM: Triangulations of point configurations and oriented matroids},
year={2002},
booktitle = {Mathematical Software},
chapter = {},
pages = {330-340},
doi = {10.1142/9789812777171_0035}
}

@inproceedings{Vaswani:2017lxt,
    author = "Vaswani, Ashish and Shazeer, Noam and Parmar, Niki and Uszkoreit, Jakob and Jones, Llion and Gomez, Aidan N. and Kaiser, Lukasz and Polosukhin, Illia",
    title = "{Attention Is All You Need}",
    booktitle = "{31st International Conference on Neural Information Processing Systems}",
    eprint = "1706.03762",
    archivePrefix = "arXiv",
    primaryClass = "cs.CL",
    month = "6",
    year = "2017"
}

@article{Adam2014,
author = {Kingma, Diederik and Ba, Jimmy},
year = {2014},
month = {12},
pages = {},
title = {Adam: A Method for Stochastic Optimization},
journal = {International Conference on Learning Representations}
}

@inproceedings{lee2019set,
  title={Set transformer: A framework for attention-based permutation-invariant neural networks},
  author={Lee, Juho and Lee, Yoonho and Kim, Jungtaek and Kosiorek, Adam and Choi, Seungjin and Teh, Yee Whye},
  booktitle={International conference on machine learning},
  pages={3744--3753},
  year={2019},
  organization={PMLR}
}

@article{Kreuzer:2000xy,
    author = "Kreuzer, Maximilian and Skarke, Harald",
    title = "{Complete classification of reflexive polyhedra in four-dimensions}",
    eprint = "hep-th/0002240",
    archivePrefix = "arXiv",
    reportNumber = "HUB-EP-00-13, TUW-00-07",
    doi = "10.4310/ATMP.2000.v4.n6.a2",
    journal = "Adv. Theor. Math. Phys.",
    volume = "4",
    pages = "1209--1230",
    year = "2000"
}

@article{Demirtas:2022hqf,
    author = "Demirtas, Mehmet and Rios-Tascon, Andres and McAllister, Liam",
    title = "{CYTools: A Software Package for Analyzing Calabi-Yau Manifolds}",
    eprint = "2211.03823",
    archivePrefix = "arXiv",
    primaryClass = "hep-th",
    month = "11",
    year = "2022"
}

@article{Batyrev:1993oya,
    author = "Batyrev, Victor V.",
    title = "{Dual polyhedra and mirror symmetry for Calabi-Yau hypersurfaces in toric varieties}",
    eprint = "alg-geom/9310003",
    archivePrefix = "arXiv",
    journal = "J. Alg. Geom.",
    volume = "3",
    pages = "493--545",
    year = "1994"
}

@article{Kreuzer:1998vb,
    author = "Kreuzer, Maximilian and Skarke, Harald",
    title = "{Classification of reflexive polyhedra in three-dimensions}",
    eprint = "hep-th/9805190",
    archivePrefix = "arXiv",
    reportNumber = "UTTG-07-98, TUW-98-13",
    doi = "10.4310/ATMP.1998.v2.n4.a5",
    journal = "Adv. Theor. Math. Phys.",
    volume = "2",
    pages = "853--871",
    year = "1998"
}

@article{Demirtas:2020dbm,
    author = "Demirtas, Mehmet and McAllister, Liam and Rios-Tascon, Andres",
    title = "{Bounding the Kreuzer-Skarke Landscape}",
    eprint = "2008.01730",
    archivePrefix = "arXiv",
    primaryClass = "hep-th",
    doi = "10.1002/prop.202000086",
    journal = "Fortsch. Phys.",
    volume = "68",
    pages = "2000086",
    year = "2020"
}

@article{MacFadden:2024him,
    author = "MacFadden, Nate and Schachner, Andreas and Sheridan, Elijah",
    title = "{The DNA of Calabi-Yau Hypersurfaces}",
    eprint = "2405.08871",
    archivePrefix = "arXiv",
    primaryClass = "hep-th",
    reportNumber = "LMU-ASC 06/24",
    month = "5",
    year = "2024"
}

@article{Berglund:2024reu,
    author = "Berglund, Per and Butbaia, Giorgi and He, Yang-Hui and Heyes, Elli and Hirst, Edward and Jejjala, Vishnu",
    title = "{Generating triangulations and fibrations with reinforcement learning}",
    eprint = "2405.21017",
    archivePrefix = "arXiv",
    primaryClass = "hep-th",
    reportNumber = "QMUL-PH-24-10",
    doi = "10.1016/j.physletb.2024.139158",
    journal = "Phys. Lett. B",
    volume = "860",
    pages = "139158",
    year = "2025"
}

@book{10.5555/1952022,
author = {De Loera, Jesus A. and Rambau, Jorg and Santos, Francisco},
title = {Triangulations: Structures for Algorithms and Applications},
year = {2010},
isbn = {3642129706},
publisher = {Springer Publishing Company, Incorporated},
edition = {1st}
}

\end{document}